\documentclass[11pt]{article}

\usepackage{jheppub}

\usepackage{slashbox,latexsym,amssymb,amsfonts,amsmath,slashed,amsthm,bm}
\usepackage{graphicx}
\usepackage{mathrsfs}  
\newcommand{\f}[2]{\frac{#1}{#2}}
\newcommand{\tf}[2]{{\textstyle\f{#1}{#2}}}

\newcommand{\la}{\langle}
\newcommand{\lla}{\la\!\la}
\newcommand{\ra}{\rangle}
\newcommand{\rra}{\ra\!\ra}

\newcommand{\slaD}{{\slashed D}}

\newcommand{\de}{\partial}

\renewcommand{\Re}{{\rm Re}\,}
\renewcommand{\Im}{{\rm Im}\,}

\newcommand{\tr}{{\rm tr}\,}
\newcommand{\Tr}{{\rm Tr}\,}

\newcommand{\Ds}{{\mathscr D}}

\newcommand{\Lc}{{\mathscr L}}

\newcommand{\Vc}{V}

\newcommand{\Ac}{A}

\newcommand{\Poc}{P}
\newcommand{\Mc}{{\mathcal M}}

\newcommand{\Sc}{S}

\newcommand{\Oc}{{\mathcal O}}

\newcommand{\Gsc}{{\mathcal G}}
\newcommand{\Gc}{G}
\newcommand{\Po}{{\mathcal R}}

\newcommand{\Ag}{B}

\newcommand{\spcorr}{C}

\newcommand{\Xsc}{{\mathscr X}}
\newcommand{\Rco}{{\rm P}}
\newcommand{\Bco}{{\rm B}}
\newcommand{\Gco}{{\rm Q}}
\newcommand{\vzero}{\vec{0}}

\makeatletter
\renewcommand{\tableofcontents}{
  \addtolength{\baselineskip}{0.1pt}
  \@starttoc{toc}
  \addtolength{\baselineskip}{-0.1pt}
}
\makeatother

\title{Localisation of Dirac modes in gauge theories and Goldstone's
  theorem at finite temperature}

\author[a]{Matteo Giordano}

\affiliation[a]{ELTE E\"otv\"os Lor\'and University, Institute for
  Theoretical Physics,\\ P\'azm\'any P\'eter s\'et\'any 1/A, H-1117,
  Budapest, Hungary}

\emailAdd{giordano@bodri.elte.hu}

\abstract{I discuss the possible effects of a finite density of
  localised near-zero Dirac modes in the chiral limit of gauge
  theories with $N_f$ degenerate fermions. I focus in particular on
  the fate of the massless quasi-particle excitations predicted by the
  finite-temperature version of Goldstone's theorem, for which I
  provide an alternative and generalised proof based on a Euclidean
  ${\rm SU}(N_f)_A$ Ward-Takahashi identity. I show that localised
  near-zero modes can lead to a divergent pseudoscalar-pseudoscalar
  correlator that modifies this identity in the chiral limit. As a
  consequence, massless quasi-particle excitations can disappear from
  the spectrum of the theory in spite of a non-zero chiral condensate.
  Three different scenarios are possible, depending on the detailed
  behaviour in the chiral limit of the ratio of the mobility edge and
  the fermion mass, which I prove to be a renormalisation-group
  invariant quantity.}


\begin{document}
\maketitle

\section{Introduction}
\label{sec:intro}

The low-energy physics of QCD at zero temperature is largely
determined by two non-perturbative phenomena, namely spontaneous
chiral symmetry breaking and confinement. The spontaneous breaking of
chiral symmetry in the limit of massless quarks and the associated
appearance of massless Goldstone bosons explain the lightness of pions
for physical quark masses and their low-energy dynamics.  The origin
of confinement can be seen instead in the opposite infinite-mass,
``quenched'' (pure gauge) limit, where a linearly rising
quark-antiquark potential forces quarks to be confined inside hadrons.

At finite temperature, confinement in SU(3) pure gauge theory is
signalled by a divergent quark free energy, inferred from the
vanishing of the Polyakov loop expectation value, which is in turn a
consequence of the centre symmetry of the theory being unbroken. For
finite quark masses this symmetry is explicitly broken but only
mildly, and confinement persists at low temperatures, signalled by a
disordered Polyakov loop in typical gauge configurations, resulting in
a small expectation value of the Polyakov loop.  At low temperatures
also the effects of the spontaneous breaking of chiral symmetry in the
massless limit are still clearly present.

At higher temperatures QCD undergoes a rapid but analytic crossover to
a phase where chiral symmetry is approximately restored, and centre
symmetry is spontaneously broken (on top of the explicit breaking
mentioned above) with the Polyakov loop getting ordered, and quarks
and gluons are liberated in the quark-gluon
plasma~\cite{Borsanyi:2010bp,Bazavov:2016uvm}. Despite the very
different origin of chiral symmetry breaking and confinement, both
chiral and confining properties change dramatically at the
transition. A similar situation is found quite generally in gauge
theories at finite temperature~\cite{Boyd:1996bx,Damgaard:1998yv,
  Karsch:1998qj,Engels:2005te,Karsch:2001nf,deForcrand:2003vyj,
  deForcrand:2008vr,Bergner:2019dim}, with deconfinement always
leading to a ``more chirally symmetric'' system, but the reasons for
this behaviour are still not fully understood.

The QCD crossover is characterised also by another drastic change,
that of the localisation properties of the low-lying modes of the
Euclidean Dirac operator. In fact, while delocalised in the
low-temperature phase, these modes become spatially localised on the
scale of the inverse temperature in the high-temperature
phase~\cite{Gockeler:2001hr,Gattringer:2001ia,
  GarciaGarcia:2005vj,GarciaGarcia:2006gr,Gavai:2008xe,
  Kovacs:2009zj,Bruckmann:2011cc,Kovacs:2012zq,Cossu:2016scb,
  Holicki:2018sms}. More precisely, above the transition temperature
the low modes are localised up to a critical point, $\lambda_c$, in
the spectrum, known as ``mobility edge'', beyond which they are again
delocalised.\footnote{Localised modes are most likely found also in
  the deep ultraviolet region of the spectrum (see footnote
  \ref{foot:UVloc}), which, however, has no physical relevance.}  For
a recent comprehensive review of localisation in gauge theories, the
interested reader can consult Ref.~\cite{Giordano:2021qav}.

The relation between localisation and deconfinement has been
extensively studied in QCD and QCD-like gauge theories in recent
years. These studies have shown that these two phenomena are indeed
intimately connected, with localised modes appearing exactly at the
critical point when the phase transition is a genuine (i.e.,
non-analytic) transition~\cite{Kovacs:2017uiz,Vig:2020pgq,
  Giordano:2019pvc,Giordano:2016nuu,Giordano:2016vhx,Kovacs:2010wx,
  Bruckmann:2017ywh,Bonati:2020lal,Baranka:2021san,Cardinali:2021fpu}. A
qualitative understanding of this connection is provided by the
``sea/islands'' picture of localisation~\cite{Bruckmann:2011cc,
  Giordano:2015vla,
  Giordano:2016cjs,Giordano:2016vhx,Giordano:2021qav}, according to
which the ``islands'' of fluctuations in the ``sea'' of ordered
Polyakov loops in the deconfined phase provide ``energetically''
favourable regions where the low Dirac eigenmodes can localise.  This
is possible since the ordering of the Polyakov loop around 1 opens a
pseudogap in the spectrum, which is then populated by a low but finite
density of localised modes.  Numerical evidence for the sea/islands
picture has been provided in
Refs.~\cite{Bruckmann:2011cc,Cossu:2016scb, Holicki:2018sms,
  Baranka:2021san}. Remarkably, the existence of an ordered phase is
all that is required for this mechanism to be at work, so that one
expects localisation of low Dirac modes to appear in the deconfined
phase of a generic gauge theory. This has been so far confirmed in a
variety of different gauge theories~\cite{Kovacs:2017uiz,
  Vig:2020pgq,Giordano:2019pvc, Giordano:2016nuu,Giordano:2016vhx,
  Kovacs:2010wx,Bruckmann:2017ywh,
  Bonati:2020lal,Baranka:2021san,Cardinali:2021fpu}, supporting the
existence of a very close relation between deconfinement and the
localisation properties of low Dirac modes.

The connection with chiral symmetry breaking has received instead less
attention in recent years, despite being the original motivation for
the study of localisation in gauge theories at finite
temperature~\cite{GarciaGarcia:2005vj,GarciaGarcia:2006gr}.  As is
well known, chiral symmetry breaking is related to the accumulation of
near-zero Dirac modes~\cite{Banks:1979yr}. In the ``disordered medium
scenario''~\cite{Diakonov:1995ea} this is explained in terms of the
mixing of topological zero modes of the Dirac operator caused by the
overlapping of instantons and anti-instantons. While the
``unperturbed'' zero modes are localised at finite temperature, they
become delocalised after mixing, with their eigenvalues broadening
into a finite band around zero. It has been observed, however, that in
the high-temperature phase the low modes of topological origin are not
sufficient to quantitatively explain the amount of localised
modes~\cite{Bruckmann:2011cc,Vig:2021oyt}. The disordered medium
scenario cannot therefore fully account for localised low modes, as is
made clear by the fact that these are found also in theories without
instantons~\cite{Giordano:2019pvc,Baranka:2021san}.  Nonetheless,
topological modes are very likely to play an important role,
complementing the sea/islands picture discussed above (see
Ref.~\cite{Vig:2021oyt}).

It is clear, in any case, that the low-lying eigenvalues of the Dirac
operator and the corresponding eigenvectors are sensitive both to
confinement and chiral symmetry breaking.  The study of localisation
could then lead not only to a better understanding of these two
phenomena individually, but also to clarify their connection, with
localisation possibly providing the mechanism through which
deconfinement improves the chiral symmetry properties of gauge systems
with fermions.

The appearance of localised modes in the spectrum of the Dirac
operator becomes somewhat less surprising if one notices its formal
analogy with the Hamiltonian of a disordered system. Such systems are
known to display eigenmode localisation since the seminal work of
Anderson~\cite{Anderson:1958vr}, and have been intensely studied in
the condensed-matter community for more than sixty years (see
Refs.~\cite{thouless1974electrons,lee1985disordered,
  kramer1993localization,Evers:2008zz,anderson50,manybody} for a
review). In fact, $-i\slaD$ can be interpreted exactly as the
Hamiltonian of a quantum system in the background of random gauge
fields, fluctuating according to the distribution determined by the
path-integral integration measure. With this insight, the critical
features of localisation observed at the mobility
edge~\cite{Giordano:2013taa, Nishigaki:2013uya,Giordano:2014qna,
  Ujfalusi:2015nha} are understood as a consequence of universality,
and of suitable symmetry considerations.  On the other hand, the
appearance of localised modes at the band centre (i.e., near zero)
only in the high-temperature phase is surprising, calling for an
explanation that is qualitatively provided by the sea/islands picture
mentioned above.

The formal analogy with disordered systems, however, does not
translate directly into an analogy between physical phenomena observed
on its gauge theory side and on its condensed matter side. The reason
for this is that, while energy levels and eigenvectors of a
Hamiltonian have a direct physical meaning in condensed matter
systems, it is not so for the Dirac eigenmodes in gauge theories,
where observables are obtained only after summing over all the
modes. The properties of individual points, and perhaps even of
regions of the Dirac spectrum, are then difficult to connect to
phenomenology. A notable exception to this state of affairs is the
chiral limit of massless quarks, where only near-zero modes have
physical relevance. This is famously exemplified by the Banks-Casher
relation~\cite{Banks:1979yr}, stating that the chiral condensate is
proportional to the density of near-zero Dirac modes in the chiral
limit. It is then possible that in this limit the localisation
properties of near-zero modes have clear observable consequences.

The investigation of this issue is made more pressing by the known
consequences of localisation for the Goldstone excitations associated
with the spontaneous breaking of a continuous
symmetry~\cite{Goldstone:1962es} also at finite
temperature~\cite{Lange:1965zz,Kastler:1966wdu,Swieca:1966wna,Morchio:1987wd,
  Strocchi:2008gsa}. It has been known for a long time that
localisation can lead to the disappearance of Goldstone modes in
non-relativistic disordered systems~\cite{McKane:1980fs}. This
phenomenon has been rediscovered more recently in the context of
relativistic lattice gauge theories at zero
temperature~\cite{Golterman:2003qe}, although for an unphysical
system, namely quenched SU(3) gauge theory with Wilson fermions, where
it explains the disappearance of Goldstone bosons in the supercritical
region outside the Aoki phase~\cite{Aoki:1983qi}.  Localisation and
the fate of quasi-particle Goldstone excitations in finite-temperature
gauge theories have been studied only recently in
Refs.~\cite{giordano_GT_lett,Giordano:2021nat}, reaching similar
conclusions: the presence of a finite density of localised near-zero
modes possibly leads to the disappearance of Goldstone excitations
from the spectrum of the theory.

In the absence of concrete models where localised near-zero modes are
explicitly shown to be sufficiently dense, the results of
Refs.~\cite{giordano_GT_lett,Giordano:2021nat} would probably be only
of academic interest, closing a loophole in the proofs of Goldstone's
theorem at zero and finite temperature. There are, however,
interesting results that could make them more than simply a
curiosity. The most intriguing one is the peak of localised near-zero
modes in QCD right above the pseudocritical temperature, observed on
the lattice studying overlap spectra in the background of HISQ
configurations for near-physical quark
masses~\cite{Dick:2015twa}. While localisation properties have not
been studied further, this peak has been observed to persist also at
lower-than-physical light-quark
masses~\cite{Ding:2020xlj,Kaczmarek:2021ser,Ding:2021gdy,Ding:2021jtn}.
This peak is usually ascribed to the topological modes mentioned
above, and it is argued~\cite{HotQCD:2012vvd,Dick:2015twa,
  Ding:2020xlj,Kaczmarek:2021ser} that it should shrink in the chiral
limit as topological excitations become a non-interacting gas, so that
its effects disappear except for what concerns the anomalous
${\rm U}(1)_A$ symmetry. However, it is not clear what mechanism
should lead to instantons and anti-instantons forming a free gas in
the chiral limit, and the evidence for the suggested behaviour of the
peak is not conclusive. While there is no conclusive evidence for the
peak surviving as a finite peak either, the possibility should not be
excluded at this stage that a finite density of localised near-zero
modes is found in the chiral limit.

Another interesting finding is the presence of two separate phase
transitions in SU(3) gauge theory with $N_f=2$ flavours of massless
adjoint quarks~\cite{Karsch:1998qj,Engels:2005te}. In this theory a
deconfining transition at temperature $T_{\rm dec}$ separates a
low-temperature, confining phase where chiral symmetry is broken by a
non-zero chiral condensate, and an intermediate, deconfined phase
where chiral symmetry is still broken but the chiral condensate is
reduced. A second transition at a higher temperature
$T_{\chi}>T_{\rm dec}$ separates this phase from the high-temperature,
deconfined and chirally-restored phase. In the intermediate phase one
surely finds a finite density of near-zero modes via the Banks-Casher
relation; since this phase is deconfined, the sea/islands picture
suggests that these modes will be localised, but no more or less
direct evidence of this exists. Moreover, the results of
Refs.~\cite{Karsch:1998qj,Engels:2005te} are consistent with the
presence of massless Goldstone excitations. It is possible that, as a
consequence of the relatively small volumes employed in those studies,
localisation and its consequences could not manifest yet, and so
larger volumes would be needed to make conclusive statements.
However, if confirmed in larger volumes, their results do not
contradict the analysis of Refs.~\cite{giordano_GT_lett,
  Giordano:2021nat}, and a detailed study of correlation functions
would be needed to test quantitatively the scenarios proposed there.

Although the basic approach is very similar to that of
Refs.~\cite{McKane:1980fs,Golterman:2003qe}, the argument of
Refs.~\cite{giordano_GT_lett,Giordano:2021nat} requires to deal with a
certain number of technical complications, mostly originating in the
loss of full O(4) invariance at finite temperature. While these
complications can be overcome, they were discussed only briefly in
Refs.~\cite{McKane:1980fs,Golterman:2003qe}, not to obscure the main
points. In this paper I present the argument in full depth, providing
details on the various aspects of the calculation. Before being able
to discuss the fate of Goldstone quasi-particles, a few intermediate
results need to be derived, that I believe are of interest in their
own right. These include:
\begin{itemize}
\item a ``Euclidean'' derivation of Goldstone's theorem at finite
  temperature, and a generalisation thereof, in the case of broken
  non-singlet axial flavour symmetry, based on the corresponding
  Ward-Takahashi identities;
\item the possible appearance in the chiral limit of a $1/m$
  divergence in the pseudoscalar-pseudoscalar correlator, where $m$ is
  the quark mass, in the presence of a finite density of localised
  near-zero modes;
\item the renormalisation of the spectral correlators appearing in the
  mode decomposition of the pseudoscalar-pseudoscalar correlator, and
  the related proof that the ratio $\lambda_c/m$ between the mobility
  edge and the quark mass is a renormalisation-group invariant
  quantity.
\end{itemize}
After a brief review of finite-temperature quantum field theory in
Section \ref{sec:FT}, and of gauge theories in Section \ref{sec:WI},
the alternative proof of Goldstone's theorem at finite temperature is
discussed in Section \ref{sec:GTFT}. The study of the
pseudoscalar-pseudoscalar correlator is reported in Section
\ref{sec:pseudochi}, including the renormalisation of the spectral
correlators and of the mobility edge, and the fate of Goldstone
excitations in the presence of localised modes.  Conclusions and
discussion of future studies are found in Section \ref{sec:concl}. To
improve readability, most of the technical details are relegated to
Appendices \ref{sec:anrealcorr} to \ref{sec:app_bound}, which include:
details on analyticity and reality properties of Euclidean correlation
functions (Appendix \ref{sec:anrealcorr}); derivation of non-singlet
axial Ward-Takahashi identities (Appendix \ref{sec:app_WTid}) and
their renormalisation (Appendix \ref{sec:app_renorm}); another
``Euclidean'' proof of Goldstone's theorem at finite temperature in
coordinate space (Appendix \ref{sec:gt_coord}); details about the
pseudoscalar-pseudoscalar correlator calculation, in particular
estimates of the contribution of exponentially localised modes
(Appendix \ref{sec:expoenv}), the study of the chiral limit (Appendix
\ref{sec:int_genarg}) and the renormalisation of the corresponding
spectral correlators (Appendix \ref{sec:lc_ren}), and the discussion
of a bound on their large-distance behaviour (Appendix
\ref{sec:app_bound}).

\section{Finite-temperature quantum field theory}
\label{sec:FT}

In this Section I review a few relevant aspects of quantum field
theory at finite temperature, mostly to set the notation.  See, e.g.,
Refs.~\cite{Kapusta:2006pm,Laine:2016hma,fulling1987temperature,
  Bros:1996mw,Cuniberti:2001hm,Meyer:2010ii,Meyer:2011gj,Strocchi:2008gsa}
for further details. The expectation value of an observable $O$ for a
system in thermal equilibrium at temperature $T=1/\beta$ is obtained
as follows from the density matrix of the canonical ensemble,
\begin{equation} 
  \label{eq:GNS}
  \lla \hat{O} \rra_\beta =  \lim_{V\to\infty} \lla \hat{O} \rra_{\beta,V} \equiv
  \lim_{V\to\infty}  \f{\Tr e^{-\beta \hat{H}_V} \hat{O}}{\Tr e^{-\beta \hat{H}_V} }
  \,, 
\end{equation}
where $\hat{H}_V$ is the finite-volume Hamiltonian, and the volume $V$
of the system is eventually sent to infinity in the thermodynamic
limit.  Here a caret denotes operators acting on the
(zero-temperature) Hilbert space of the system, and ${\rm Tr}$ the
trace over this space.

For (possibly composite) local bosonic field operators
$\hat{\phi}_{1,2}(x)$, $x=(t,\vec{x})$, one defines the thermal
(real-time) two-point correlation functions
\begin{equation}
  \label{eq:corrfunc}
  \begin{aligned}
    \Gc^{(+)}_{\phi_1 \phi_2}(t,\vec{x}) &\equiv \lla
    \hat{\phi}_1(t,\vec{x})\hat{\phi}_2(0) \rra_{\beta} \equiv
    \lim_{V\to\infty} \lla e^{it \hat{H}_V}\hat{\phi}_1(0,\vec{x})
    e^{-it \hat{H}_V} \hat{\phi}_2(0)\rra_{\beta,V}
    \,,\\
    \Gc^{(-)}_{\phi_1 \phi_2}(t,\vec{x}) &\equiv \lla
    \hat{\phi}_2(0)\hat{\phi}_1(t,\vec{x}) \rra_{\beta} \equiv
    \lim_{V\to\infty} \lla \hat{\phi}_2(0) e^{it
      \hat{H}_V}\hat{\phi}_1(0,\vec{x}) e^{-it \hat{H}_V}
    \rra_{\beta,V} \,.
  \end{aligned}
\end{equation}
Throughout this paper, an arrow denotes collectively the three spatial
components of a four-vector. Here and everywhere else in this paper,
the temporal evolution of $\hat{\phi}_{1,2}(t,\vec{x})$ in the
infinite-volume limit (also for complex time argument) is understood
in the sense of the weak limit of $H_V$ implied by
Eq.~\eqref{eq:corrfunc}. Under suitable convergence conditions on the
infinite-volume limit of the temporal evolution, the correlation
functions Eq.~\eqref{eq:corrfunc} satisfy~\cite{Haag:1967sg} the KMS
condition~\cite{Kubo:1957mj,Martin:1959jp}. Together with the
relativistic locality condition,
$[\hat{\phi}_{i}(x),\hat{\phi}_{j}(y)]=0$ for spacelike Minkowski
separation $(x-y)^2<0$, this implies that
$\Gc^{(\pm)}_{\phi_1 \phi_2}(t,\vec{x})$ are the boundary values of an
analytic function $\Gc_{\phi_1 \phi_2}(z,\vec{x})$, analytic in the
cut complex plane
${\cal D} \equiv \mathbb{C}\backslash\{z ~|~|\Re z|\ge |\vec{x}|,\,\Im
z = n\beta,\,n\in\mathbb{Z}\}$, and furthermore periodic in the
imaginary $z$ direction with period $\beta$,
$\Gc_{\phi_1 \phi_2}(z+in\beta,\vec{x})=\Gc_{\phi_1
  \phi_2}(z,\vec{x})$ (see Refs.~\cite{fulling1987temperature,
  Bros:1996mw,Cuniberti:2001hm,Strocchi:2008gsa}). The correlation
functions $\Gc^{(\pm)}_{\phi_1 \phi_2}(t,\vec{x})$ for
$t\in\mathbb{R}$ are then recovered as
\begin{equation}
  \label{eq:corrfunc4}
  \begin{aligned}
    \Gc^{(+)}_{\phi_1 \phi_2}(t,\vec{x})& = \Gc_{\phi_1
      \phi_2}(t-i\epsilon,\vec{x})
    \,,
    \\
    \Gc^{(-)}_{\phi_1 \phi_2}(t,\vec{x})& = \Gc_{\phi_1
      \phi_2}(t+i\epsilon,\vec{x}) =\Gc_{\phi_1
      \phi_2}(t-i(\beta-\epsilon),\vec{x})
    \,,
  \end{aligned}
\end{equation}
where the limit $\epsilon\to 0^+$ at the end of the calculation is
understood.

In particular, $\Gc^{(\pm)}_{\phi_1 \phi_2}$ can be obtained by
analytic continuation from the Euclidean correlation function
$\Gsc_{\phi_1 \phi_2}(t,\vec{x})$,
\begin{equation}
  \label{eq:corrfunc9}
  \begin{aligned}
    \Gsc_{\phi_1 \phi_2}(t,\vec{x}) &\equiv \lla
    \theta(t)\hat{\phi}_1(-it,\vec{x})\hat{\phi}_2(0) + \theta(-t)
    \hat{\phi}_2(0)\hat{\phi}_1(-it,\vec{x}) \rra_\beta \\ &\equiv
    \lla T\{\hat{\phi}_{E1}(t,\vec{x})\hat{\phi}_{E2}(0)\} \rra_\beta
    \,,
  \end{aligned}
\end{equation}
where
$\hat{\phi}_{E1,2}(t,\vec{x})\equiv \hat{\phi}_{1,2}(-it,\vec{x})$,
and $T$ stands for time-ordering in Euclidean (imaginary) time
$t$. This function is the restriction to real $z$ of
$\Gsc_{\phi_1 \phi_2}(z,\vec{x})=\Gc_{\phi_1 \phi_2}(-iz,\vec{x})$,
analytic in the cut complex plane $i{\cal D}$, and periodic in the
real direction,
$\Gsc_{\phi_1 \phi_2}(z+n\beta,\vec{x})= \Gsc_{\phi_1
  \phi_2}(z,\vec{x})$.  One then has
\begin{equation}
  \label{eq:corrfunc6}
  \begin{aligned}
    \Gc^{(+)}_{\phi_1 \phi_2}(t,\vec{x})&= \Gsc_{\phi_1
      \phi_2}(\epsilon+it,\vec{x}) \,, &&& \Gc^{(-)}_{\phi_1
      \phi_2}(t,\vec{x})&= \Gsc_{\phi_1 \phi_2}(-\epsilon+it,\vec{x})
    \,.
  \end{aligned}
\end{equation}
In particular, the thermal expectation value of the commutator
$[\hat{\phi}_1(x),\hat{\phi}_2(0)]$ equals the discontinuity of
$\Gsc_{\phi_1 \phi_2}$ along the imaginary axis (corresponding to real
time),
\begin{equation}
  \label{eq:corrfunc7}
  \begin{aligned}
    \lla [\hat{\phi}_1(t,\vec{x}),\hat{\phi}_2(0)] \rra_{\beta} & =
    \Gc^{(+)}_{\phi_1 \phi_2}(t,\vec{x})-\Gc^{(-)}_{\phi_1
      \phi_2}(t,\vec{x}) = \Gsc_{\phi_1 \phi_2}(\epsilon+it,\vec{x})-
    \Gsc_{\phi_1 \phi_2}(-\epsilon+it,\vec{x}) \\ & = \Gsc_{\phi_1
      \phi_2}(\epsilon+it,\vec{x})- \Gsc_{\phi_1
      \phi_2}(\beta-\epsilon+it,\vec{x})\,,
  \end{aligned}
\end{equation}
where $t\in\mathbb{R}$ and the second line follows from periodicity.
As a further consequence of periodicity in Euclidean time,
$\Gsc_{\phi_1 \phi_2}(t,\vec{x})$ with $t\in\mathbb{R}$ can be written
as a mixed Fourier sum/Fourier transform as follows,
\begin{equation}
  \label{eq:corrfunc9bis}
  \Gsc_{\phi_1 \phi_2}(t,\vec{x})=\f{1}{\beta}\sum_n\int
  \f{d^3k}{(2\pi)^3}\,e^{-i(\omega_nt+\vec{k}\cdot\vec{x})}\tilde{\Gsc}_{\phi_1
    \phi_2}(\omega_n,\vec{k})\,,  
\end{equation}
where $\omega_n$ are the bosonic Matsubara frequencies,
$\omega_n=\f{2\pi n}{\beta}$, $n\in\mathbb{Z}$, and
\begin{equation}
  \label{eq:corrfunc9ter}
  \tilde{\Gsc}_{\phi_1 \phi_2}(\omega_n,\vec{k}) \equiv   \int_0^\beta dt\int
  d^3x\,e^{i(\omega_nt+\vec{k}\cdot\vec{x})}\Gsc_{\phi_1 \phi_2}(t,\vec{x})
  \equiv \int_\beta
  d^4x\,e^{i(\omega_nt+\vec{k}\cdot\vec{x})}\Gsc_{\phi_1 \phi_2}(t,\vec{x})
  \,,
\end{equation}
with the subscript $\beta$ denoting compactification of the temporal
direction.

A central role in this paper is played by the spectral function,
\begin{equation}
  \label{eq:spec_func_def1}
  \begin{aligned}
    \tilde{\rho}_{\phi_1\phi_2}(\omega,\vec{k}) &\equiv \int d^4x
    \,e^{i(\omega t -\vec{k}\cdot\vec{x})}
    \lla[\hat{\phi}_1(t,\vec{x}),\hat{\phi}_2(0)]\rra_\beta \\ &= \int
    d^4x \,e^{i(\omega t -\vec{k}\cdot\vec{x})}\left( \Gsc_{\phi_1
        \phi_2}(\epsilon+it,\vec{x})- \Gsc_{\phi_1
        \phi_2}(-\epsilon+it,\vec{x}) \right)\,,
  \end{aligned}
\end{equation}
and by the closely related retarded and advanced propagators,
\begin{equation}
  \label{eq:GT19}
  \begin{aligned}
    \tilde{r}_{\phi_1\phi_2}(\omega,\vec{k})&\equiv i\int d^4x
    \,e^{i(\omega t -\vec{k}\cdot\vec{x})}\theta(t)
    \lla[\hat{\phi}_1(t,\vec{x}),\hat{\phi}_2(0)]\rra_\beta\,,\\
    \tilde{a}_{\phi_1\phi_2}(\omega,\vec{k})&\equiv -i\int d^4x
    \,e^{i(\omega t -\vec{k}\cdot\vec{x})}\theta(-t)
    \lla[\hat{\phi}_1(t,\vec{x}),\hat{\phi}_2(0)]\rra_\beta\,.
  \end{aligned}
\end{equation}
These are the boundary values of analytic functions, analytic
respectively for $\Im \omega >0$ and $\Im \omega <0$, from which one
obtains the spectral function as follows,
\begin{equation}
  \label{eq:spec_func_def1_bis}
  \begin{aligned}
    \tilde{\rho}_{\phi_1\phi_2}(\omega,\vec{k}) =
    -i\left(\tilde{r}_{\phi_1\phi_2}(\omega+i\epsilon,\vec{k})
      -\tilde{a}_{\phi_1\phi_2}(\omega-i\epsilon,\vec{k})\right)\,.
  \end{aligned}
\end{equation}
More directly, one can use the analytic continuation relations between
the retarded and advanced propagators and the Fourier coefficients of
the Euclidean correlator,
$\tilde{\Gsc}_{\phi_1 \phi_2}(\omega_n,\vec{k})$, to reconstruct
$\tilde{r}_{\phi_1\phi_2}$ and $\tilde{a}_{\phi_1\phi_2}$, and so
$ \tilde{\rho}_{\phi_1\phi_2}$, by analytic interpolation in the sense
of Carlson's theorem, under reasonable hypotheses of moderate
asymptotic growth in the complex time
variable~\cite{Bros:1996mw,Cuniberti:2001hm}.  For $n\neq 0$ one
has~\cite{Bros:1996mw,Cuniberti:2001hm,Meyer:2010ii,Meyer:2011gj}
\begin{equation}
  \label{eq:GT20}
  \tilde{\Gsc}_{\phi_1\phi_2}(\omega_n,\vec{k}) =
  \left\{\begin{aligned}
      &    \tilde{r}_{\phi_1\phi_2}(i\omega_n,-\vec{k})\,, &&& n&>0\,,\\
      &
      \tilde{a}_{\phi_1\phi_2}(i\omega_n,-\vec{k})\,, &&& n&<0\,.
  \end{aligned}\right.
\end{equation}
These results apply also to the case $n=0$ if the spectral function is
regular at the origin. If instead there is a transport peak, i.e., if
\begin{equation}
  \label{eq:spec_func_def2}
  \tilde{\rho}_{\phi_1\phi_2}(\omega,\vec{k}) =
  2\pi  A_{\phi_1\phi_2}(\vec{k})\omega\delta(\omega) +
  B_{\phi_1\phi_2}(\omega,\vec{k})\,,  
\end{equation}
with $B_{\phi_1\phi_2}$ regular at $\omega=0$, one
has~\cite{Meyer:2010ii,Meyer:2011gj}
\begin{equation}
  \label{eq:spec_func_def3}
  \tilde{\Gsc}_{\phi_1\phi_2}(0,\vec{k})  - \tilde{r}_{\phi_1\phi_2}(i\epsilon,-\vec{k}) =
  \tilde{\Gsc}_{\phi_1\phi_2}(0,\vec{k})  - \tilde{a}_{\phi_1\phi_2}(-i\epsilon,-\vec{k}) =
  A_{\phi_1\phi_2}(-\vec{k})\,.
\end{equation}
Further details are given in Appendix \ref{sec:anrealcorr}.

Time-ordered Euclidean field correlators at finite temperature can be
expressed in terms of a path integral (see, e.g.,
Refs.~\cite{Kapusta:2006pm,Laine:2016hma}),
\begin{equation}
  \label{eq:eucl0}
  \lla  T\{\hat{\phi}_{E1}(x_1)\ldots \hat{\phi}_{En}(x_n)\}
  \rra_\beta  
  =
  \la \phi_1(x_1)\ldots \phi_n(x_n)\ra_\beta 
  \equiv
  \f{\int_\beta [\Ds\chi] \,e^{-S_E[\chi]}\phi_1[\chi(x_1)]\ldots
    \phi_n[\chi(x_n)]}{    \int_\beta [\Ds\chi] \,e^{-S_E[\chi]}}\,,
\end{equation}
where $S_E$ is a suitable Euclidean action, and the path integration
$\int_\beta [\Ds\chi]$ is over sets of bosonic ($c$-number) and
fermionic (Grassmann) field variables that are respectively periodic or
antiperiodic in the time direction.  Thermal correlation functions are
then reconstructed by means of analytic continuation in the time
coordinate according to Eq.~\eqref{eq:corrfunc6}.

In the axiomatic setting, the analyticity properties of the real-time
correlation functions follow from the general properties expected of
quantum fields, and so the properties of the imaginary-time
correlators are a consequence of the basic assumptions (see, e.g.,
Ref.~\cite{Bros:1996mw}). If one instead takes the Euclidean theory
defined by Eq.~\eqref{eq:eucl0} as the starting point, these
properties become necessary conditions that the theory must satisfy in
order to be able to ultimately obtain a local relativistic quantum
field theory. An example is discussed below in Section
\ref{sec:reg_cond}.

\section{Gauge theories with Dirac fermions}
\label{sec:WI}

In this Section I briefly describe gauge theories with Dirac fermions
quantised in the path-integral approach. The discussion is quite
general, and applies to four-dimensional gauge theories of compact Lie
groups, minimally coupled to $N_f$ degenerate ``flavours'' of Dirac
fermions of mass $m$, transforming in some representation of the
group, at finite temperature and in the imaginary-time formalism. The
fermionic part of the Euclidean action is
$S_{\rm F}=\int_\beta d^4x\,\Lc_{\rm F}$, with $\Lc_{\rm F}$ the
following Euclidean Lagrangean density,
\begin{equation}
  \label{eq:dirlag}
  \Lc_{\rm F} =  \bar{\psi} (\slaD + m) \psi\,,
  \qquad
  \slaD = \gamma_\mu D_\mu\,, \qquad
  D_\mu=\de_\mu + ig\Ag_\mu\,.
\end{equation}
Here $\psi$ and $\bar{\psi}$ denote collectively two independent sets
of Grassmann variables $\psi_{f \eta c}(x)$ and
$\bar{\psi}_{f \eta c}(x)$, with flavour index $f=1,\ldots,N_f$,
``Dirac'' index $\eta=1,\ldots,4$, and ``colour'' index
$c=1,\ldots,N_c$ corresponding to the gauge group representation.  In
Eq.~\eqref{eq:dirlag} and in the following, summation over the
suppressed discrete indices, as well as over repeated explicit
indices, is understood, unless explicitly stated otherwise.  The
(generally non-Abelian) Hermitean gauge fields
$\Ag_\mu(x)=\Ag_\mu^\dag(x)$ read $\Ag_\mu = \Ag_\mu^a T^a$, with
$\Ag_\mu^a$ real fields and $T^a=T^{a\dag}$ a set of $N_c\times N_c$
matrices providing a representation of the gauge group generators.
Both $\Ag_\mu$ and the Dirac operator $\slaD$ in the background of the
gauge field act trivially on flavour space. The Euclidean Hermitean
gamma matrices $\gamma_\mu=\gamma_\mu^\dag$, $\mu=1,\ldots, 4$,
satisfy the anticommutation relations
$\{\gamma_\mu,\gamma_\nu\}=2\delta_{\mu\nu}$; the fifth gamma matrix
$\gamma_5\equiv -\gamma_1\gamma_2\gamma_3\gamma_4$ satisfies
$\gamma_5=\gamma_5^\dag$, $\gamma_5^2=\mathbf{1}$, and
$\{\gamma_5,\gamma_\mu\}=0$.

Expectation values are formally defined in terms of a path integral as
\begin{equation}
  \label{eq:gauge_corr_func1}
  \begin{aligned}
    \la \Oc \ra_\beta &\equiv Z_\beta^{-1} \int_\beta [\Ds
    \Ag]\,e^{-S_{\rm G}[\Ag]}\int_\beta [\Ds \psi]
    [\Ds\bar{\psi}]\,e^{-S_{\rm F}[\psi,\bar{\psi},\Ag]}\,\Oc[\psi,\bar{\psi},\Ag]\,,\\
    Z_\beta &\equiv \int_\beta [\Ds \Ag]\,e^{-S_{\rm
        G}[\Ag]}\int_\beta [\Ds \psi] [\Ds\bar{\psi}]\,e^{-S_{\rm
        F}[\psi,\bar{\psi},\Ag]}\,.
  \end{aligned}
\end{equation}
Following for definiteness the Faddeev-Popov-De \kern-1pt Witt
gauge-invariant approach as in Ref.~\cite{Bernard:1974bq}, $S_{\rm G}$
is the gauge part of the action, including the usual Yang-Mills action
and a gauge-fix\-ing term, while the Faddeev-Popov determinant
necessary to restore gauge invariance is included in the integration
measure $\int_\beta[\Ds \Ag]$.\footnote{Strictly speaking, the
  gauge-invariant continuum integration measure is not well defined
  beyond perturbation theory due to the existence of Gribov
  copies~\cite{Gribov:1977wm,Singer:1978dk}. This issue is dealt with
  by first formulating the theory in a gauge-invariant way on a
  lattice and taking eventually the continuum limit. The symbol
  $\int_\beta[\Ds \Ag]$ should then be understood as a shorthand for
  this procedure. Here I prefer to avoid the technical complications
  of the lattice formulation for the sake of clarity, as they pose no
  obstacle to the development of the main argument (see Section
  \ref{sec:reno} for further comments).} Here the subscript $\beta$
denotes the periodicity condition
$\Ag_\mu(\beta,\vec{x})=\Ag_\mu(0,\vec{x})$ to be imposed on the gauge
fields.  Integration over the fermion fields is done in the sense of
Berezin integration of Grassmann variables, over field configurations
satisfying the antiperiodicity conditions
$\psi(\beta,\vec{x})=-\psi(0,\vec{x})$ and
$\bar{\psi}(\beta,\vec{x})=-\bar{\psi}(0,\vec{x})$, denoted again by
the subscript $\beta$.  Since the fermionic action is quadratic, one
formally has
\begin{equation}
  \label{eq:gauge_corr_func2}
  \begin{aligned}
    \int_\beta [\Ds \psi] [\Ds\bar{\psi}]\,e^{-S_{\rm
        F}[\psi,\bar{\psi},\Ag]} = {\rm Det}(\slaD[\Ag] + m)\,,
  \end{aligned}
\end{equation}
where ${\rm Det}$ denotes the functional determinant, and the
dependence on the gauge fields has been made explicit. In general,
after integrating out the fermion fields one is left with
\begin{equation}
  \label{eq:gauge_corr_func1_ferm}
  \begin{aligned}
    \la \Oc \ra_\beta &= Z_\beta^{-1} \int_\beta [\Ds \Ag]\,e^{-S_{\rm
        G}[\Ag]} {\rm Det}(\slaD[\Ag] + m)\Oc_{\rm G}[\Ag]\equiv \la
    \Oc_{\rm G} \ra _\beta\,,\\
    \Oc_{\rm G}[\Ag] &\equiv \left({\rm Det}(\slaD[\Ag] +
      m)\right)^{-1}\int_\beta [\Ds \psi] [\Ds\bar{\psi}]\,e^{-S_{\rm
        F}[\psi,\bar{\psi},\Ag]}\,\Oc[\psi,\bar{\psi},\Ag]\,,
  \end{aligned}
\end{equation}
with $\Oc_G$ built out of gauge fields and of fermionic propagators in
a fixed gauge-field background,
$(\slaD[\Ag] + m)^{-1}$.

\subsection{Dirac eigenmodes and localisation}
\label{sec:dirac_modes_loc}

The eigenmodes of the Euclidean Dirac operator play an essential role
in this paper. The Dirac operator is anti-Hermitean, with purely
imaginary eigenvalues.  At finite temperature and in a finite spatial
box of volume $V$, imposing periodic spatial boundary conditions on
the gauge fields to preserve translation invariance, these eigenvalues
form a discrete set $\{i\lambda_n\}$, $\lambda_n\in\mathbb{R}$, with
corresponding eigenvectors $\psi_n$, $\slaD\psi_n = i\lambda_n\psi_n$,
obeying antiperiodic boundary conditions in the temporal direction and
periodic boundary conditions in the spatial directions. Since the
Dirac operator is trivial in flavour space, one treats the eigenmodes
as carrying only Dirac and colour indices on top of the spacetime
coordinate, $\psi_n=\psi_{n\,\eta c}(x)$. For future utility I
introduce the following notation,
\begin{equation}
  \label{eq:scprod_def}
  (\psi_{n'}(x),\Gamma\psi_n(x))
  \equiv  \sum_{\eta',c',\eta,c} \psi_{n'\,\eta' c'}(x)^*
  \Gamma_{\eta' c'\, \eta c}\psi_{n\,\eta c}(x) \,, \qquad
  \Vert \psi_n(x)\Vert^2 \equiv (\psi_n(x),\psi_n(x))\,,
\end{equation}
i.e., the scalar product $(\cdot,\cdot)$ is restricted to Dirac and
colour space, while the coordinate $x$ is kept fixed.  Throughout this
paper I will always assume that Dirac modes have been orthonormalised,
i.e.,
\begin{equation}
  \label{eq:orthonorm}
  \int_\beta d^4x \,  (\psi_{n'}(x),\psi_n(x)) = \delta_{n'n}\,.
\end{equation}
Due to the chiral property $\{\gamma_5,\slaD\}=0$, nonzero eigenvalues
appear in pairs $\pm i\lambda_n$, with corresponding eigenvectors
$\psi_n$ and $\gamma_5\psi_n$.\footnote{In case of degenerate
  eigenvalues one can always choose suitable eigenvectors that satisfy
  this relation.}  Moreover, one can choose the zero modes
$\psi_{n_0}$, $\slaD\psi_{n_0} = 0$, to have definite chirality, i.e.,
to obey $\gamma_5\psi_{n_0} = \xi_{n_0} \psi_{n_0}$ with
$\xi_{n_0}=\pm 1$. For the fermionic determinant,
Eq.~\eqref{eq:gauge_corr_func2}, one formally has in terms of the
Dirac eigenvalues
${\rm Det}(\slaD +m)=\prod_n(i\lambda_n +m)^{N_f} = m^{N_0
  N_f}\prod_{n,\lambda_n>0}(\lambda_n^2 +m^2)^{N_f}$, with $N_0$ the
number of exact zero modes.

Up to an unimportant factor of $i$, $\slaD[\Ag]$ for a fixed
background field can be seen as the Hamiltonian of a four-dimensional
quantum-mechanical system evolving in an additional, fictitious
time. This system is effectively three-dimensional due to the
compactification of the temporal direction.  Moreover, for purely
fermionic observables $\Oc$ the corresponding $\Oc_{\rm G}$ in
Eq.~\eqref{eq:gauge_corr_func1_ferm} can be expressed in terms of the
eigenvalues and eigenvectors of this Hamiltonian, with the remaining
integration being in practice a (gauge-invariant) average over the
background gauge fields. This is formally identical to the ensemble
average of a disordered system with (Hermitean) Hamiltonian
$-i\slaD[\Ag]$ with energy levels $\lambda_n[\Ag]$ and eigenvectors
$\psi_n[\Ag]$, with gauge field configurations providing different
realisations of disorder, distributed according to the probability
distribution determined by the path-integral measure after fermions
have been integrated out, Eq.~\eqref{eq:gauge_corr_func1_ferm}.

It is well known that for disordered systems the eigenmodes can become
localised, as was first realised by Anderson in his seminal
paper~\cite{Anderson:1958vr}. Typically, localised and delocalised
modes are found in disjoint spectral regions, separated by so-called
{\it mobility edges} where a second-order phase transition takes
place, known as {\it Anderson transition}. The subject of Anderson
localisation and Anderson transitions has been and still is a very
active area of research in condensed matter physics (see
Refs.~\cite{thouless1974electrons,lee1985disordered,
  kramer1993localization,Evers:2008zz,anderson50,manybody} for a
review), and has recently become of interest also in high-energy
physics after the observation of localised modes in the
high-temperature phase of gauge theories on the
lattice~\cite{Gockeler:2001hr,Gattringer:2001ia,GarciaGarcia:2005vj,
  GarciaGarcia:2006gr,Gavai:2008xe,Kovacs:2009zj, Bruckmann:2011cc,
  Kovacs:2012zq,Cossu:2016scb,Holicki:2018sms,Giordano:2021qav,
  Kovacs:2010wx,Kovacs:2017uiz,Giordano:2019pvc,Giordano:2016nuu,
  Giordano:2016vhx,Vig:2020pgq,Bruckmann:2017ywh,Bonati:2020lal,
  Baranka:2021san,Cardinali:2021fpu, Giordano:2015vla,
  Giordano:2016cjs,Vig:2021oyt,Giordano:2013taa,Nishigaki:2013uya,
  Giordano:2014qna,Ujfalusi:2015nha}.

For the purposes of this paper, very little information is needed
about localisation, besides the generic characterisation of localised
and delocalised modes. Qualitatively, localised modes are supported
essentially only in a finite spatial region whose size remains
basically unchanged as the system size is increased. Delocalised
modes, instead, extend over the whole system and keep spreading out as
the system size grows, although not necessarily at the same rate. More
precisely, the localisation properties of modes in a given spectral
region are determined quantitatively by the scaling with the volume of
the {\it inverse participation ratio},
\begin{equation}
  \label{eq:IPR}
    {\rm IPR}_n = 
  \int_\beta d^4x\,  \Vert \psi_n(x)\Vert^4\,, 
\end{equation}
averaged over gauge configurations (i.e., realisations of disorder).
Notice that since $\gamma_5^2=\mathbf{1}$, one has that
$\Vert \psi_n(x)\Vert^2 = \Vert \gamma_5\psi_n(x)\Vert^2$, and so
$ \psi_n$ and $ \gamma_5\psi_n$ have the same IPR.  Working in a box
of fixed temporal size $\beta$ and varying spatial volume $V$, if
modes in the given spectral region are typically non-negligible only
in a spatial region of size $O(V^\alpha)$ one finds
\begin{equation}
  \label{eq:IPR_qual}
  {\rm IPR}\sim V^{\alpha}  \left(V^{-\alpha}\right)^2 = V^{-\alpha}
\end{equation}
where the exponent $\alpha$ is the {\it fractal dimension} of the
modes.\footnote{While other definitions of fractal dimension can be
  adopted (e.g., the ``infrared dimension'' of
  Refs.~\cite{Alexandru:2021pap,Alexandru:2021xoi,Horvath:2021zjk}),
  it is the one obtained from the IPR that turns out to be important
  for our purposes (see Section \ref{sec:pplargev}).} For localised
modes $\alpha=0$, while delocalised modes have
$0<\alpha\le 1$.\footnote{Since
  $$
  1 = \left| \int_\beta d^4x \, \Vert \psi_n(x)\Vert^2\right|^2 \le
  \int_\beta d^4x \, \Vert \psi_n(x)\Vert^4 \int_\beta d^4x \, 1 =
  \beta V \cdot {\rm IPR}_n\,,
  $$
  one necessarily has $1-\alpha \ge 0$. In the condensed matter
  literature the term ``delocalised'' is usually reserved to the case
  $\alpha=1$, while modes with $0<\alpha<1$ are called ``critical''.
  For our purposes there is no need to distinguish these two cases
  (see Section \ref{sec:pplargev}).}

\subsection{Flavour symmetries and Ward-Takahashi identities}
\label{sec:gt_WI}

Besides local gauge symmetry, the fermionic action is manifestly
invariant under a set of spacetime symmetries, namely translations,
spatial rotations, and reflections through the hyperplanes orthogonal
to the temporal and spatial directions. In the context of the
Faddeev-Popov-De Witt approach, these are manifest also in the gauge
action if one uses a covariant gauge, such as Lorenz gauge.  The
fermionic action also enjoys flavour symmetries related to
transformations in the internal flavour space. In particular, the
fermionic Lagrangean is invariant under a group of
${\rm U}(N_f)={\rm U}(1)_B\times {\rm SU}(N_f)_V$ transformations,
where ${\rm U}(1)_B$ corresponds to a common change of phase of all
the different flavours, and ${\rm SU}(N_f)_V$ corresponds to special
unitary rotations of the flavour components.  For massless fermions,
$m=0$, the two chiral components of the fermionic fields are
decoupled, and the symmetry is further enhanced to the chiral symmetry
${\rm U}(N_f)_L\times {\rm U}(N_f)_R={\rm U}(1)_B\times {\rm U}(1)_A
\times {\rm SU}(N_f)_L \times {\rm SU}(N_f)_R$.  The
${\rm SU}(N_f)_L \times {\rm SU}(N_f)_R$ factor contains the subgroups
of ${\rm SU}(N_f)_V$ (vector) and ${\rm SU}(N_f)_A$ (axial)
transformations, given respectively by
\begin{align}
  \label{eq:suv}
  \psi &\to e^{i\alpha_a t^a}\psi\,, &
                                       \bar\psi &\to \bar{\psi}e^{-i\alpha_a t^a}\,, & \alpha_a &\in \mathbb{R}\,,\\
  \label{eq:sua}
  \psi &\to e^{i\beta_a t^a\gamma_5}\psi\,, &
                                              \bar\psi &\to \bar{\psi}e^{i\beta_a t^a\gamma_5}\,,& \beta_a &\in \mathbb{R}\,,
\end{align}
where $t^a$, $a=1,\ldots N_f^2-1$ are the Hermitean and traceless
generators of ${\rm SU}(N_f)$, obeying $[t^a,t^b]=if^{abc}t^c$ with
totally antisymmetric and real structure constants $f^{abc}$.
Normalisation is chosen so that $f^{abc}f^{abd}= N_f \delta^{cd}$ and
$\tr t^a t^b = \f{1}{2}\delta^{ab}$. The ${\rm U}(1)_B$ and
${\rm SU}(N_f)_V$ symmetries are expected not to break spontaneously,
also in the massless limit~\cite{Vafa:1983tf}.  Spontaneous breaking
of the ${\rm SU}(N_f)_A$ symmetry is instead possible. The
${\rm U}(1)_A$ symmetry is known to be
anomalous~\cite{Adler:1969gk,Bell:1969ts} due to non-invariance of the
functional integration measure~\cite{Fujikawa:1979ay,Fujikawa:1980eg}
and will not be considered in this paper.

The symmetry under the transformations Eq.~\eqref{eq:sua} implies an
infinite set of Ward-Takahashi
identities~\cite{Ward:1950xp,Takahashi:1957xn}. Their derivation is
rather standard, and is briefly reviewed for completeness in Appendix
\ref{sec:app_WTid} for the case at hand; here I only report the
results, which are also not new (see, e.g., Ref.~\cite{Brandt:2014qqa}
for $N_f=2$). Defining the infinitesimal, $x$-dependent
transformation
\begin{equation}
  \label{eq:sua_inf}
  \delta_A\psi(x) = i\epsilon_a(x) t^a\gamma_5\psi(x)\,, \qquad
  \delta_A\bar{\psi}(x) = i\epsilon_a(x)\bar{\psi}(x) t^a\gamma_5\,,
\end{equation}
one obtains for any observable $\Oc$ the identity
\begin{equation}
  \label{eq:WI4}
  \left\la \left( -\de_\mu \Ac^a_\mu(x) + 2m\Poc^a(x)   \right) \Oc\right\ra_\beta
  = \left\la -i\f{\delta_A \Oc}{\delta\epsilon_a(x)}\right\ra_\beta\,,
\end{equation}
where $\Ac^a_\mu$ are the flavour non-singlet axial-vector currents,
and $\Poc^a$ are the flavour non-singlet pseudoscalar densities,
\begin{equation} 
  \label{eq:bil}
  \Ac^a_\mu(x) \equiv\bar{\psi}(x) \gamma_\mu\gamma_5 t^a    \psi(x)
  \,, \qquad  \Poc^a(x) \equiv\bar{\psi}(x) \gamma_5 t^a    \psi(x)        \,.
\end{equation}
Of particular interest here is the case $\Oc = \Poc^b(y)$. A
straightforward calculation leads to
\begin{equation}
  \label{eq:WI6}
  -\de_\mu \la   \Ac^a_\mu(x)  \Poc^b(0)\ra_\beta
  +  2m \la \Poc^a(x)  \Poc^b(0)\ra_\beta
  = \delta^{(4)}(x)\delta^{ab}\Sigma\,, 
\end{equation}
where $\Sigma$ is the chiral condensate, defined by
$\la\bar{\psi}_f \psi_g\ra_\beta \equiv \delta_{fg} \Sigma$, which
follows from vector flavour symmetry. The four-dimensional Dirac delta
in Eq.~\eqref{eq:WI6} is understood to be periodic in time,
\begin{equation}
  \label{eq:Ddelta}
  \delta^{(4)}(x)=    \delta_P(t)\delta^{(3)}(\vec{x})\,,
  \qquad
  \delta_P(t)=\sum_{n=-\infty}^\infty\delta(t-n\beta)\,.
\end{equation}
Exploiting vector flavour symmetry further, one finds
\begin{equation}
  \label{eq:sym_2point_def}
  \begin{aligned}
    \la \Ac^a_\mu(x)\Poc^b(0) \ra_\beta &\equiv
    \delta^{ab}\Gsc_{AP\,\mu}(x) \,, &&& \la \Poc^a(x) \Poc^b(0)
    \ra_\beta &\equiv \delta^{ab}\Gsc_{PP}(x) \,,
  \end{aligned}
\end{equation}
and Eq.~\eqref{eq:WI6} can be recast as
\begin{equation}
  \label{eq:WI8}
  -\de_\mu \Gsc_{AP\,\mu}(x)
  +  2m \Gsc_{PP}(x)
  = \delta^{(4)}(x)  \Sigma\,.
\end{equation}
The momentum-space version of this identity, obtained through a
Fourier transform [see Eq.~\eqref{eq:corrfunc9ter}], reads
\begin{equation}
  \label{eq:WI8_mom}
  i\omega_n \tilde{\Gsc}_{AP\,4}(\omega_n,\vec{k})
  +     i\vec{k}\cdot
  \vec{\widetilde{\mkern 0mu\Gsc}}_{AP}(\omega_n,\vec{k})
  +  2m \tilde{\Gsc}_{PP}(\omega_n,\vec{k})
  =   \Sigma\,.
\end{equation}

\subsection{Time-reflection symmetry}
\label{sec:syman}

The fermionic action is invariant under the following ``time reflection'' 
transformation,
\begin{equation}
  \label{eq:trevferm}
  \begin{aligned}
    \psi(t,\vec{x}) &\to \gamma_4\gamma_5 \psi(\beta-t,\vec{x})\,,&&&
    \bar{\psi}(t,\vec{x}) &\to
    \bar{\psi}(\beta-t,\vec{x})\gamma_5\gamma_4      \,,\\
    \Ag_\mu(t,\vec{x}) &\to \zeta_\mu \Ag_\mu(\beta-t,\vec{x})\,,
  \end{aligned}
\end{equation}
with $\zeta_4 =-1$ and $\zeta_{1,2,3}=1$. No summation over $\mu$ is
implied here and in the following equations.  Using the
Faddeev-Popov-De Witt procedure in a covariant gauge, the gauge action
is also invariant under the transformation Eq.~\eqref{eq:trevferm}, so
this leaves the full action invariant.  Under time reflection one has
for the pseudoscalar densities and axial-vector currents
\begin{equation}
  \label{eq:bil_T}
  \begin{aligned}
    \Poc^a(t,\vec{x}) &\to - \Poc^a(\beta-t,\vec{x})\,, &&&
    \Ac^a_\mu(t,\vec{x}) &\to -\zeta_\mu \Ac^a_\mu(\beta-t,\vec{x})\,.
  \end{aligned}
\end{equation}
For the correlators $\Gsc_{AP\,\mu}$ and $\Gsc_{PP}$ [see
Eq.~\eqref{eq:sym_2point_def}] one then finds
\begin{align}
  \label{eq:sym_2point}
  \begin{split}
    \Gsc_{AP\,\mu}(t,\vec{x})&= \zeta_\mu \la
    \Ac^a_\mu(\beta-t,\vec{x}) \Poc^a(\beta,\vec{0}) \ra_\beta 
    = \zeta_\mu \la \Ac^a_\mu(\beta-t,\vec{x}) \Poc^a(0,\vec{0})
    \ra_\beta= \zeta_\mu\Gsc_{AP\,\mu}(\beta-t,\vec{x}) \,,
  \end{split}
\\[0.5em]
  \label{eq:sym_2point3}
  \Gsc_{PP}(t,\vec{x})&= 
                        \la \Poc^a(\beta-t,\vec{x}) \Poc^a(\beta,\vec{0})\ra_\beta = 
                        \la \Poc^a(\beta-t,\vec{x}) \Poc^a(0,\vec{0}) \ra_\beta=\Gsc_{PP}(\beta-t,\vec{x})\,,
\end{align}
where antiperiodicity of $\psi$ and $\bar{\psi}$ has been used to
replace $ \Poc^a(\beta,\vec{0}) \to \Poc^a(0,\vec{0})$, and no
summation over $a$ is implied. The symmetry properties of the
coordinate-space correlators translate into the following relations
for the momentum-space correlators,
\begin{equation}
  \label{eq:GT33}
  \tilde{\Gsc}_{AP\,\mu}(\omega_n,\vec{k}) =  
  \zeta_\mu\tilde{\Gsc}_{AP\,\mu}(-\omega_n,\vec{k})\,,  \qquad
  \tilde{\Gsc}_{PP}(\omega_n,\vec{k}) =  
  \tilde{\Gsc}_{PP}(-\omega_n,\vec{k})\,.
\end{equation}

\subsection{Analytic continuation}
\label{sec:ancont}

Since they will be used repeatedly, it is convenient to summarise the
relevant analytic continuation relations needed to reconstruct the
real-time, Minkowskian thermal expectation values from the
imaginary-time, Euclidean correlation functions. The relevant
Minkowskian operators are the axial-vector current and pseudoscalar
density operators,
\begin{equation}
  \label{eq:anp5_bis_main}
  \begin{aligned}
    \hat{\Ac}^a_{\mu}(x) \equiv \bar{\hat{\psi}}(x)
    \tilde{\gamma}_{\mu}\tilde{\gamma}_{5}t^a\hat{\psi}(x)\,,\qquad
    \hat{\Poc}^a(x) \equiv \bar{\hat{\psi}}(x)
    \tilde\gamma_{5}t^a\hat{\psi}(x)\,.
  \end{aligned}
\end{equation}
Here $\tilde{\gamma}^\mu$, $\mu=0,\ldots, 3$ are the Minkowskian gamma
matrices, obeying
$\{\tilde{\gamma}^\mu,\tilde{\gamma}^\nu\}=2\eta^{\mu\nu}$ with
$\eta^{\mu\nu}={\rm diag}(1,-1,-1,-1)$, and
$\tilde{\gamma}^5=i\tilde{\gamma}^0\tilde{\gamma}^1\tilde{\gamma}^2\tilde{\gamma}^3$.
These are related with the Euclidean gamma matrices $\gamma_\mu$ and
$\gamma_5$ as $\gamma_{4}=\tilde{\gamma}^0$,
$\gamma_{j}=-i\tilde{\gamma}^j$, and $\gamma_{5}=\tilde{\gamma}^5$.
In Eq.~\eqref{eq:anp5_bis_main},
$\bar{\hat{\psi}}=\hat{\psi}^\dag \tilde{\gamma}^0$, as usual.  Using
the general analytic continuation relation Eq.~\eqref{eq:corrfunc6},
one finds that Euclidean and Minkowskian two-point correlation
functions are related as follows,
\begin{equation}
  \label{eq:GT19ext_1_main}
  \begin{aligned}
    \lla \hat{\Ac}^{a0}(t,\vec{x})\hat{\Poc}^a(0)\rra_\beta &=
    \Gsc_{AP\,4}(\epsilon+it,\vec{x}) \,, &&& \lla \hat{\Poc}^a(0)
    \hat{\Ac}^{a0}(t,\vec{x})\rra_\beta &=
    \Gsc_{AP\,4}(-\epsilon+it,\vec{x}) \,,
    \\
    \lla \hat{\Ac}^{aj}(t,\vec{x})\hat{\Poc}^a(0)\rra_\beta &= i
    \Gsc_{AP\,j}(\epsilon+it,\vec{x}) \,, &&& \lla\hat{\Poc}^a(0)
    \hat{\Ac}^{aj}(t,\vec{x})\rra_\beta &= i
    \Gsc_{AP\,j}(-\epsilon+it,\vec{x}) \,,\\
    \lla \hat{\Poc}^{a}(t,\vec{x})\hat{\Poc}^a(0)\rra_\beta &=
    \Gsc_{PP}(\epsilon+it,\vec{x}) \,, &&& \lla \hat{\Poc}^a(0)
    \hat{\Poc}^{a}(t,\vec{x})\rra_\beta &=
    \Gsc_{PP}(-\epsilon+it,\vec{x}) \,. 
  \end{aligned}
\end{equation}
Here and in the rest of this subsection no summation over the flavour
index $a$ is implied. Of particular interest is the reconstruction of
the spectral function, Eq.~\eqref{eq:spec_func_def1}, through that of
the retarded and advanced propagators, Eq.~\eqref{eq:GT19}. In this
work I need the spectral functions $\tilde{\rho}_{\Ac^{0a}\Poc^{a}}$
and $\tilde{\rho}_{\Poc^a \Poc^a}$, that are independent of $a$ thanks
to vector flavour invariance. For brevity I will denote them as
follows,
\begin{equation}
  \label{eq:GT39ext2_def_0}
  \begin{aligned}
     \tilde{c}(\omega,\vec{k})&\equiv
    \tilde{\rho}_{\Ac^{0a}\Poc^{a}}(\omega,\vec{k})\,,
    &&&     \tilde{c}^P(\omega,\vec{k})
    &\equiv \tilde{\rho}_{\Poc^a
  \Poc^a}(\omega,\vec{k})\,. 
\end{aligned}
\end{equation}
These spectral functions can be obtained by analytic continuation
using directly Eq.~\eqref{eq:GT19ext_1_main} [see
Eq.~\eqref{eq:spec_func_def1}], or from the retarded and advanced
propagators, which in turn can be reconstructed by analytic
interpolation from the Fourier coefficients of the Euclidean
correlator~\cite{Bros:1996mw,Cuniberti:2001hm}. Setting for brevity
($\mu=0,\ldots,3$)
\begin{equation}
  \label{eq:GT19_all_together}
  \begin{aligned}
    \tilde{r}^\mu(\omega,\vec{k})&\equiv \tilde{r}_{\Ac^{\mu
        a}\Poc^a}(\omega,\vec{k})\,, &&&
    \tilde{a}^\mu(\omega,\vec{k})&\equiv \tilde{a}_{\Ac^{\mu
        a}\Poc^a}(\omega,\vec{k})\,,
    \\
    \tilde{r}^P(\omega,\vec{k})&\equiv
    \tilde{r}_{\Poc^a\Poc^a}(\omega,\vec{k})\,, &&&
    \tilde{a}^P(\omega,\vec{k})&\equiv
    \tilde{a}_{\Poc^a\Poc^a}(\omega,\vec{k})\,,
  \end{aligned}
\end{equation}
one has from Eq.~\eqref{eq:GT20} that for $n\neq 0$
\begin{equation}
  \label{eq:GT20_again_0}
  \begin{aligned}
    \tilde{\Gsc}_{AP\,4}(\omega_n,\vec{k})
    &=\tilde{r}^0(i\omega_n,-\vec{k})\,, &&& n&>0\,, &&&
    \tilde{\Gsc}_{AP\,4}(\omega_n,\vec{k}) &=
    \tilde{a}^0(i\omega_n,-\vec{k})\,,       &&& n&<0 \,,\\
    i\tilde{\Gsc}_{AP\,j}(\omega_n,\vec{k}) &=
    \tilde{r}^j(i\omega_n,-\vec{k})\,, &&& n&>0\,, &&&
    i\tilde{\Gsc}_{AP\,j}(\omega_n,\vec{k}) &=
    \tilde{a}^j(i\omega_n,-\vec{k})\,, &&& n&<0\,,\\
    \tilde{\Gsc}_{PP}(\omega_n,\vec{k}) &=
    \tilde{r}^P(i\omega_n,-\vec{k})\,, &&& n&>0\,, &&&
    \tilde{\Gsc}_{PP}(\omega_n,\vec{k})
    &=\tilde{a}^P(i\omega_n,-\vec{k})\,, &&& n&<0\,. 
  \end{aligned}
\end{equation}
These relations hold also for $n=0$ if a transport peak is absent.

\subsection{Regularisation and renormalisation}
\label{sec:reno}

The discussion so far has been entirely formal, ignoring the
ill-defined nature of path integrals. As is well known, these require
a suitable regularisation to become mathematically well defined, and
an appropriate renormalisation procedure to remove the divergences
appearing when the regularisation is removed. Since regularisation
usually breaks some of the symmetries, their recovery after
renormalisation is carried out is not guaranteed in the general case,
and this can spoil the formal results discussed above. In particular,
Ward-Takahashi identities will be violated in the regulated theory if
the regularisation breaks the corresponding symmetry, and it is not
obvious that they can be recovered in the same form after
renormalisation.

Before drawing any conclusion from the symmetry properties discussed
in the previous Subsections, it is important to make sure that they
can be enforced in the finite, renormalised theory.  The best way to
enforce a symmetry is obviously to choose a regularisation that does
not break it, in which case renormalisation will not spoil it. The
best known non-perturbative regularisation of path integrals is the
lattice regularisation (see, e.g., Refs.~\cite{Rothe:1992nt,
  Montvay:1994cy,Gattringer:2010zz}), which is especially convenient
when dealing with gauge theories, since it allows one to mantain
manifest gauge invariance. In the lattice approach, the formal
functional integral is replaced with a well-defined finite-dimensional
integral over fields defined only on the discrete elements of a finite
lattice, eventually taking the limits of infinite volume and zero
lattice spacing. This approach clearly breaks most of the spacetime
symmetry; using a hypercubic lattice with periodic boundary conditions
one can still retain symmetry under translations by multiples of the
lattice spacing, a discrete subgroup of the SO(4) group, and
reflections [so that Eqs.~\eqref{eq:sym_2point} and
\eqref{eq:sym_2point3} hold also in the regulated theory].
Nonetheless, for asymptotically free theories it is widely believed,
and supported by a vast amount of numerical evidence, that a continuum
limit exists (after a suitable renormalisation procedure) where
spacetime symmetries are fully restored.

While vector flavour symmetry can be implemented exactly on the
lattice, the axial flavour symmetry is problematic due to the known
difficulties of implementing exact chiral symmetry for Dirac operators
discretised on the lattice~\cite{Nielsen:1981hk,Nielsen:1980rz,
  Nielsen:1981xu}.  Nonetheless, for lattice Dirac operators
satisfying the Ginsparg-Wilson relation~\cite{Ginsparg:1981bj}, such
as the fixed-point action~\cite{Hasenfratz:1993sp,DeGrand:1995ji},
domain-wall fermions~\cite{Kaplan:1992bt,Shamir:1993zy}, and overlap
fermions~\cite{Narayanan:1993sk,Narayanan:1993ss,Neuberger:1997fp,
  Neuberger:1998wv}, one has an exact chiral-type symmetry that holds
on any finite lattice~\cite{Luscher:1998pqa}, and that reduces to the
usual chiral symmetry in the formal continuum limit. This implies
exact Ward-Takahashi identities for suitably defined lattice currents
and densities, that hold for any lattice spacing and tend to the
continuum identities as the spacing goes to
zero~\cite{Luscher:1998pqa,Hasenfratz:1998jp,Kikukawa:1998py,
  Hasenfratz:2002rp}, and guarantee that the desired symmetry can be
enforced in the renormalised theory. In particular, renormalised
continuum correlation functions will satisfy the continuum
Ward-Takahashi identities -- of course, assuming that such a limit
exists.

Taking the continuum limit and the associated restoration of spacetime
symmetries for granted, one can use the continuum Ward-Takahashi
identities as fully meaningful relations between renormalised
quantities, and ignore where they came from. This will suffice for the
discussion of the finite-temperature version of Goldstone's theorem in
Section \ref{sec:GTFT}. On the other hand, when studying the
pseudoscalar-pseudoscalar correlator in detail in Section
\ref{sec:pseudochi} one has to keep track of the effects of
renormalisation. Instead of dealing with the technical complications
of the lattice approach, it is simpler to discuss the issue of
renormalisation directly in the continuum limit. In fact, if this
limit exists, then a renormalised theory with the desired symmetries
can also be obtained directly in the continuum, starting from a
regularised theory where the representation of fermionic observables
in terms of sums over the eigenmodes of the continuum Dirac operator
is cut off symmetrically at some ultraviolet scale
$\Lambda$.\footnote{A suitable regularisation of the integration over
  gauge fields is also required. However, this does not affect the
  argument.}  Ward-Takahashi identities are not exact anymore in this
case, but their violations should disappear after appropriate
renormalisation and removal of the cutoff.

A detailed discussion of the renormalisation issues related to the
Ward-Takahashi identity Eq.~\eqref{eq:WI6}, in continuum language, is
provided in Appendix \ref{sec:app_renorm}.  As mentioned above, using
Ginsparg--Wilson fermions~\cite{Ginsparg:1981bj} in the lattice
regularisation of the theory one can use the lattice Ward-Takahashi
identities implied by the exact lattice chiral
symmetry~\cite{Luscher:1998pqa} to show
nonperturbatively~\cite{Hasenfratz:1998jp} that $m$ renormalises only
multiplicatively; that the composite operators $\Ac^a_\mu$ require no
renormalisation after the usual mass and coupling renormalisations
have been carried out; and that the multiplicative renormalisation
constants $Z_P$ and $Z_S$ of the non-singlet pseudoscalar and singlet
scalar densities satisfy $Z_P=Z_S=Z_m^{-1}$ with $Z_m$ the mass
renormalisation constant. Combined with the properties of bilinear
correlators under the ``${\cal R}_5$-parity''
transformation~\cite{Frezzotti:2003ni}, the lattice chiral symmetry
implies that all additive divergent contact terms drop from the
lattice analogue of Eq.~\eqref{eq:WI6} in the chiral limit.  Based on
the argument given above, it is then safe to use the continuum
identity Eq.~\eqref{eq:WI6} in its {\it regularised} version to
discuss renormalisation issues in a simpler continuum language.  After
renormalisation, Eq.~\eqref{eq:WI6} in its {\it renormalised} version
can be used as the starting point for an alternative derivation of the
finite temperature version of Goldstone's
theorem~\cite{Lange:1965zz,Kastler:1966wdu,Swieca:1966wna,
  Morchio:1987wd, Strocchi:2008gsa}.  While renormalisation is an
essential part of the construction of the theory itself, and deserves
a careful discussion as such, it will become clear that it plays a
limited role in the arguments of this paper concerning the chiral
limit, which is in fact dominated by the low-end, infrared part of the
Dirac spectrum.

\section{Goldstone's theorem at finite temperature}
\label{sec:GTFT}

In this Section I discuss an alternative derivation of the finite
temperature analogue of Goldstone's theorem~\cite{Lange:1965zz,
  Kastler:1966wdu,Swieca:1966wna,Morchio:1987wd,Strocchi:2008gsa},
including a slight but useful generalisation, based on the
Ward-Takahashi identity Eq.~\eqref{eq:WI6}. In this Section all
Euclidean quantities are understood to be renormalised.

\subsection{Review of the standard derivation}
\label{sec:GTFT_standard}

For quantum field theories at finite temperature, the analogue of
Goldstone's theorem~\cite{Goldstone:1962es} proved in
Refs.~\cite{Lange:1965zz,Kastler:1966wdu,Swieca:1966wna,Morchio:1987wd}
states that the spontaneous breaking of a continuous symmetry in a
theory invariant under spatial translations leads to a gapless
spectrum of ``quasi-particle'' excitations (see
Ref.~\cite{Strocchi:2008gsa} for a detailed discussion and a full list
of references).  A heuristic proof of this ``Goldstone's theorem at
finite temperature'' is based on the following observations. Let
$\hat{J}^\mu$ be the conserved Noether current associated with the
symmetry, $\de_\mu\hat{J}^\mu=0$, and let
\begin{equation}
  \label{eq:GT5}
\hat{Q}_V(t)\equiv  \int_V d^3x\,\hat{J}^{0}(t,\vec{x})\,,
\end{equation}
be the corresponding charge, regularised by restricting spatial
integration to a finite volume $V$. If a nonzero expectation value is
found for the commutator
\begin{equation}
  \label{eq:gt_standard0}
\lim_{V\to\infty}  \lla  [i\hat{Q}_V(0),\hat{\Oc}]  \rra_{\beta}=b\neq 0\,,
\end{equation}
for some local observable $\Oc$, then current conservation and
relativistic locality imply
\begin{equation}
  \label{eq:gt_standard1}
\lim_{V\to\infty}  \lla [i\hat{Q}_V(t),\hat{\Oc}]
\rra_{\beta}=b\,,~~\forall t\,.
\end{equation}
Taking the Fourier transform (in the sense of distributions) of
Eq.~\eqref{eq:gt_standard1} one then finds
\begin{equation}
  \label{eq:gt_standard2}
\lim_{\vec{k}\to 0} i\tilde{\rho}_{J^0\Oc}(\omega,\vec{k}) =
  \lim_{\vec{k}\to 0} 
\int d^4x\,e^{i(\omega t  -\vec{k}\cdot\vec{x})} \lla
[i\hat{J}^{0}(t,\vec{x}),\hat{\Oc}]\rra_\beta =   2\pi b \delta(\omega) \,, 
\end{equation}
from which one infers the existence of massless quasi-particle
excitations, i.e., such that their energy vanishes and their lifetime
becomes infinite in the zero-momentum limit.\footnote{The same
  conclusions hold more generally for non-relativistic systems,
  replacing the requirement of relativistic locality with Swieca's
  condition~\cite{Swieca:1966wna} for the commutator of the spatial
  part of the current with the relevant observable (see
  Ref.~\cite{Strocchi:2008gsa}).} I will refer to these as Goldstone
excitations or quasi-particles throughout this paper.

In the case at hand, the relevant symmetry is the non-singlet axial
part of chiral symmetry. After analytic continuation to Minkowski
spacetime, Eq.~\eqref{eq:WI4} expresses conservation of the
non-singlet axial currents $\hat{\Ac}_\mu^a$ in the chiral limit
$m\to 0$, under the usual assumption that the second term on the
left-hand side can be dropped. Equations \eqref{eq:WI6} and
\eqref{eq:WI8} further show that axial flavour symmetry is
spontaneously broken, in the sense of Eq.~\eqref{eq:gt_standard0}, if
$\Sigma\neq 0$.  In fact, integrating Eq.~\eqref{eq:WI8} in the
$m\to 0$ limit over space and over the infinitesimal time interval
$[-\epsilon,\epsilon]$ one gets as $\epsilon\to 0$,
\begin{equation}
  \label{eq:GT2}
  \Sigma_* = -  \int d^3x \,
  \left[\Gsc_{AP\,4}(\epsilon,\vec{x})-\Gsc_{AP\,4}(-\epsilon,\vec{x})\right]
  =   \int d^3x\, \lla [i\hat{\Ac}^{a0}(0,\vec{x}),i\hat{\Poc}^a(0)] \rra_{\beta} \,,
\end{equation}
where no summation over $a$ is implied and $\Sigma_*$ is the chiral
condensate in the chiral limit. Here I used continuity of
$\int d^3x \, \vec{\nabla}\cdot\vec{\Gsc}_{AP}(t,\vec{x})$ at
$t=0$,\footnote{This follows from
  $\int d^3x \,\vec{\nabla}\cdot\vec{\Gsc}_{AP}(t,\vec{x})=
  \lim_{R\to\infty}\int_{\de B_R}
  d^2\vec{\Sigma}\cdot\vec{\Gsc}_{AP}(t,\vec{x})$, with $B_R$ the ball
  of radius $R$ and $d^2\vec{\Sigma}$ the corresponding infinitesimal
  surface element oriented outwards, and from continuity in $t$ of
  $\vec{\Gsc}_{AP}(t,\vec{x})$ for $t<|\vec{x}|$ (see Section
  \ref{sec:FT}).  It is assumed that $\vec{\Gsc}_{AP}(t,\vec{x})$
  vanishes sufficiently fast at spatial infinity (at least like
  $|\vec{x}|^{-2}$) so that the surface integral is convergent (but
  not necessarily zero).}  and the analytic continuation relations
Eq.~\eqref{eq:GT19ext_1_main}. Clearly, Eq.~\eqref{eq:GT2} is nothing
but Eq.~\eqref{eq:gt_standard0} at $t=0$ with
$\hat{Q}_V(0)= \int_V d^3x\,\hat{\Ac}^{a0}(0,\vec{x})$ the
finite-volume axial charge and $\hat{\Oc}=i\hat{\Poc}^a(0)$. One can
now use current conservation and relativistic locality to complete the
argument, obtaining Eq.~\eqref{eq:gt_standard2} with $-\tilde{c}$ [see
Eq.~\eqref{eq:GT39ext2_def_0}] on the left-hand side, and $b=\Sigma_*$
on the right-hand side, and infer the existence of pseudoscalar
Goldstone quasi-particles, i.e., ``quasi-pions'' [see
Eq.~\eqref{eq:GT39} below].

The standard argument outlined above makes essential use of current
conservation as an operator equation to infer
Eq.~\eqref{eq:gt_standard1} from Eq.~\eqref{eq:gt_standard0}. In the
following I discuss a more direct argument that requires only
knowledge of the Ward-Takahashi identity Eq.~\eqref{eq:WI6} in its
energy-momentum-space form, Eq.~\eqref{eq:WI8_mom}, works directly
with Euclidean quantities, and allows for a simple but useful
generalisation. This argument has been presented briefly in
Ref.~\cite{giordano_GT_lett}; here I provide a more detailed
discussion. In Appendix \ref{sec:gt_coord} I discuss the
coordinate-space version of the argument, which is new.  Of course,
appropriate analyticity conditions must be satisfied in order to be
able to reconstruct the physical, Minkowskian correlation
functions. Moreover, a suitable regularity condition must also be
satisfied to guarantee relativistic locality of the reconstructed
theory. This regularity condition plays an important role and will be
discussed next.

\subsection{Regularity condition}
\label{sec:reg_cond}

In this Subsection I ``reverse-engineer'' a condition that has to be
imposed on $\vec{\mkern 0mu\widetilde{\Gsc}}_{AP}$ in order to obtain
the desired locality properties of quantum field theory in Minkowski
space. Starting from the relativistic locality condition,
$[\vec{\mkern 0mu\hat{\Ac}}^{a}(x),\hat{\Poc}^b(0)]=0$ for $x^2< 0$,
one finds (in the sense of distributions) that
\begin{equation}
  \label{eq:GT17}
  \begin{aligned}
    & \lim_{k\to 0} \vec{k}\cdot \int d^3x \,
    e^{-i\vec{k}\cdot\vec{x}} \lla[ \vec{\mkern
      0mu\hat{\Ac}}^a(t,\vec{x}), \hat{\Poc}^b(0)]\rra_\beta =
    \lim_{k\to 0} i\int d^3x \,
    \left\{\vec{\nabla}e^{-i\vec{k}\cdot\vec{x}}\right\} \cdot \lla[
    \vec{\mkern 0mu\hat{\Ac}}^a(t,\vec{x}), \hat{\Poc}^b(0)]\rra_\beta
    \\
    =& \lim_{k\to 0} i\int d^3x \, \vec{\nabla}\cdot \left\{
      e^{-i\vec{k}\cdot\vec{x}} \lla[ \vec{\mkern
        0mu\hat{\Ac}}^a(t,\vec{x}), \hat{\Poc}^b(0)]\rra_\beta\right\}
    -i\int d^3x \, e^{-i\vec{k}\cdot\vec{x}}\, \vec{\nabla}\cdot \lla[
    \vec{\mkern 0mu\hat{\Ac}}^a(t,\vec{x}), \hat{\Poc}^b(0)]\rra_\beta
    \\ =& -i\int d^3x \, \vec{\nabla}\cdot \lla[ \vec{\mkern
      0mu\hat{\Ac}}^a(t,\vec{x}), \hat{\Poc}^b(0)]\rra_\beta =0\,,
  \end{aligned}
\end{equation}
since the first term on the second line and the term on the third line
vanish due to finiteness of the support of the commutator at fixed
$t$.\footnote{More formally, after smearing over $t$ with a function
  of compact support $h(t)$, the quantity
  $\vec{C}(\vec{x}\,) =\int dt \,h(t)\lla[\vec{\mkern
    0mu\hat{\Ac}}^a(t,\vec{x}),\hat{\Poc}^a(0)]\rra_\beta$ is a
  distribution in $\vec{x}$ of compact support, and so by the
  Paley-Wiener theorem for distributions its Fourier transform
  $\vec{\mkern 0mu\widetilde{C}}(\vec{k})=\int d^3x \,
  e^{-i\vec{k}\cdot\vec{x}}\vec{C}(\vec{x}\,)
  $ is an entire function, so that
  $\vec{k}\cdot \vec{\mkern 0mu\widetilde{C}}(\vec{k}) \to 0$ as
  $\vec{k}\to 0$ follows.}  This holds independently of the quark mass
$m$, and should be true in particular in the chiral limit if one is to
obtain a decent quantum field theory.  In this limit, the final
equality implies the time-independence of the regularised charge
commutator if the Noether current is conserved, so that
Eq.~\eqref{eq:gt_standard1} follows from Eq.~\eqref{eq:gt_standard0}
(see the discussion in Ref.~\cite{Strocchi:2008gsa}, Section 15.2.II).

To work out the implications of Eq.~\eqref{eq:GT17} in the Euclidean
setting, one uses the relations Eq.~\eqref{eq:GT20_again_0} between
the retarded and advanced propagators $\tilde{r}^j$ and $\tilde{a}^j$
and the Euclidean correlator $\tilde{\Gsc}_{AP\,j}$. Plugging them
into Eq.~\eqref{eq:GT17}, one finds ($n> 0$, no summation over $a$)
\begin{equation}
  \label{eq:GT21}
  \begin{aligned}
    \lim_{\vec{k}\to 0} \vec{k}\cdot \vec{\mkern
      0mu\widetilde{\Gsc}}_{AP}(\omega_{\pm n},\vec{k}) &= \pm
    \lim_{\vec{k}\to 0} \vec{k}\cdot \int_{-\infty}^\infty dt
    \,\theta(\pm t)e^{-\omega_{\pm n} t} \int d^3x
    \,e^{i\vec{k}\cdot\vec{x}} \lla[\vec{\mkern
      0mu\hat{\Ac}}^{a}(t,\vec{x})
    ,\hat{\Poc}^a(0)]\rra_\beta \\
    &= \pm \int_{-\infty}^\infty dt \,\theta(\pm t)e^{-\omega_{n} |t|}
    \lim_{\vec{k}\to 0}\vec{k}\cdot\int d^3x
    \,e^{i\vec{k}\cdot\vec{x}} \lla[\vec{\mkern
      0mu\hat{\Ac}}^{a}(t,\vec{x}) ,\hat{\Poc}^a(0)]\rra_\beta =0 \,,
\end{aligned}
\end{equation}
where the exchange of the limit $\vec{k}\to 0$ and integration over
$t$ is justified by the exponential damping factor.  From the
Euclidean perspective, this is a necessary condition that the
Euclidean correlators must obey in order to reconstruct a decent
Minkowskian theory.  I then assume that
$\vec{\mkern 0mu\widetilde{\Gsc}}_{AP}$ obeys the {\it regularity
  condition}
$ \vec{k}\cdot\vec{\mkern
  0mu\widetilde{\Gsc}}_{AP}(\omega_{n},\vec{k}) \to 0$ as
$\vec{k}\to 0$ for $n\neq 0$.

\subsection{Euclidean proof of Goldstone's theorem in momentum space}
\label{sec:egt_ms}

To obtain a Euclidean proof of Goldstone's theorem at finite
temperature, one starts from the Ward-Takahashi identity in
energy-momentum-space Eq.~\eqref{eq:WI8_mom}. Setting
\begin{equation}
  \label{eq:GT32}
  \Po(x) \equiv
  2m    {\Gsc}_{PP}(x)\,,
\qquad
  \widetilde{\Po}(\omega_n,\vec{k}) \equiv
  2m    \widetilde{\Gsc}_{PP}(\omega_n,\vec{k})\,,
\end{equation}
this identity becomes
\begin{equation}
  \label{eq:GT31}
  i\omega_n   \widetilde{\Gsc}_{AP\,4}(\omega_n,\vec{k}) + i\vec{k}\cdot
  \vec{\mkern 0mu\widetilde{\Gsc}}_{AP}(\omega_n,\vec{k}) +
    \widetilde{\Po}(\omega_n,\vec{k}) 
= \Sigma\,.
\end{equation}
The symmetry properties Eq.~\eqref{eq:GT33} imply that
$\widetilde{\Gsc}_{AP\,4}(\omega_n,\vec{k}) =
-\widetilde{\Gsc}_{AP\,4}(-\omega_n,\vec{k})$, so in particular
$\widetilde{\Gsc}_{AP\,4}(0,\vec{k}) =0$, while the regularity
condition Eq.~\eqref{eq:GT21} requires
$\vec{k}\cdot \vec{\mkern
  0mu\widetilde{\Gsc}}_{AP}(\omega_n,\vec{k})\to 0$ as $\vec{k}\to 0$
for $n\neq 0$ (also in the chiral limit).  Notice also
$\widetilde{\Po}(-\omega_n,\vec{k})
=\widetilde{\Po}(\omega_n,\vec{k})$.  One then obtains from
Eq.~\eqref{eq:GT31}
\begin{align}
  \label{eq:GT34}
  i\vec{k}\cdot    \vec{\mkern 0mu\widetilde{\Gsc}}_{AP}(0,\vec{k}) + 
  \widetilde{\Po}(0,\vec{k}) &= \Sigma\,,\\
  \label{eq:GT34_2}
  \lim_{\vec{k}\to 0} \left\{  i\omega_n   \widetilde{\Gsc}_{AP\,4}(\omega_n,\vec{k}) + 
  \widetilde{\Po}(\omega_n,\vec{k})\right\} &= \Sigma\,, \quad
                                              n\neq 0\,.
\end{align}
It is usually (although perhaps implicitly) assumed that the
pseudoscalar-pseudoscalar correlator is sufficiently regular as a
function of $m$ in the chiral limit, so that
$\widetilde{\Po}(\omega_n,\vec{k})\to 0$ as $m\to 0$. I will refer to
this as the standard scenario. As I show below in Section
\ref{sec:pseudochi}, this may not be the case if a finite density of
localised near-zero Dirac modes is present. More precisely, in the
presence of localised near-zero modes the pseudoscalar-pseudoscalar
correlator can develop a $1/m$ divergence, that exactly cancels out
the factor of $m$ and leaves behind a finite $\widetilde{\Po}$ in the
chiral limit. I will refer to this as the non-standard scenario.

Denoting by a subscript $*$ the quantities obtained in the chiral
limit, taking now $m\to 0$ in Eq.~\eqref{eq:GT31} followed by
$\vec{k}\to 0$ one obtains in full generality
\begin{align}
  \label{eq:GT34_chiral}
  i\vec{k}\cdot    
  \vec{\mkern 0mu\widetilde{\Gsc}}_{AP\,*}(0,\vec{k})  +
  \widetilde{\Po}_{*}(0,\vec{k})
  &= \Sigma_*\,,\\
  \label{eq:GT34_2_chiral}
  \lim_{\vec{k}\to 0} \left\{  i\omega_n
  \widetilde{\Gsc}_{AP\,4\,*}(\omega_n,\vec{k}) + \widetilde{\Po}_{*}(\omega_n,\vec{k}) \right\} &= \Sigma_*\,, \quad
                                                                                                   n\neq
                                                                                                   0\,. 
\end{align}
Exploiting rotation invariance, one finds from the equation for zero
Matsubara frequency, Eq.~\eqref{eq:GT34_chiral}, that
\begin{equation}
  \label{eq:GT35_first}
  \vec{\mkern 0mu\widetilde{\Gsc}}_{AP\,*}(0,\vec{k})
  = -i\vec{k}\left(\f{\Sigma_*-\widetilde{\Po}_{*}(0,\vec{k})}{\vec{k}^2}\right)
  \mathop\to_{\vec{k}\to 0}
  -i\left(\Sigma_*-\widetilde{\Po}_{*}(0,\vec{0})\right)\f{\vec{k}}{\vec{k}^2}\,, 
\end{equation}
and so it follows that
$\vec{\mkern 0mu\widetilde{\Gsc}}_{AP\,*}(0,\vec{k})$ has a pole at
$\vec{k}=0$ if $\Sigma_*-\widetilde{\Po}_{*}(0,\vec{0})\neq 0$, so in
particular in the standard scenario, where
$\widetilde{\Po}_{*}(0,\vec{0})=0$, if $\Sigma_*\neq 0$.  However, the
existence of this pole does not imply {\it a priori} that massless
Goldstone quasi-particles are present. This is different from the
zero-temperature case, where one would find
$\widetilde{\Gsc}_{AP\,\mu\, *} \propto p_\mu/p^2$ due to O(4)
invariance. In that case, after analytic continuation
$(p_4,\vec{p}\,)\to (-ip^0,\vec{p}\,)$ to Minkowski space one finds a
pole at $(p^0)^2- \vec{p}^{\,2}=0$ in the axial-vector-pseudoscalar
correlator, which directly implies the presence of massless particles
in the spectrum. At finite temperature full O(4) invariance is lost,
and the connection with the spectrum is encoded in the
axial-vector-pseudoscalar spectral function, whose reconstruction
requires the analytic interpolation of the discrete Fourier components
of the Euclidean correlator. The presence of a pole in
$\vec{\mkern 0mu\widetilde{\Gsc}}_{AP\,*}$ at zero frequency is
therefore by itself not sufficient to infer the relevant properties of
the spectral function at zero frequency.

To make progress one needs to exploit Eq.~\eqref{eq:GT34_2_chiral}. To
this end, one sets
\begin{equation}
  \label{eq:GT_neweqs_2}
  {\rm G}_*(\omega_n) \equiv \lim_{\vec{k}\to
    0}\widetilde{\Gsc}_{AP\,4\, *}(\omega_n,\vec{k})\,, \qquad
  {\rm R}_*(\omega_n) \equiv \lim_{\vec{k}\to 0}
\widetilde{\Po}_{*}(\omega_n,\vec{k})\,, 
\end{equation}
and using Eq.~\eqref{eq:GT34_2_chiral} one finds
\begin{equation}
  \label{eq:GT35bis}
  \begin{aligned}
    {\rm G}_*(\omega_n) &=\f{\Sigma_*-{\rm R}_*(\omega_n)}{ i\omega_n}
    \,, ~~~~ n\neq 0\,,
  \end{aligned}
\end{equation}
while ${\rm G}_*(0) = 0$.  It is instructive to discuss first the
standard scenario in which $\widetilde{\Po}_*$ vanishes and so
${\rm R}_*=0$. In this case Eq.~\eqref{eq:GT35bis} entirely determines
${\rm G}_*(\omega_n)$, up to the value of $\Sigma_*$. It is then easy
to obtain its unique analytic interpolation (in the sense of Carlson's
theorem) to a function $\bar{\rm G}_*(\Omega)$ of a complex variable
$\Omega\in\mathbb{C}$, and reconstruct the relevant Minkowskian
quantities by analytic
continuation~\cite{Bros:1996mw,Cuniberti:2001hm}.  One
finds\footnote{Formally, one defines
  $F(\Omega) = \f{\Sigma}{i\Omega} - \bar{\rm G}_*(\Omega)$,
  $\Omega\neq 0$, and $F(0)=0$, and looks for analytic interpolations
  $\bar{\rm G}_*(\Omega)$ obeying Eq.~\eqref{eq:GT35bis}
  $\forall n\neq 0$. This function obeys $F(\omega_n)=0~\forall n$,
  and Carlson's theorem implies that its unique interpolation analytic
  in the upper half of the complex plane is $F(\Omega) \equiv 0$,
  which further extends uniquely to $F(\Omega) \equiv 0$ on the whole
  complex plane. This leads to Eq.~\eqref{eq:GT39_ext0}. Notice that
  $\lim_{\Omega\to 0}\bar{\rm G}_*(\Omega)\neq {\rm G}_*(0)=0$. }
\begin{equation}
  \label{eq:GT39_ext0}
\bar{\rm G}_*(\Omega)=
\f{\Sigma_*}{i\Omega}\,,  
\end{equation}
which is analytic in the whole complex plane except for a pole at
$\Omega=0$.  This is enough to reconstruct the spectral function in
the zero-momentum limit. In fact, using Eq.~\eqref{eq:GT20_again_0},
one sees that the interpolation Eq.~\eqref{eq:GT39_ext0} for
$\Im \Omega>0$ and $\Im\Omega<0$ corresponds respectively to the
retarded and advance propagators $\tilde{r}^0$ and $\tilde{a}^0$ in
the chiral and zero-momentum limit,
\begin{equation}
  \label{eq:GT39ext3_bis}
  \begin{aligned}
    \lim_{\vec{k}\to
      0}\tilde{r}_*^0(\Omega,\vec{k})&=\f{\Sigma_*}{\Omega}
    ~\text{for}~ \Im \Omega>0 \,, &&& \lim_{\vec{k}\to
      0}\tilde{a}_*^0(\Omega,\vec{k}) &=\f{\Sigma_*}{\Omega}
    ~\text{for}~ \Im \Omega<0 \,.
  \end{aligned}
\end{equation}
Since the spectral density is given by
\begin{equation}
  \label{eq:GT39ext2_2}
  i  \tilde{c}(\omega,\vec{k}) = 
  \tilde{r}^0(\omega+i\epsilon,\vec{k})-
  \tilde{a}^0(\omega-i\epsilon,\vec{k})    \,, \qquad \omega\in
  \mathbb{R} \,,
\end{equation}
one finds in the chiral limit
\begin{equation}
  \label{eq:GT39}
  \lim_{\vec{k}\to 0}
  i\tilde{c}_*(\omega,\vec{k})=
  \f{\Sigma_*}{\omega+i\epsilon} -   \f{\Sigma_*}{\omega-i\epsilon} =
  -2\pi i\Sigma_* \delta(\omega)\,,
\end{equation}
and so if $\Sigma_*\neq 0$ one finds massless Goldstone excitations in
the spectrum. This completes this alternative proof of Goldstone's
theorem at finite temperature under the usual assumptions on the
symmetry breaking term.

In the non-standard scenario where $\widetilde{\Po}_*\neq 0$,
$\bar{\rm G}_*(\Omega)$ is not fully determined, but one can still use
Eq.~\eqref{eq:GT35bis} to relate it with the analytic interpolation of
${\rm R}_*(\omega_n)$, which will be denoted with
$\bar{\rm R}_*(\Omega)$.  One finds
\begin{equation}
  \label{eq:GTrev4_0}
  \begin{aligned}
 \bar{\rm G}_*(\Omega)=
 \f{\Sigma_*-\bar{\rm R}_*(\Omega)}{i\Omega}\,. 
  \end{aligned}
\end{equation}
Using Eq.~\eqref{eq:GT39ext2_2} and the well known result
$\f{1}{\omega\pm i\epsilon} = {\rm PV}\f{1}{\omega} \mp
i\pi\delta(\omega)$, together with the symmetry property
$\bar{\rm R}_*(-\Omega)=\bar{\rm R}_*(\Omega)$ that follows from
Eq.~\eqref{eq:GT33} by analytic continuation, one finds for the
spectral function in the chiral limit
\begin{equation}
  \label{eq:GTrev5}
  \begin{aligned}
    \lim_{\vec{k}\to 0} i\tilde{c}_*(\omega,\vec{k}) &= \bar{\rm
      G}_*(\epsilon-i\omega)-\bar{\rm G}_*(-\epsilon-i\omega) \\ &= -
    i\pi\delta(\omega)\left[ 2\Sigma_* -\bar{\rm R}_*(\epsilon)
      -\bar{\rm R}_*(-\epsilon)\right] - {\rm
      PV}\f{1}{\omega}\left[\bar{\rm R}_*(\epsilon-i\omega)
      - \bar{\rm R}_*(-\epsilon-i\omega)\right]\\
    &= - i 2\pi\delta(\omega)\left[ \Sigma_* -\bar{\rm
        R}_*(\epsilon)\right]- {\rm PV}\f{1}{\omega}\left[\bar{\rm
        R}_*(\epsilon-i\omega) - \bar{\rm
        R}_*(\epsilon+i\omega)\right] \,.
  \end{aligned}
\end{equation}
The quantity in the second square bracket on the right-hand side of
Eq.~\eqref{eq:GTrev5} is manifestly antisymmetric, so that one can
drop the principal-value prescription.  To discuss its regularity
properties it is convenient to express it in terms of the pseudoscalar
spectral function. Recalling Eq.~\eqref{eq:GT20_again_0}, one sees
that
\begin{equation}
  \label{eq:GT39ext3_PP_bis}
      \begin{aligned}
        \bar{\rm R}_*(\epsilon-i\omega) &= \lim_{\vec{k}\to
          0}\lim_{m\to 0}2m\tilde{r}^P(i\epsilon+\omega,-\vec{k})\,, \\
        \bar{\rm R}_*(-\epsilon-i\omega) &= \lim_{\vec{k}\to
          0}\lim_{m\to 0} 2m\tilde{a}^P(-i\epsilon
        +\omega,-\vec{k})\,,
      \end{aligned}
\end{equation}
and so
\begin{equation}
  \label{eq:GT39ext3_PP_ter}
  \begin{aligned}
    \bar{\rm R}_*(\epsilon-i\omega)- \bar{\rm R}_*(-\epsilon-i\omega)
    &= \lim_{\vec{k}\to 0} i \int d^4x \,e^{i(\omega t
      -\vec{k}\cdot\vec{x})}
    \lim_{m\to 0} 2m     \lla[\hat{\Poc}^{a}(t,\vec{x}),\hat{\Poc}^a(0)]\rra_\beta\\
    &= i\lim_{\vec{k}\to 0} \lim_{m\to 0} 2m \tilde{c}^P
    (\omega,\vec{k})\,.
  \end{aligned}
\end{equation}
In the last passage it is assumed that the chiral limit can be
exchanged with the Fourier transform at finite $\vec{k}$.

The pseudoscalar spectral function at $\vec{k}=0$ is an antisymmetric
function of $\omega$ [see Eq.~\eqref{eq:anp11_bis} in Appendix
\ref{sec:anrealcorr}], expected to be regular at $\omega=0$ and so
vanishing at least like $\omega$ (see Ref.~\cite{Meyer:2011gj}).
Moreover, no transport peak $\propto \omega\delta(\omega)$ is expected
in the pseudoscalar channel. These expectations are supported both by
analytic perturbative results (also in the chiral limit) and by
numerical lattice
calculations~\cite{Karsch:2003wy,Aarts:2005hg,Burnier:2017bod}.  It is
then reasonable to assume that the transport peak is absent, and that
$\tilde{c}^P $ is regular at $\omega=0$, and remains so as
$m\to 0$.\footnote{These assumptions can be weakened. The effects of a
  transport peak can be taken into account, see footnote
  \ref{foot:transpeak}. Concerning the behaviour at $\omega=0$, it is
  sufficient to assume that $\tilde{c}^P$ has an integrable
  singularity, which can further be demanded only in the relevant
  chiral and zero-momentum limits.}  Under these assumptions one finds
from Eq.~\eqref{eq:spec_func_def3}
\begin{equation}
  \label{eq:remn_at_zero}
  \bar{\rm R}_*(\epsilon)=
  \lim_{\vec{k}\to 0}\lim_{m\to 0} 2m  \tilde{r}^P(i\epsilon,-\vec{k})=
  \lim_{\vec{k}\to 0} \lim_{m\to 0} 2m  \widetilde{\Gsc}_{PP}(0,\vec{k}) =
  \lim_{\vec{k}\to 0} \widetilde{\Po}_{*}(0,\vec{k}) =
  {\rm R}_*(0)\,,
\end{equation}
so that Eq.~\eqref{eq:GTrev5} reduces to
\begin{equation}
  \label{eq:GTrev5_quater}
\lim_{\vec{k}\to 0}    i\tilde{c}_*(\omega,\vec{k})
    = -2\pi i\delta(\omega)[ \Sigma_* -{\rm R}_*(0)]
-\f{1}{\omega}[\bar{\rm R}_*(\epsilon-i\omega)
- \bar {\rm R}_*(\epsilon+i\omega)]\,,
\end{equation}
with the second term regular as
$\omega\to 0$.\footnote{\label{foot:transpeak}In the presence of a
  transport peak
  $\tilde{c}^P(\omega,\vec{k})|_{\rm tp} = 2\pi
  A_{PP}(\vec{k})\omega\delta(\omega)$, the principal value
  prescription must be kept in Eq.~\eqref{eq:GTrev5_quater}, and since
  ${\rm PV}\f{1}{\omega} \omega\delta(\omega) =
  \f{\omega^2}{\omega^2+\epsilon^2}\delta(\omega) =0$ one finds no
  contribution to $\tilde{c}_*(\omega,\vec{k}\to 0)$ from the second
  term.  Setting
  $A_*=\lim_{\vec{k}\to 0}\lim_{m\to 0}2m A_{PP}(\vec{k})$, from
  Eq.~\eqref{eq:spec_func_def3} it follows
  $ \lim_{\epsilon\to 0}\bar{\rm R}_*(\epsilon)= {\rm R}_*(0) - A_*$,
  and so the delta term is changed to
  $-2\pi i\delta(\omega)[ \Sigma_* -{\rm R}_*(0)+A_*]$.}  Only the
first term affects the presence of massless Goldstone excitations in
the spectrum, which exist if $\Sigma_* -{\rm R}_*(0)\neq 0$. This
completes the proof of the generalised Goldstone's theorem at finite
temperature in the presence of a nonzero ${\rm R}_*(0)$.

It is worth commenting on the result above, especially in relation
with the usual Goldstone's theorem at finite temperature. As already
discussed above, a nonzero ${\rm R}_*(0)$ can only appear if the
pseudoscalar-pseudoscalar correlator develops a $1/m$ divergence as
$m\to 0$, cancelling out the factor of $m$ appearing in the
Ward-Takahashi identity and leaving a finite contribution in the
chiral limit. This mechanism is reminiscent of the formation of
anomalies, although here one is sensitive to the infrared rather than
the ultraviolet regime of the theory. For this reason, I will refer to
a nonzero ${\rm R}_*(0)$ as an {\it anomalous remnant}.

It is clear from Eq.~\eqref{eq:GTrev5_quater} that in principle
${\rm R}_*(0)$ could cancel $\Sigma_*$, so that in spite of the
apparent spontaneous breaking of a continuous symmetry by a nonzero
expectation value of an order parameter one would find no Goldstone
excitations, as one would expect from the usual Goldstone's theorem at
finite temperature. However, an anomalous remnant radically modifies
the usual Ward-Takahashi identity in the chiral limit, signalling that
the non-singlet axial currents are {\it not} conserved in this
limit. More precisely, the anomalous remnant makes the non-singlet
axial flavour symmetry {\it explicitly} broken even in the chiral
limit. As a consequence, one evades Goldstone's theorem at finite
temperature, since this has current conservation as one of its main
hypotheses. The presence of massless quasi-particle excitations is not
guaranteed by a nonzero condensate alone, and it is rather the
difference between the amount of spontaneous breaking, measured by
$\Sigma_*$, and of explicit breaking, measured by ${\rm R}_*(0)$, that
determines the fate of Goldstone excitations.\footnote{I show below in
  Section \ref{sec:fate_goldstone} that
  $|\Sigma_*| -|{\rm R}_*(0)|= -(\Sigma_* -{\rm R}_*(0))\ge 0$.}

It is interesting to notice that the presence of a pole at zero
spatial momentum in the correlator
$\vec{\mkern 0mu\widetilde{\Gsc}}_{AP}(0,\vec{k})$ is after all
sufficient to infer the existence of massless Goldstone
excitations. In fact, the residue at this pole equals
$\Sigma_* - \widetilde{\Po}_{*}(0,\vec{0})= \Sigma_* -{\rm R}_*(0)$ up
to a constant factor [see Eq.~\eqref{eq:GT35_first}], so that the pole is
present if and only if a Dirac-delta term is present in the
axial-vector-pseudoscalar spectral function.

As a final remark, I mention that one can extend the calculation done
above to the case of finite fermion mass without any difficulty. For the
singular part of the spectral function $\tilde{c}(\omega,\vec{k})$ one
finds
\begin{equation}
  \label{eq:GTrev_mass}
\lim_{\vec{k}\to 0} i \tilde{c}(\omega,\vec{k})|_{\rm sing} = -2\pi
i\delta(\omega) [\Sigma - {\rm R}(0)]\,,  
\end{equation}
analogously to Eq.~\eqref{eq:GTrev5_quater}. At finite quark mass one
generally finds a nonzero
${\rm R}(0)=\lim_{\vec{k}\to 0} \widetilde{\Po}(0,\vec{k}) = \int
d^4x\, 2m {\Gsc}_{PP}(x) = 2m\chi_\pi$, where $\chi_\pi$ is the
pseudoscalar susceptibility, and a nonvanishing condensate
$\Sigma$. However, one can show that at finite $m$
\begin{equation}
  \label{eq:GTrev_mass2}
  \Sigma = {\rm R}(0)= 2m\chi_\pi\,, 
\end{equation}
and so the singular part of $\tilde{c}$ vanishes and no massless
quasi-particle excitation appears, as one expects when $m\neq 0$.
This requires one to show that there is no pole in
$\vec{\mkern 0mu\widetilde{\Gsc}}_{AP}(0,\vec{k})$ at $\vec{k}=0$ [see
Eq.~\eqref{eq:GT34}]. Equivalently, one notices that
Eq.~\eqref{eq:GTrev_mass2} is just the usual integrated Ward-Takahashi
identity relating the chiral condensate and the pseudoscalar
susceptibility, which holds if one can drop the boundary term when
integrating Eq.~\eqref{eq:WI8} over spacetime. Both the absence of a
pole and the vanishing of the boundary term follow if
$\vec{\Gsc}_{AP}(x)$ falls off sufficiently fast at large
distances. This can be shown to be the case. An argument by Vafa and
Witten~\cite{Vafa:1983tf,Nussinov:1999sx} establishes an exponential
bound on the two-point correlation function of any flavour non-singlet
gauge-invariant local operator, at any nonzero $m$ in theories with a
positive path-integral measure. This applies in particular to the
axial-vector currents $A^a_\mu$, and to the finite-temperature gauge
theories considered here. Using the Ward-Takahashi identity
\begin{equation}
  \label{eq:aa_WI}
  \de_\mu
  \langle \Ac^a_\mu(x) \Ac^b_\nu(0) \rangle_\beta = 2m \langle
  \Poc^a(x)\Ac^b_\nu(0)\rangle_\beta =  2m \delta^{ab} {\Gsc}_{AP\,\nu}(x) \,,
\end{equation}
obtained by setting $\Oc=\Ac^b_\nu(0)$ in Eq.~\eqref{eq:WI4}, this
implies an exponential bound on $\vec{\Gsc}_{AP}$, and so the desired
result follows.\footnote{In fact, the bound of Ref.~\cite{Vafa:1983tf}
  is easily extended to the two-point function of any pair of flavour
  non-singlet local bilinear operators like $\Ac^a_\mu$ and $\Poc^a$,
  so it applies directly to $\vec{\Gsc}_{AP}(x)$.}  Notice that if
$\Sigma_*\neq 0$, Eq.~\eqref{eq:GTrev_mass2} implies a divergent
pseudoscalar susceptibility in the chiral limit, and a finite limit
for ${\rm R}(0)\to \Sigma_*$ as $m\to 0$.  This limit in general does
{\it not} coincide with ${\rm R}_*(0)$, which is obtained by taking
first the chiral limit followed by the zero-momentum limit [see
comment before Eq.~\eqref{eq:GT34_chiral}], since the two limits
generally do not commute.  While the localisation properties of Dirac
modes play no role in establishing the integrated Ward-Takahashi
identity Eq.~\eqref{eq:GTrev_mass2}, I will show below in Section
\ref{sec:pseudochi} that they are crucial in the determination of
${\rm R}_*(0)$.

\section{Pseudoscalar correlator in the chiral limit}
\label{sec:pseudochi}

As discussed in the previous Section, it is usually assumed that the
contribution ${\Po}$ [see Eq.~\eqref{eq:GT32}] of the
pseudoscalar-pseudoscalar correlator to the non-singlet axial
Ward-Takahashi identity Eq.~\eqref{eq:WI6} vanishes in the chiral
limit. As anticipated, I argue now that this is not the case if a
finite density of localised modes is found near the origin of the
Dirac spectrum: Under certain technical conditions, such modes lead to
the development of a $1/m$ infrared divergence, that compensates the
factor of $m$ and thus gives a finite contribution to the
Ward-Takahashi identity also in the chiral limit. As discussed above
at the end of Section \ref{sec:egt_ms}, this leads to an anomalous
remnant ${\rm R}_*(0)$ [see Eq.~\eqref{eq:GT_neweqs_2}] that competes
with the chiral condensate to determine the fate of Goldstone
excitations.

In this Section I initially work with ``bare'', unrenormalised
quantities (denoted by a subscript $B$), appearing in a suitably
regularised version of the path integral,
Eq.~\eqref{eq:gauge_corr_func1}.  This could be, e.g., a lattice
regularisation with Ginsparg-Wilson
fermions~\cite{Ginsparg:1981bj,Hasenfratz:1993sp,DeGrand:1995ji,
  Kaplan:1992bt,Shamir:1993zy,Narayanan:1993sk,Narayanan:1993ss,
  Neuberger:1997fp,Neuberger:1998wv}, that guarantees control over the
chiral properties of the theory~\cite{Luscher:1998pqa,
  Hasenfratz:1998jp,Kikukawa:1998py, Hasenfratz:2002rp}. However, as
explained in Section \ref{sec:reno}, it is justified to work directly
in the continuum, which allows for a simpler and clearer treatment of
the main issues, without having to deal with the technicalities of the
lattice approach.

The starting point is the decomposition of the bare pseudoscalar
correlator $\la \Poc_B^a(x)\Poc_B^b(0)\ra_{\beta}$ in terms of the
eigenmodes of the Dirac operator.  For infrared (IR) regularisation
purposes one works in a finite spatial volume $V$, imposing periodic
boundary conditions in the spatial directions. Antiperiodic boundary
conditions are imposed instead in the (compact) time direction due to
the antiperiodicity condition on fermion fields, see Section
\ref{sec:WI}. In this setting the spectrum of $\slaD$ becomes
discrete, and so the eigenvalues $i\lambda_n$ and the corresponding
eigenvectors $\psi_n$ will be labeled by an index $n$ taking integer
values. Moreover, as discussed above in Section \ref{sec:reno}, for
ultraviolet (UV) regularisation purposes one cuts off the spectrum at
some scale $\Lambda$, including in the mode decompositions only modes
with $|\lambda_n|\le \Lambda$. Cutting off modes in this way makes the
chiral condensate and the relevant correlation functions finite,
introducing violations in the Ward-Takahashi identity that, however,
disappear as $\Lambda\to\infty$. These violations, as well as UV
divergences, are of no concern here, since they originate in the UV
part of the Dirac spectrum, while in the chiral limit only the
low-end, IR part of the spectrum plays a role, as will become clear
below. Nonetheless, in spite of the fact that they have no physical
effect, UV modes should be carefully handled to obtain physically
meaningful, renormalised quantities.

\subsection{Mode decomposition}
\label{sec:ppmoddec}

A straightforward calculation gives the following result for the
pseudoscalar-pseudoscalar correlator expressed in terms of Dirac
modes,
\begin{equation}
  \label{eq:mod_dec3}
  -  \la \Poc_B^a(x)\Poc_B^b(0)\ra_{\beta}   = \lim_{V\to\infty}
  \f{\delta^{ab}}{2} \left\la \sum_{n,n'} \f{\Oc^{\gamma_5}_{n'n}(x)
      \Oc^{\gamma_5}_{nn'}(0) }{(i\lambda_n +m_B)(i\lambda_{n'} +m_B)}
  \right\ra_{\beta}
  \equiv - \delta^{ab} \Gsc_{PP\,B}(x)
  \,.
\end{equation}
Here and in the following equations, only expectation values of
functionals $\Oc_{\rm G}[B_B]$ of the (bare) gauge fields appear.  In
Eq.~\eqref{eq:mod_dec3} $m_B$ denotes the bare fermion mass, and $\tr$
denotes the trace over Dirac and colour indices. The dependence of
$\Gsc_{PP\,B}$ on $m_B$ and on the temperature $T=1/\beta$ is left
implicit. Moreover [see Eq.~\eqref{eq:scprod_def}],
\begin{equation}
  \label{eq:mod_dec3bis}
  \Oc_{n'n}^\Gamma(x) \equiv     (\psi_{n'}(x),\Gamma\psi_n(x))
  \,. 
\end{equation}
In this paper I will be concerned only with
$\Gamma=\mathbf{1},\gamma_5$, in which case the following properties
hold,
\begin{equation}
  \label{eq:ogamma_prop}
  \begin{aligned}
    \Oc_{-n'-n}^\Gamma(x) & = \Oc_{n'n}^{\Gamma}(x)\,, &&&
    \Oc_{n'n}^{\Gamma}(x)^*&=\Oc_{nn'}^{\Gamma}(x)\,, &&& \Oc_{-n'
      n}^{\Gamma}(x) &= \Oc_{n' n}^{\gamma_5\Gamma}(x) \,,
  \end{aligned}
\end{equation}
where the notation $-n$ indicates that the mode
$\psi_{-n} \equiv \gamma_5\psi_n$ is involved. As mentioned above, the
sums over modes in Eq.~\eqref{eq:mod_dec3}, as well as the product in
the determinant ${\rm Det}(\slaD +m)=\prod_n(i\lambda_n +m)^{N_f}$
appearing in Eq.~\eqref{eq:gauge_corr_func1_ferm}, are restricted to
$|\lambda_n|\le \Lambda$.

Taking into account the symmetry of the spectrum, the relation between
eigenvectors implied by the chiral property discussed in Section
\ref{sec:dirac_modes_loc}, and the properties
Eq.~\eqref{eq:ogamma_prop}, one can recast Eq.~\eqref{eq:mod_dec3} as
\begin{equation}
  \label{eq:mod_dec3_1}
  - \Gsc_{PP\,B}(x) 
  =  \lim_{V\to\infty}
  \f{1}{2}\int_{-\Lambda}^{\Lambda}d\lambda\int_{-\Lambda}^{\Lambda}d\lambda'\, 
  \f{(m_B^2-\lambda\lambda')\spcorr^{\gamma_5}_{V,\,\Lambda}(\lambda,\lambda';x;m_B)}{(\lambda^2+m_B^2)(\lambda^{\prime\, 
      2}+m_B^2)}\,,
\end{equation}
where I introduced the {\it spectral correlators}
\begin{equation}
  \label{eq:spcorr_1}
  \spcorr^{\Gamma}_{V,\,\Lambda}(\lambda,\lambda';x;m_B) \equiv \bigg\la \sum_{n,n'}
  \delta(\lambda-\lambda_n) \delta(\lambda'-\lambda_{n'})
  \Re\{    \Oc_{n'n}^{\Gamma}(x) \Oc_{nn'}^{\Gamma}(0)\}
  \bigg\ra_{\beta}\,,
\end{equation}
and made explicit their dependence on the bare fermion mass, the
volume, and the UV regulator.  It is convenient to separate the cases
$\lambda_n=\pm \lambda_{n'}$ from the rest, and
set\footnote{\label{foot:degmodes0}Strictly speaking, in the
  definition of $\bar{\spcorr}^{\,\Gamma}_{V,\,\Lambda}$ the condition
  $n\neq \pm n'$ should read $\lambda_n\neq \pm \lambda_{n'}$, while
  contributions from distinct but exactly degenerate modes,
  $\lambda_n = \lambda_{n'}$ with $n\neq n'$, should be included in
  $\spcorr_{s\,V,\,\Lambda}^\Gamma$. However, zero modes do not
  contribute in the thermodynamic limit (see below), and degenerate
  nonzero modes are expected to appear only on a set of configurations
  of zero measure, and can be ignored. See footnote
  \ref{foot:degmodes} for further comments.}
\begin{equation}
  \label{eq:spcorr_more_definitions}
  \begin{aligned}
    \spcorr^{\Gamma}_{s\,V,\,\Lambda}(\lambda;x;m_B) &\equiv \bigg\la
    \sum_{n} \delta(\lambda-\lambda_n)
    \Oc_{nn}^{\Gamma}(x) \Oc_{nn}^{\Gamma}(0) \bigg\ra_{\beta}\,, \\
    \bar{\spcorr}^{\,\Gamma}_{V,\,\Lambda}(\lambda,\lambda';x;m_B)
    &\equiv \bigg\la \sum_{\substack{n,n'\\ n\neq \pm n'}}
    \delta(\lambda-\lambda_n) \delta(\lambda'-\lambda_{n'}) \Re\{
    \Oc_{n'n}^{\Gamma}(x) \Oc_{nn'}^{\Gamma}(0)\} \bigg\ra_{\beta}\,,
  \end{aligned}
\end{equation}
in terms of which one has
\begin{equation}
  \label{eq:spcorr_1_reprise}
  \begin{aligned}
    \spcorr^{\gamma_5}_{V,\,\Lambda}(\lambda,\lambda';x;m_B) &=
    \delta(\lambda+\lambda')
    \spcorr^{\mathbf{1}}_{s\,V,\,\Lambda}(\lambda;x;m_B) +
    \delta(\lambda-\lambda')
    \spcorr^{\gamma_5}_{s\,V,\,\Lambda}(\lambda;x;m_B) \\ &\phantom{=}
    + \bar{\spcorr}^{\,\gamma_5}_{V,\,\Lambda}(\lambda,\lambda';x;m_B)
    \,.
  \end{aligned}
\end{equation}
In the language of random Hamiltonians (see Section
\ref{sec:dirac_modes_loc}), the quantities defined in
Eqs.~\eqref{eq:spcorr_1} and \eqref{eq:spcorr_more_definitions} are a
type of Green's functions measuring the correlation between
eigenmodes.  Using Eq.~\eqref{eq:ogamma_prop} one finds the following
symmetry relations,
\begin{equation}
  \label{eq:sym_rel_sp_corr}
  \begin{aligned}
    \bar{\spcorr}^{\,\Gamma}_{V,\,\Lambda}(-\lambda,-\lambda';x;m_B)&=
    \bar{\spcorr}^{\,\Gamma}_{V,\,\Lambda}(\lambda,\lambda';x;m_B)\,,
    &&&
    \bar{\spcorr}^{\,\Gamma}_{V,\,\Lambda}(\lambda',\lambda;x;m_B)&=
    \bar{\spcorr}^{\,\Gamma}_{V,\,\Lambda}(\lambda,\lambda';x;m_B)\,,
    \\
    \bar{\spcorr}^{\,\Gamma}_{V,\,\Lambda}(-\lambda,\lambda';x;m_B)&=
    \bar{\spcorr}^{\,\gamma_5\Gamma}_{V,\,\Lambda}(\lambda,\lambda';x;m_B)\,,
    &&& \spcorr^{\Gamma}_{s\,V,\,\Lambda}(-\lambda;x;m_B)&=
    \spcorr^{\Gamma}_{s\,V,\,\Lambda}(\lambda;x;m_B)\,.
  \end{aligned}
\end{equation}
As they hold in any volume, these relations will hold also in the
thermodynamic limit $V\to\infty$.

\subsection{Large-volume limit}
\label{sec:pplargev}

In this Subsection I discuss the large-volume limit of the spectral
correlators, and show that here the localisation properties of the
Dirac eigenmodes play a crucial role: while they do not affect the
large-volume behaviour of $\bar{\spcorr}^{\,\Gamma}_{V,\,\Lambda}$,
they strongly affect whether $\spcorr^\Gamma_{s\,V,\,\Lambda}$
survives the infinite-volume limit or not. 

In order to see how the localisation properties of the modes affect
their contribution to the spectral correlators,
Eq.~\eqref{eq:spcorr_more_definitions}, notice first the exact bounds
\begin{equation}
  \label{eq:largev0}
  |\Oc^\Gamma_{n'n}(x)|^2 = |  (\psi_{n'}(x),\Gamma \psi_n(x) ) |^2
  \le \Vert\Gamma\Vert^2 \Vert \psi_{n'}(x)\Vert^2 
  \,\Vert\psi_n(x) \Vert^2\,, 
\end{equation}
where the matrix norm $\Vert \Gamma \Vert =1$ for
$\Gamma=\mathbf{1},\gamma_5$, and
\begin{equation}
  \label{eq:largev_corrbound}
  \begin{aligned}
    \left| \la \Oc^\Gamma_{n'n}(x) \Oc^\Gamma_{nn'}(0)
      \ra_\beta\right| &\le \la \left|\Oc^\Gamma_{n'n}(x) \right|
    \left|\Oc^\Gamma_{nn'}(0) \right|\ra_\beta \\ &
    \le
    \Vert\Gamma\Vert^2 \la \Vert \psi_{n'}(x)\Vert \,\Vert\psi_n(x)
    \Vert \, \Vert \psi_{n'}(0)\Vert \,\Vert\psi_n(0) \Vert \ra_\beta
    \\ & \le \Vert\Gamma\Vert^2 \sqrt{ \la \Vert \psi_{n'}(x)\Vert^2
      \, \Vert \psi_{n}(x)\Vert^2 \ra_\beta}\sqrt{\la
      \Vert\psi_{n'}(0) \Vert^2\,\Vert\psi_n(0) \Vert^2 \ra_\beta} \\
    & = \Vert\Gamma\Vert^2 \la \Vert \psi_{n'}(x)\Vert^2 \, \Vert
    \psi_{n}(x)\Vert^2 \ra_\beta\,.
  \end{aligned}
\end{equation}
These follow from the Cauchy-Schwarz inequality, on the third line
applied to the expectation value $\la\ldots\ra_\beta$ which is defined
by a path-integral with positive-definite integration
measure. Translation invariance is used in the last passage.  In a
first approximation, one can treat these bounds more loosely as
estimates of the magnitude of $\Oc^\Gamma_{n'n}$ and their correlation
functions, in order to obtain the volume dependence of the individual
contributions to Eq.~\eqref{eq:spcorr_more_definitions}.  While these
estimates should be supplemented by suitable factors in order to take
into account the dependence on $\Gamma$ and, more importantly, on the
correlation between modes $n$ and $n'$ and on the distance between $0$
and $x$, they should suffice if all one is interested in is their
volume dependence.

Consider first the case $n\neq \pm n'$.  If modes $n$ near $\lambda$
have fractal dimension $\alpha(\lambda)$, i.e., are mostly supported
in regions whose size scales like $V^{\alpha(\lambda)}$, then
$\Vert\psi_n(x) \Vert^2 \sim V^{-\alpha(\lambda)}$ inside the
supporting region while being negligible outside. Translation
invariance implies that the probability of finding a given spacetime
point inside the support of the mode is $V^{\alpha(\lambda)}/V$.
Finally, the correlation between modes is expected to decrease as
$|\lambda_n-\lambda_{n'}|$ increases, and so it will be small for most
pairs of modes. One can then estimate
\begin{equation}
  \label{eq:largev_corrbound2}
  \la  \Vert\psi_{n'}(x) \Vert^2\,\Vert\psi_n(x) \Vert^2 \ra_\beta \sim
  \left(\f{1}{V^{\alpha(\lambda')}}\f{V^{\alpha(\lambda')}}{V}\right)
  \left(\f{1}{V^{\alpha(\lambda)}} \f{V^{\alpha(\lambda)}}{V}\right)
  = \f{1}{V^2}\,,
\end{equation}
irrespectively of the localisation properties of the modes.  From the
last line in Eq.~\eqref{eq:largev_corrbound} one then estimates
\begin{equation}
  \label{eq:large_v_estimate_diff}
  \left| \la \Oc^\Gamma_{n'n}(x) \Oc^\Gamma_{nn'}(0) \ra_\beta\right|
  \sim \f{1}{V^2}\,.
\end{equation}
Since $\bar{\spcorr}_{V,\,\Lambda}^{\,\Gamma}$ involves a double sum
over modes, and since the number of modes per unit spectral interval
typically scales like the volume,\footnote{The case in which the
  growth is faster than $O(V)$, leading to points in the spectrum
  where the spectral density has an integrable singularity in the
  infinite-volume limit (van Hove singularity), is not considered
  here.} one has $O(V^2)$ contributions of order $O(1/V^2)$. One then
expects
$\bar{\spcorr}_\Lambda^{\,\Gamma}(\lambda,\lambda';x;m_B)
\equiv\lim_{V\to \infty}
\bar{\spcorr}_{V,\,\Lambda}^{\,\Gamma}(\lambda,\lambda';x;m_B) $ to be
nonzero as long as both spectral densities $\rho_B(\lambda)$ and
$\rho_B(\lambda')$ are nonzero, independently of the localisation
properties of modes near $\lambda$ and $\lambda'$. The \textit{bare
  spectral density} is defined as
\begin{equation}
  \label{eq:spec_dens_def}
  \rho_{B,V}(\lambda) \equiv \f{1}{\beta V}\left\la
    \sideset{}{^\prime}\sum_{n}
    \delta(\lambda-\lambda_n)\right\ra_\beta\,,
  \qquad
  \rho_{B}(\lambda) \equiv \lim_{V\to\infty}   \rho_{B,V}(\lambda) \,,
\end{equation}
with $\rho_{B,V}$ the bare spectral density in a finite volume.  Here
$ \sum_{n}^\prime$ denotes the sum over nonzero modes only.  Exact
zero modes are explicitly excluded even though they would not
contribute in the thermodynamic limit, since their number scales only
like $N_0\sim \sqrt{V}$ at large volume. As a further consequence,
since the estimate Eq.~\eqref{eq:large_v_estimate_diff} applies in
particular to pairs of distinct zero modes, and pairs of a zero and a
nonzero mode, one finds that the corresponding total contributions
scale respectively like $V/V^2$ and $V\sqrt{V}/V^2$, and so
contributions involving zero modes can be dropped from
$\bar{\spcorr}_{V,\,\Lambda}^{\,\Gamma}$ in
Eq.~\eqref{eq:spcorr_more_definitions}.

The situation is different in the case $n=\pm n'$, where correlations
cannot be neglected. Using the last line in
Eq.~\eqref{eq:largev_corrbound}, one can estimate for modes near
$\lambda$ that
\begin{equation}
  \label{eq:large_v_estimate_same}
  \left| \la \Oc^\Gamma_{nn}(x) \Oc^\Gamma_{nn}(0) \ra_\beta\right|
  \sim
  \la  \Vert\psi_n(x) \Vert^4\ra_\beta = \f{1}{\beta V} \la {\rm IPR}_n
  \ra_\beta
  \sim \f{1}{VV^{\alpha(\lambda)}}\,,
\end{equation}
where translation invariance was used. Since a single sum over modes
appears in $\spcorr_{s\,V,\,\Lambda}^\Gamma$, there are $O(V)$ such
contributions in any spectral region with a finite density of modes,
and so one expects $\spcorr_{s\,V,\,\Lambda}^\Gamma(\lambda;x)\to 0$
in the large-volume limit {\it except if modes near $\lambda$ are
  localised}, in which case one expects to find a finite
value. Equation~\eqref{eq:large_v_estimate_same} can be turned into a
rigorous bound showing that
$\spcorr_{s\,V,\,\Lambda}^\Gamma(\lambda;x)$ vanishes in the
large-volume limit if the fractal dimension of modes near $\lambda$ is
nonzero, i.e., if they are delocalised. From the last line in
Eq.~\eqref{eq:largev_corrbound}, in fact, one has\footnote{The Dirac
  deltas in Eq.~\eqref{eq:spcorr_bound1} can be handled rigorously,
  without changing the result, by integrating first over small
  intervals around $\lambda,\lambda'$, using then
  Eq.~\eqref{eq:largev_corrbound}, and finally taking the limit of
  infinitesimal intervals.}
\begin{equation}
  \label{eq:spcorr_bound1}
  \begin{aligned}
    |\spcorr_{s\,V,\,\Lambda}^\Gamma(\lambda;x;m_B)| &\le \bigg\la
    \sum_{n} \delta(\lambda-\lambda_n) | \Oc_{nn}^{\Gamma}(x)| |
    \Oc_{nn}^{\Gamma}(0)| \bigg\ra_{\beta} \le \bigg\la \sum_{n}
    \delta(\lambda-\lambda_n) \Vert \psi_{n}(x)\Vert^4
    \bigg\ra_{\beta}\\ & = \f{1}{\beta V} \bigg\la \sum_{n}
    \delta(\lambda-\lambda_n) {\rm IPR}_n \bigg\ra_{\beta} =
    \rho_{B,V}(\lambda) \overline{\rm IPR}_V(\lambda) +
    \delta(\lambda)\f{N_0}{\beta V} \overline{\rm IPR}_V^0 \,,
  \end{aligned}
\end{equation}
where I used translation invariance, and where
\begin{equation}
  \label{eq:spcorr_bound2}
  \overline{\rm IPR}_V(\lambda) \equiv \f{1}{\beta V \rho_{B,V}(\lambda)}\bigg\la
  \displaystyle\sideset{}{^\prime}\sum_{n}
  \delta(\lambda-\lambda_n) {\rm IPR}_n \bigg\ra_{\beta}
  \,,  \qquad
  \overline{\rm    IPR}_V^0 \equiv 
  \f{1}{N_0}\bigg\la
  \sum_{\substack{n_0\\\lambda_{n_0}=0}} {\rm IPR}_{n_0} \bigg\ra_{\beta}
  \,,
\end{equation}
is the average ${\rm IPR}$ computed locally in the
spectrum.\footnote{The dependence of $\overline{\rm IPR}_V$ on
  $\Lambda$ and $m_B$ is irrelevant here and has been suppressed. In
  computing $\overline{\rm IPR}_V^0$ one should average over the
  degenerate zero-mode subspace using the procedure of
  Ref.~\cite{Baranka:2021san}, treating separately the two
  chiralities, but this would not change the fact that zero modes do
  not contribute here (see below).} Assuming that modes have fractal
dimension $\alpha(\lambda)$ near $\lambda$, one has
$\overline{\rm IPR}_V(\lambda)\sim V^{-\alpha(\lambda)}$. If modes are
delocalised near $\lambda\neq 0$, i.e., $\alpha(\lambda)>0$, one finds
that $\spcorr_{s\,V,\,\Lambda}^\Gamma$ vanishes in the thermodynamic
limit. On the other hand, if modes are localised then
$\alpha(\lambda)=0$ and $\spcorr_{s\,V,\,\Lambda}^\Gamma$ need not
vanish, and as shown above one expects that it does
not.\footnote{While capturing correctly the overall volume dependence,
  the simple estimate Eq.~\eqref{eq:large_v_estimate_same} entirely
  misses the $x$-dependence of the spectral correlators. Taking
  localisation literally, one would find
  $ \left| \la \Oc^\Gamma_{nn}(x) \Oc^\Gamma_{nn}(0) \ra_\beta\right|
  \sim \f{1}{\ell^3}$ for spatial separation smaller than the typical
  localisation length $\ell$, and zero otherwise.  A more precise
  analysis is carried out in Appendix \ref{sec:expoenv} using a more
  realistic exponential envelope of localised modes, that leads to
  expect an exponential suppression in $|\vec{x}|$ rather than
  strictly finite support.}  One then expects
$\spcorr_{{\rm loc}\,\Lambda}^\Gamma\equiv\lim_{V\to
  \infty}\spcorr_{s\,V,\,\Lambda}^\Gamma$ to be finite where a finite
density of localised modes is present and only there, as signalled by
the subscript ``${\rm loc}$''.  An explicit calculation shows that
this is the case (see Section \ref{sec:anomalous_rem} below).  Notice
that since the last term in Eq.~\eqref{eq:spcorr_bound1} scales like
$\sqrt{V}/V$, exact zero modes give no contribution to
$\spcorr_{s\,V,\,\Lambda}^\Gamma$ in the thermodynamic limit
independently of their localisation
properties.\footnote{\label{foot:degmodes}Exactly degenerate but
  distinct nonzero modes could also contribute to
  $\spcorr_{s\,V,\,\Lambda}^\Gamma$.  For these modes the estimate
  Eq.~\eqref{eq:large_v_estimate_diff} applies so that their
  contribution is suppressed like
  $N_{\rm deg} V/V^2=N_{\rm deg} V^{-1}$ in the thermodynamic limit,
  with $N_{\rm deg}$ their typical (possibly $\lambda$-dependent)
  degeneracy, unless they are strongly spatially correlated. In this
  case, the estimate Eq.~\eqref{eq:large_v_estimate_same} applies
  instead, leading to a total contribution of order
  $N_{\rm deg}V^{-\alpha}$.  These modes can then be relevant only if
  localised, strongly spatially correlated, and appearing on a set of
  configurations of finite measure. This seems very unlikely to
  happen.} 

Summarising, in the thermodynamic limit one finds that
$\bar{\spcorr}_\Lambda^{\,\Gamma}$ is nonzero where the spectral
density is nonzero, while $\spcorr_{{\rm loc}\,\Lambda}^\Gamma$ is
nonzero only in spectral regions where modes are
localised.\footnote{\label{foot:coex}As pointed out in Section
  \ref{sec:WI}, localised and delocalised modes usually do not coexist
  in the same spectral region. However, the qualitative estimate given
  above under Eq.~\eqref{eq:spcorr_bound2} does not depend on this
  non-coexistence assumption. If for some system one could separate
  localised and delocalised modes in a given spectral region and
  define the density $\rho_{\rm loc}(\lambda)$ of localised modes,
  then $\spcorr_{{\rm loc}\,\Lambda}^\Gamma$ would be nonzero where
  $\rho_{\rm loc}\neq 0$.}  In the infinite-volume limit
Eq.~\eqref{eq:mod_dec3_1} then reduces to
\begin{equation}
  \label{eq:mod_dec7_reduced}
  \begin{aligned}
    - \Gsc_{PP\,B}(x) &= \int_0^\Lambda
    d\lambda\,\left(\f{\spcorr_{{\rm
            loc}\,\Lambda}^{\mathbf{1}}(\lambda;x;m_B)}{\lambda^2
        +m_B^2} + \f{(m_B^2 - \lambda^2)\spcorr_{{\rm
            loc}\,\Lambda}^{\gamma_5}(\lambda;x;m_B)}{(\lambda^2 +m_B^2)^2}
    \right)\\
    &\phantom{=}+ \f{1}{2} \int_{-\Lambda}^\Lambda
    d\lambda\int_{-\Lambda}^\Lambda d\lambda'\, \f{(m_B^2 -
      \lambda\lambda^\prime)
      \bar{\spcorr}_{\Lambda}^{\gamma_5}(\lambda,\lambda';x;m_B)
    }{(\lambda^2 +m_B^2)(\lambda^{\prime\, 2} +m_B^2)} \,,
  \end{aligned}
\end{equation}
having used the symmetry relations Eq.~\eqref{eq:sym_rel_sp_corr}.

\subsection{Renormalisation}
\label{sec:pprenorm}

The bare pseudoscalar-pseudoscalar correlator needs to be suitably
renormalised before the UV cutoff $\Lambda$ is removed. As discussed
in detail in Appendix \ref{sec:app_renorm}, both additive and
multiplicative renormalisation are required, and the renormalised
pseudoscalar-pseudoscalar correlator reads
\begin{equation}
  \label{eq:reno_PP_corr}
  \Gsc_{PP}(x)   = \lim_{\Lambda\to \infty} Z_m^2\left[
    \Gsc_{PP\,B}(x) - {\rm CT}_{PP}(x)\right]\,, 
\end{equation}
where ${\rm CT}_{PP}(x)$ are divergent contact terms and $Z_m$ is the
mass renormalisation constant, with the renormalised mass defined as
$m_B=Z_m m$. Divergent contact terms originate from large Dirac
eigenvalues in Eq.~\eqref{eq:mod_dec7_reduced}. However, they are
polynomial in the fermion mass and so drop from $\Po$ in the chiral
limit, and therefore can essentially be ignored as far as the
Ward-Takahashi identity, Eqs.~\eqref{eq:WI6} and \eqref{eq:WI8}, is
concerned. More generally, all the contributions to $\Po$ coming from
large eigenvalues vanish in the chiral limit, including any finite
term remaining after the subtraction procedure, since they yield at
most a constant term in the pseudoscalar-pseudoscalar correlator as
$m\to 0$.  Multiplicative renormalisation, on the other hand, is
required also in the chiral limit.  To identify the divergent contact
terms, one splits the integrals in Eq.~\eqref{eq:mod_dec7_reduced} at
a suitably chosen $m$-independent subtraction scale. To disentangle
additive and multiplicative divergences, it is convenient to work with
renormalised spectral correlators. This also allows one to see more
clearly how the remaining finite terms behave in the chiral limit.

The procedure is most easily illustrated in the case of the chiral
condensate.  The bare chiral condensate $\Sigma_B$ is obtained from
the spectral density, Eq.~\eqref{eq:spec_dens_def}, via the
Banks-Casher relation~\cite{Banks:1979yr},
\begin{equation}
  \label{eq:mod_dec2}
  -\Sigma_B   =
  \int_0^\Lambda d\lambda\,\rho_B(\lambda,m_B)\f{2m_B}{\lambda^2 +m_B^2}\,.
\end{equation}
Here the dependence of $\rho_B$ on the bare mass has been made
explicit.  The spectral density grows like $\lambda^3$ at large
$\lambda$, thus leading to quadratic and logarithmic additive
divergences, which originate from the mixing of the scalar density
with the identity operator (see Appendix \ref{sec:app_renorm}).  A
possible renormalisation scheme to take care of them is the following
(see, e.g, Ref.~\cite{Leutwyler:1992yt}).  One splits the integral at
some (mass-independent) subtraction scale $\mu_B$, and for
$\lambda>\mu_B$ one expands the denominator in powers of
$m_B^2/\lambda^2$, obtaining
\begin{equation}
  \label{eq:mod_dec2_bis}
  \begin{aligned}
    -\Sigma_B & = \int_0^{\mu_B}
    d\lambda\,\rho_B(\lambda,m_B)\f{2m_B}{\lambda^2 +m_B^2} +
    \int_{\mu_B}^\Lambda d\lambda\,\rho_B(\lambda,m_B)
    \f{2m_B^5}{\lambda^4(\lambda^2+m_B^2)}
    \\
    &\phantom{=}+ \int_{\mu_B}^\Lambda d\lambda\,\rho_B(\lambda,m_B)
    \left( \f{2m_B}{\lambda^2} - \f{2m_B^3}{\lambda^4} \right) =
    \Sigma_B^{(1)} + \Sigma_B^{(2)} \,.
  \end{aligned}
\end{equation}
The integrals on the first line, that define $\Sigma_B^{(1)}$, are
formally convergent as $\Lambda\to\infty$, i.e., if one ignores that
the bare spectral density is also $\Lambda$-dependent, and furthermore
are made finite by multiplicative renormalisation. In fact,
multiplying by $Z_m$ one finds after a change of variables
\begin{equation}
  \label{eq:mod_dec2_ter}
  \begin{aligned}
    Z_m\Sigma_B^{(1)} &= \int_0^{\f{\mu_B}{Z_m}}
    d\lambda\,Z_m\rho_B(Z_m\lambda,Z_mm)\f{2m}{\lambda^2 +m^2} +
    \int_{\f{\mu_B}{Z_m}}^{\f{\Lambda}{Z_m}}
    d\lambda\,Z_m\rho_B(Z_m\lambda,Z_mm)
    \f{2m^5}{\lambda^4(\lambda^2+m^2)} \,.
  \end{aligned}
\end{equation}
As it was shown in Refs.~\cite{DelDebbio:2005qa,Giusti:2008vb}, the quantity
\begin{equation}
  \label{eq:rho_ren}
  \rho(\lambda,m) \equiv
  \lim_{\Lambda\to\infty}Z_m\rho_B(Z_m\lambda,Z_mm)
\end{equation}
is finite, and so Eq.~\eqref{eq:mod_dec2_ter} has a finite limit as
$\Lambda\to\infty$ if $\mu\equiv \mu_B/Z_m$ is kept fixed.  Notice
that since $Z_m$ depends on $\Lambda$ only logarithmically, the (bare)
subtraction scale $\mu_B=Z_m \mu$ has to depend logarithmically on the
cutoff, while $\Lambda/Z_m$ still diverges as $\Lambda\to\infty$.  The
terms on the second line of Eq.~\eqref{eq:mod_dec2_bis}, that define
$\Sigma_B^{(2)}$, remain divergent also after multiplication by $Z_m$,
and need to be subtracted. One then defines the renormalised
condensate as
\begin{equation}
  \label{eq:sub_cond}
  \begin{aligned}
    -\Sigma \equiv \lim_{\Lambda\to\infty} -Z_m\Sigma_B^{(1)} =
    \int_0^{\mu} d\lambda\,\rho(\lambda,m)\f{2m}{\lambda^2 +m^2} +
    \int_{\mu}^{\infty} d\lambda\,\rho(\lambda,m)
    \f{2m^5}{\lambda^4(\lambda^2+m^2)}\,.
  \end{aligned}
\end{equation}
The second term is at least of order $O(m^5)$ and vanishes in the
chiral limit.  For the first term one finds instead the well-known
result~\cite{Banks:1979yr}
\begin{equation}
  \label{eq:sub_cond_bis}
  \begin{aligned}
    -\Sigma_* & = \lim_{m\to 0} \rho(0,m) \int_0^{\f{\mu}{m}}
    dz\,\f{2}{1 + z^2 } + \lim_{m\to 0} \int_0^{\mu} d\lambda \, \f{2m
      f(\lambda)}{\lambda^2 +m^2} =\pi\rho(0,0)\,,
  \end{aligned}
\end{equation}
where $\rho(0,m)\equiv\lim_{\lambda\to 0} \rho(\lambda,m)$ and
$\rho(0,0)\equiv\lim_{m\to 0} \rho(0,m)$, and where
$f(\lambda)=\rho(\lambda,m)-\rho(0,m)$ is assumed to vanish at least
as fast as some power law as $\lambda\to 0$, i.e.,
$\lambda^{-\gamma} f(\lambda)\to 0$ as $\lambda\to 0$ for some
$\gamma>0$. In fact, in this case, the second term in
Eq.~\eqref{eq:sub_cond_bis} is of order at most $m^{\f{1}{2^N}}$ for
some sufficiently large $N$ such that $2^N\gamma>1$, as shown in
Appendix \ref{sec:int_genarg} [see Eq.~\eqref{eq:bound1}].

The same procedure can now be repeated for the
pseudoscalar-pseudoscalar correlator. In this case, after splitting
the integrals and multiplying by the renormalisation constant $Z_m^2$,
one finds
\begin{equation}
  \label{eq:mod_dec7_resplit_mult_reno0}
  \Gsc_{PP\,B}(x)   = \sum_{i,j=1}^2\Gsc^{(ij)}_{PP\,B}(x)\,,
\end{equation}
with
\begin{equation}
  \label{eq:mod_dec7_resplit_mult_reno}
  \begin{aligned}
    - Z_m^2\Gsc_{PP\,B}^{(ij)}(x) &= \delta_{ij}\int_{I_i}
    d\lambda\,\f{Z_m\spcorr_{{\rm
          loc}\,\Lambda}^{\mathbf{1}}(Z_m\lambda;x;Z_m m)}{\lambda^2
      +m^2} \\
    &\phantom{=}+ \delta_{ij}\int_{I_i} d\lambda\,
    \f{(m^2-\lambda^2)Z_m \spcorr_{{\rm loc}\,\Lambda}^{\gamma_5}(Z_m
      \lambda;x;Z_m m)}{(\lambda^2
      +m^2)^2}\\
    &\phantom{=}+ \int_{I_i} d\lambda\int_{I_j} d\lambda'\, \f{
      (m^2+\lambda\lambda^\prime )
      Z_m^2\bar{\spcorr}_{\Lambda}^{\mathbf{1}}(Z_m \lambda,Z_m
      \lambda';x;Z_m m)
    }{(\lambda^2 +m^2)(\lambda^{\prime\, 2} +m^2)} \\
    &\phantom{=}+ \int_{I_i} d\lambda\int_{I_j} d\lambda'\, \f{ (m^2-
      \lambda\lambda^\prime) Z_m^2
      \bar{\spcorr}_{\Lambda}^{\gamma_5}(Z_m \lambda,Z_m
      \lambda';x;Z_m m) }{(\lambda^2 +m^2)(\lambda^{\prime\, 2} +m^2)}
    \,,
  \end{aligned}
\end{equation}
where $I_{1}=[0,\mu]$ and $I_{2}=[\mu,\Lambda/Z_m]$.  Here the
properties Eq.~\eqref{eq:sym_rel_sp_corr} have been used to restrict
the double integral on the second line of
Eq.~\eqref{eq:mod_dec7_reduced} to the positive part of the spectrum.
It is clear that additive divergences can originate only from
$\Gsc_{PP\,B}^{(12)}$, $\Gsc_{PP\,B}^{(21)}$, and
$\Gsc_{PP\,B}^{(22)}$. The symmetry of the integrand under
$\lambda\leftrightarrow\lambda'$ implies
$\Gsc^{(12)}_{PP\,B}=\Gsc^{(21)}_{PP\,B}$.  Using an argument similar
to that of Refs.~\cite{DelDebbio:2005qa,Giusti:2008vb}, I show in
Appendix \ref{sec:lc_ren} that the functions
\begin{equation}
  \label{eq:spec_corr_reno2}
  \begin{aligned}
    \spcorr_{{\rm loc}}^\Gamma(\lambda;x;m) &\equiv
    \lim_{\Lambda\to\infty} Z_m\spcorr_{{\rm
        loc}\,\Lambda}^\Gamma(Z_m\lambda;x;Z_m m)\,,\\
    \bar{\spcorr}^{\,\Gamma}(\lambda,\lambda';x;m) &
    \equiv\lim_{\Lambda\to\infty}
    Z_m^2\bar{\spcorr}_{\Lambda}^{\,\Gamma}(Z_m\lambda,Z_m\lambda';x;Z_m
    m)\,,
  \end{aligned}
\end{equation}
are renormalised, finite quantities. This makes
$\Gsc_{PP}^{(11)}\equiv
\lim_{\Lambda\to\infty}Z_m^2\Gsc_{PP\,B}^{(11)}$ finite in the
large-$\Lambda$ limit. It remains only to identify and subtract the
additively divergent contributions to $\Gsc_{PP\,B}^{(12)}$ and
$\Gsc_{PP\,B}^{(22)}$.  For the purposes of this paper this need not
be done explicitly: all that matters is that the remaining finite
terms stay finite also in the chiral limit, which is easy to
show. Details are provided in Appendix \ref{sec:int_genarg}.  One then
concludes that
\begin{equation}
  \label{eq:CT_chiral}
  Z_m^2\left( \Gsc_{PP\,B}^{(12)}(x)+\Gsc_{PP\,B}^{(21)}(x)+\Gsc_{PP\,B}^{(22)}(x)\right)
  =  Z_m^2{\rm CT}_{PP}(x) + F(x;m) + \ldots\,, 
\end{equation}
with omitted terms vanishing as $\Lambda\to \infty$, and $F$ a finite
$\Lambda$-independent quantity that remains finite also as $m\to 0$.
One has then $\Gsc_{PP}= \Gsc_{PP}^{(11)} + F$, and for the quantity
of interest, namely $\Po_*=\lim_{m\to 0}\Po$ [see Eq.~\eqref{eq:GT32}]
one finds $\Po_*=\lim_{m\to 0} 2m \Gsc_{PP}^{(11)}$, and the following
spectral representation,
\begin{equation}
  \label{eq:CT_chiral2_spec}
  \begin{aligned}
    - \Po_*(x) &= \lim_{m\to 0} 2m \int_0^{\mu} d\lambda\,\left(
      \f{\spcorr_{{\rm loc}}^{\mathbf{1}}(\lambda;x;m)}{\lambda^2
        +m^2} + \f{(m^2-\lambda^2)\spcorr_{{\rm
            loc}}^{\gamma_5}(\lambda;x;m)}{(\lambda^2 +m^2)^2}\right)\\
    &\phantom{= \lim_{m \to 0}} + 2m\int_0^{\mu} d\lambda\int_0^{\mu}
    d\lambda'\,\f{(m^2+\lambda\lambda')\bar{\spcorr}^{\mathbf{1}}(\lambda,\lambda';x;m)+
      (m^2-\lambda\lambda')\bar{\spcorr}^{\gamma_5}(\lambda,\lambda';x;m)}{(\lambda^2
      +m^2)(\lambda^{\prime\,2} +m^2)}\,.
  \end{aligned}
\end{equation}

\subsection{Renormalisation of the mobility edge}

The result Eq.~\eqref{eq:spec_corr_reno2} has an important consequence
for the renormalisation properties of the mobility edges.  As
discussed above in Section \ref{sec:pplargev}, the unrenormalised
spectral correlators
$\spcorr_{{\rm loc}\,\Lambda}^\Gamma(\lambda;x;m_B)$ have support only
in disjoint regions where modes are localised, separated by mobility
edges $\lambda^{(i)}_{c\,B}$ from regions where modes are instead
delocalised.  Since renormalisation consists only in a multiplicative
factor and a rescaling of $\lambda$ and $m_B$, the support of
$\spcorr_{{\rm loc}}^\Gamma(\lambda;x;m)$ is still made of disjoint
regions delimited by the renormalised mobility edges
$\lambda^{(i)}_{c}= \lim_{\Lambda \to\infty}
Z_m^{-1}\lambda^{(i)}_{c\,B}$.  In other words, the mobility edges
renormalise like the fermion mass, so that the ratios
$\lambda_{c\,B}^{(i)}/m_B$ are
re\-nor\-mal\-i\-sa\-tion-group-invariant quantities free from UV
divergences, i.e., the equality
\begin{equation}
  \label{eq:ren_mob_edge}
  \f{\lambda_{c\,B}^{(i)}}{m_B} = \f{\lambda_{c}^{(i)}}{m} + 
  o(\Lambda^0)
\end{equation}
holds up to corrections that vanish as the UV regulator is removed.
This had been suggested before~\cite{Kovacs:2012zq}, and was supported
by numerical results on the lattice~\cite{Kovacs:2012zq,
  Cardinali:2021fpu}, but had not been proved yet. Details are
provided in Appendix \ref{sec:lc_ren}.

\subsection{Chiral limit}
\label{sec:ppchlim}

The final step in order to obtain $ \Po_*$ is to determine the
behaviour of the various terms appearing in
Eq.~\eqref{eq:CT_chiral2_spec} as $m\to 0$.  To proceed it is
necessary to make assumptions about the position of the mobility
edge(s), if present.  It is assumed from now on that modes are
localised in the range $[0,Z_m\lambda_c]$ in the UV-regulated theory,
and so in the range $[0,\lambda_c]$ of the ``renormalised'' spectrum,
with $\lambda_c=\lambda_c(m)$ the renormalised mobility edge.  It is
also assumed that, if other regions with localised modes are present,
then these are found above some renormalised lower mobility edge
$\lambda_c'$ and remain separated from $\lambda=0$ in the chiral
limit.\footnote{\label{foot:UVloc}Localised modes have been observed
  at the high end ($|\lambda|>\lambda_c'$) of the spectrum of
  staggered fermions in pure gauge $\mathbb{Z}_2$ gauge theory in 2+1
  dimensions on the lattice~\cite{Baranka:2021san}, and it is likely
  that this feature is found also in other theories. However, these
  are ultraviolet modes that should not affect the continuum
  physics. In particular, it is unlikely that $\lambda_c'$ reaches
  down to the origin in the chiral limit.} With these assumptions,
Eq.~\eqref{eq:CT_chiral2_spec} becomes
\begin{equation}
  \label{eq:CT_chiral2_spec_again_0}
  \Po_*(x) =     \Po_*^{(1)}(x)+     \Po_*^{(2)}(x)+
  \Po_*^{(3)}(x)\,,
\end{equation}
where
\begin{equation}
  \label{eq:CT_chiral2_spec_again_1}
  - \Po_*^{(1)}(x) = \lim_{m\to 0} \phantom{+} 2m \int_0^{\lambda_c}
  d\lambda\,\left( \f{\spcorr_{{\rm
          loc}}^{\mathbf{1}}(\lambda;x;m)}{\lambda^2 +m^2} +
    \f{(m^2-\lambda^2)\spcorr_{{\rm
          loc}}^{\gamma_5}(\lambda;x;m)}{(\lambda^2
      +m^2)^2}\right)
\end{equation}
and\footnote{The upper limit of integration is set to $\mu$ for
  generality, and does not mean that modes are localised in the whole
  interval $[\lambda_c',\mu]$. }
\begin{equation}
  \label{eq:CT_chiral2_spec_again_2}
  -  \Po_*^{(2)}(x)  = \lim_{m \to 0} 2m \int_{\lambda_c'}^\mu
  d\lambda\,\left( \f{\spcorr_{{\rm
          loc}}^{\mathbf{1}}(\lambda;x;m)}{\lambda^2 +m^2} +
    \f{(m^2-\lambda^2)\spcorr_{{\rm
          loc}}^{\gamma_5}(\lambda;x;m)}{(\lambda^2 +m^2)^2}\right)
\end{equation}
receive contributions only from localised modes, while
\begin{equation}
  \label{eq:CT_chiral2_spec_again}
  - \Po_*^{(3)}(x) = \lim_{m\to 0}  2m\int_0^{\mu} d\lambda\int_0^{\mu}
  d\lambda'\,\f{(m^2+\lambda\lambda')\bar{\spcorr}^{\mathbf{1}}(\lambda,\lambda';x;m)+
    (m^2-\lambda\lambda')\bar{\spcorr}^{\gamma_5}(\lambda,\lambda';x;m)}{(\lambda^2
    +m^2)(\lambda^{\prime\,2} +m^2)}
\end{equation}
receives contributions from both localised and delocalised modes.  It
is shown in Appendix \ref{sec:int_genarg} that in the chiral limit the
integrals appearing in $\Po_*^{(1)}$,
Eq.~\eqref{eq:CT_chiral2_spec_again_1}, behave as follows as functions
of $m$,
\begin{equation}
  \label{eq:chlim_terms1}
  \begin{aligned}
    \int_0^{\lambda_c} d\lambda\, \f{\spcorr_{{\rm
          loc}}^{\mathbf{1}}(\lambda;x;m)}{\lambda^2 +m^2} &= \f{1}{m}
    \left(\arctan\left(\f{\lambda_c}{m}\right) \,\spcorr_{{\rm
          loc}}^{\mathbf{1}}(0;x;0) + o(m^0)\right)\,,
    \\
    \int_0^{\lambda_c} d\lambda\, \f{(m^2-\lambda^2)\spcorr_{{\rm
          loc}}^{\gamma_5}(\lambda;x;m)}{(\lambda^2 +m^2)^2} &=
    \f{1}{m} \left(
      \f{\f{\lambda_c}{m}}{1+\f{\lambda_c^2}{m^2}}\spcorr_{{\rm
          loc}}^{\gamma_5}(0;x;0) + o(m^0) \right)\,,
  \end{aligned}
\end{equation}
so possibly diverging like $1/m$ and leading to a finite $\Po_*$,
while the integrals appearing in $\Po_*^{(2,3)}$ tend to constants in
the chiral limit, and so give no contribution to $\Po_*$. These
results are valid under the rather mild technical assumptions that
finite limits exist for the spectral correlators as $\lambda\to 0$
and/or $\lambda'\to 0$, and that such limits are approached at least
as fast as some power law, and moreover that the resulting quantities
have a finite limit as $m\to 0$.  One then obtains the main result of
this Subsection,
\begin{equation}
  \label{eq:ppcor_chilim_final}
  \begin{aligned}
    - \Po_*(x) &= \pi\xi \spcorr_{{\rm loc}}^{\mathbf{1}}(0;x;0)+
    \eta\spcorr_{{\rm loc}}^{\gamma_5}(0;x;0)\,,
  \end{aligned}
\end{equation}
where
\begin{equation}
  \label{eq:ppcor_chilim_final2}
  \xi \equiv \f{2}{\pi}\arctan\kappa \,, \qquad
  \eta \equiv \f{2\kappa}{1+\kappa^2}\,,\qquad
  \kappa \equiv \lim_{m\to 0} \f{\lambda_c(m)}{m} \,.
\end{equation}
Notice that $\kappa$ is renormalisation-group invariant.  Notice also
that $\spcorr_{{\rm loc}}^{\Gamma}(0;x;0)$ are obtained using the
following order of limits, starting from the bare, finite-volume
spectral correlators,
\begin{equation}
  \label{eq:ppcor_chilim_final3}
  \spcorr_{{\rm loc}}^{\Gamma}(0;x;0) = \lim_{m\to
    0}\lim_{\lambda\to 0} \lim_{\Lambda\to\infty}\lim_{V\to\infty}
  Z_m\spcorr_{s\,V,\,\Lambda}^{\Gamma}(Z_m\lambda;x;Z_mm_B)\,.
\end{equation}
Finally, notice that the subtraction scale $\mu$ does not affect the final
result, as it should be. 

Assuming that $\spcorr_{{\rm loc}}^{\Gamma}(0;x;0)$ is nonzero in the
chiral limit, there are three possible scenarios depending on how the
mobility edge scales with $m$ in the chiral limit.

\paragraph{(i)} If $\lambda_c$ vanishes faster than $m$, then
$\kappa\to 0$.  In this case one finds $\Po_*=0$, so that the
presence of localised near-zero modes does not affect the
Ward-Takahashi identity Eq.~\eqref{eq:WI6}. If
$\lambda_c\sim m^{\delta+1}$, then $\kappa \sim m^{\delta}$, and
\textbf{(i-a)} if $\delta\ge 1$, the pseudoscalar-pseudoscalar
correlator remains finite in the chiral limit, while \textbf{(i-b)} if
$0<\delta<1$ it develops an infrared divergence
$1/m^{1-\delta}$.\footnote{If $\delta\ge 1$, no divergence can appear
  in the terms omitted in Eq.~\eqref{eq:chlim_terms1}, which are
  subleading as long as $\delta>0$, see
  Eq.~\eqref{eq:chi_contrib2_extra} and subsequent discussion in
  Appendix \ref{sec:int_genarg}.  }

\paragraph{(ii)} If $\lambda_c$ vanishes as fast as $m$, then
$\kappa\to \text{constant}$. In this case the
pseudoscalar-pseudo\-scal\-ar correlator develops an infrared
divergence $1/m$, and one finds $\Po_*\neq 0$. One finds for the
coefficient of the first term $\pi\xi<\pi$, and the second term is
finite.

\paragraph{(iii)} If $\lambda_c$ vanishes more slowly than $m$,
including not vanishing at all,\footnote{The quite unlikely case of
  $\lambda_c$ diverging in the chiral limit is contained in case
  (iii).  In fact, below some value of $m$ one would find
  $\lambda_c>\mu$, and so $\lambda_c$ would be replaced by the cutoff
  $\mu$ in the calculations above.} then $\kappa\to \infty$.  Also in
this case the pseudoscalar-pseudoscalar correlator develops an
infrared divergence $1/m$, and one finds $\Po_*\neq 0$.  The
coefficient of the first term is maximal and equal to $\pi$ in this
case, while the second term vanishes.

\subsection{Anomalous remnant}
\label{sec:anomalous_rem}

The anomalous remnant ${\rm R}_*(0)$ is obtained by integrating
$\Po_*(x)$ over Euclidean spacetime. If one can exchange the order of
integration and of the sequence of limits appearing in
Eq.~\eqref{eq:ppcor_chilim_final3}, then the calculation is
straightforward. Set
\begin{equation}
  \label{eq:ano_rem1}
  \begin{aligned}
    - {\rm R}_*(0) &= -\int_\beta d^4x\,\Po_*(x) = \pi\xi
    I_*^{\mathbf{1}} + \eta I_*^{\gamma_5}\,,
  \end{aligned}
\end{equation}
where [see Eqs.~\eqref{eq:spcorr_more_definitions} and \eqref{eq:spec_corr_reno2}]
\begin{equation}
  \label{eq:ano_rem2}
  \begin{aligned}
    I_*^{\Gamma} &= \int_\beta d^4x\,\lim_{m\to
      0}\lim_{\lambda\to 0} \spcorr_{{\rm loc}}^{\Gamma}(\lambda;x;m)\\
    &= \int_\beta d^4x\, \lim_{m\to 0}\lim_{\lambda\to 0}
    \lim_{\Lambda\to\infty}\lim_{V\to\infty} Z_m \sum_n\left \la
      \delta( Z_m \lambda-\lambda_n) \Oc_{nn}^{\Gamma}(x)
      \Oc_{nn}^{\Gamma}(0) \right\ra_{\beta}\,,
  \end{aligned}
\end{equation}
and denote
$\textrm{Lim}\equiv \lim_{m\to 0}\lim_{\lambda\to 0}
\lim_{\Lambda\to\infty}\lim_{V\to\infty}$. Under the
interchangeability assumption made above one finds
\begin{equation}
  \label{eq:ano_rem3}
  \begin{aligned}
    I_*^{\mathbf{1}} &= \textrm{Lim}\, Z_m \sum_n\left \la \delta( Z_m
      \lambda-\lambda_n) \int_\beta d^4x\,\Oc_{nn}^{\mathbf{1}}(x)
      \Oc_{nn}^{\mathbf{1}}(0)
    \right\ra_{\beta}\\
    &= \textrm{Lim}\, Z_m \sum_n\left \la \delta( Z_m
      \lambda-\lambda_n) \Oc_{nn}^{\mathbf{1}}(0) \right\ra_{\beta} =
    \textrm{Lim}\, Z_m \f{1}{\beta V}\sum_n\left \la \delta( Z_m
      \lambda-\lambda_n)
    \right\ra_{\beta}\\
    &= \textrm{Lim}\, Z_m \rho_{B,V}(Z_m\lambda,Z_m m) = \lim_{m\to
      0}\lim_{\lambda\to 0} \lim_{\Lambda\to\infty}
    Z_m \rho_B(Z_m\lambda,Z_m m)\\
    &= \lim_{m\to 0}\lim_{\lambda\to 0}\rho(\lambda,m)= \lim_{m\to
      0}\rho(0,m) = \rho(0,0) \,,
  \end{aligned}
\end{equation}
where $\int_\beta d^4x\, \Oc_{n'n}^{\mathbf{1}}(x) = \delta_{n'n}$ was
used [see Eq.~\eqref{eq:orthonorm}] and exact zero modes were dropped,
while
\begin{equation}
  \label{eq:ano_rem4}
  \begin{aligned}
    I_*^{\gamma_5} &=\textrm{Lim}\, Z_m \sum_n\left \la \delta( Z_m
      \lambda-\lambda_n) \int_\beta d^4x\,\Oc_{nn}^{\gamma_5}(x)
      \Oc_{nn}^{\gamma_5}(0) \right\ra_{\beta} = 0 \,,
  \end{aligned}
\end{equation}
since
$\int_\beta d^4x\, \Oc_{n'n}^{\gamma_5}(x) = \int_\beta d^4x\,
\Oc_{n'-n}^{\mathbf{1}}(x) = \delta_{n'-n}$. In conclusion,
\begin{equation}
  \label{eq:ano_rem5}
  -{\rm R}_*(0) = \pi \xi    \rho(0,0)  \,.
\end{equation}
This is the main result of this Section: in the presence of a finite
density of localised near-zero modes, and if $\xi\neq 0$, one finds a
nonvanishing anomalous remnant ${\rm R}_*(0)$.

Of course, one can consider the spacetime integral in
Eq.~\eqref{eq:ano_rem2} also away from the chiral and zero-eigenvalue
limits. Under the assumption that the other limits can be suitably
interchanged, the calculation above shows that
\begin{equation}
  \label{eq:ano_rem2_noch}
  I^{\Gamma}(\lambda;m) =   \int_\beta d^4x\,\spcorr_{{\rm
      loc}}^{\Gamma}(\lambda;x;m) =
  \delta_{\Gamma,\mathbf{1}}\rho(\lambda,m)\,,
\end{equation}
if $\alpha(\lambda)=0$, and zero otherwise, and so
$\spcorr_{{\rm loc}}^{\mathbf{1}}$ must be nonzero in the presence of
a finite density of localised modes, as anticipated in Section
\ref{sec:pplargev}.

The result Eq.~\eqref{eq:ano_rem5} crucially depends on the
possibility of integrating over spacetime before taking the various
limits. For localised modes this is justified as follows.
\begin{itemize}

\item In the finite-volume and UV-regularised theory (e.g., on a
  finite lattice), the mode sum is over a finite number of modes, and
  can certainly be exchanged with integration.

\item Averaging over gauge fields in the regularised setting should
  cause no problem.  For example, in the lattice regularisation the
  discretised gauge field is represented in terms of link variables
  which are elements of the gauge group. Averaging then consists in a
  multiple compact Haar integral over the gauge group, that can
  certainly be exchanged with spacetime integration, i.e., with
  summation over lattice sites.

\item The infinite-volume limit and the removal of the UV cutoff are
  difficult to control analytically.  Nonetheless, it is expected that
  at finite fermion mass the pseudoscalar-pseudoscalar correlator is
  bounded exponentially as a function of $|\vec{x}|$ (or at least by
  some integrable function), independently of the localisation
  properties of the modes,\footnote{\label{foot:bound}For localised
    modes this is shown below in Appendix \ref{sec:expoenv}. For
    delocalised modes, averaging over gauge fields leads to
    destructive interference effects among modes, and one expects an
    exponential decay of the correlator.}  uniformly in $V$ and
  $\Lambda$. This is confirmed by numerical experience on the lattice
  (see, e.g., Ref.~\cite{Bazavov:2019www}). One can then use the
  dominated convergence theorem (see, e.g., Ref.~\cite{liebloss2001})
  to justify exchanging integration with these limits.

\item The crucial step is the chiral limit. In this limit, the
  exponential (or, more generally, integrable) bound mentioned above
  may be lost. This is what one expects if near-zero modes, which are
  the only ones of physical relevance, are delocalised: indeed, in the
  standard scenario ($\Po_*=0$) the finite-temperature Goldstone
  theorem leads to massless pseudoscalar excitations and a
  non-integrable algebraic decay $1/|\vec{x}|$ of the
  pseudoscalar-pseudoscalar correlator.  On the other hand, for
  localised modes an integrable bound is expected to be inherited by
  the correlator from the modes, and one can use again dominated
  convergence to justify the exchange of integration and chiral limit.
  A more detailed argument showing that the $1/m$-divergent part of
  the pseudoscalar-pseudoscalar correlator should inherit the fast
  decay properties of the localised near-zero modes from the spectral
  correlators is provided in Appendix \ref{sec:app_bound}.
  
\end{itemize}

As a final remark, notice that integration over spacetime is
equivalent to the zero-momentum limit of the Fourier transform. This
is not expected to commute with the chiral limit in general: for
example, it does not if massless Goldstone bosons are present.  These
limits should, however, commute if near-zero modes are localised.

\subsection{Fate of Goldstone excitations}
\label{sec:fate_goldstone}

The main result of the previous Subsection, Eq.~\eqref{eq:ano_rem5},
can now be used to discuss the fate of Goldstone excitations in the
chiral limit of the theory. The singular part in the spectral
function, Eq.~\eqref{eq:GTrev5_quater}, reads
\begin{equation}
  \label{eq:fate1}
  \lim_{\vec{k}\to 0}    \tilde{c}_*(\omega,\vec{k})|_{\rm singular}
  = -2\pi \delta(\omega)[ \Sigma_* -{\rm R}_*(0)]
  = 2\pi \delta(\omega)\rho(0,0)( 1 -\xi)\,,
\end{equation}
where I have used the Banks-Casher relation~\cite{Banks:1979yr} in the
chiral limit, $\Sigma_* = -\pi\rho(0,0)$.  Notice that the coefficient
of the Dirac delta is larger than or equal to 0. There are three
possible scenarios, corresponding to scenarios (i)--(iii) for the
behaviour of the pseudoscalar correlator.

\paragraph{(i)} If $\lambda_c$ vanishes faster than $m$, then $\xi =
0$, and the presence of massless quasi-particles depends exclusively
on $\rho(0,0)$ being non-zero. This is identical to the standard
scenario encountered in the usual finite-temperature Goldstone's theorem.

\paragraph{(ii)} If $\lambda_c$ vanishes as fast as $m$, then
$0 < \xi < 1$, and massless quasi-particles are present if
$\rho(0,0)\neq 0$. This is similar to the standard scenario, although
the coefficient of the singular term is reduced.  This results in a
reduction of the residue
$-i[\Sigma_*-{\rm R}_*(0)]=2\pi i \rho(0,0)( 1 -\xi)$ of the pole at
zero momentum in $\vec{\mkern 0mu\widetilde{\Gsc}}_{AP}(0,\vec{k})$
[see Eq.~\eqref{eq:GT35_first}].

\paragraph{(iii)} If $\lambda_c$ vanishes more slowly than $m$,
including if it remains finite, then $\xi=1$ and Goldstone excitations
disappear from the spectrum, even if the spectral density is finite at
the origin.\footnote{If localised and delocalised modes coexist in the
  same spectral region (see footnote \ref{foot:coex}), then
  $\rho(0,0)$ should be replaced by the density of localised near-zero
  modes, $\rho_{{\rm loc}}(0,0)<\rho(0,0)$, in
  Eq.~\eqref{eq:ano_rem5}; and $\xi$ should be replaced by
  $\xi\f{\rho_{{\rm loc}}(0,0)}{\rho(0,0)}<1$ in Eq.~\eqref{eq:fate1}
  and in the discussion above. Notice that Goldstone excitations
  cannot disappear in this case.}

\section{Conclusions and outlook}
\label{sec:concl}

Besides its connection with deconfinement, supported by an increasing
amount of evidence, the physical consequences of the localisation of
low Dirac modes in high-temperature gauge
theories~\cite{GarciaGarcia:2005vj,GarciaGarcia:2006gr,Kovacs:2009zj,
  Kovacs:2012zq, Cossu:2016scb,Holicki:2018sms,Giordano:2021qav,
  Kovacs:2010wx,Kovacs:2017uiz,Giordano:2019pvc,Giordano:2016nuu,
  Giordano:2016vhx,Vig:2020pgq,Bruckmann:2017ywh,
  Bonati:2020lal,Baranka:2021san,Cardinali:2021fpu,
  Bruckmann:2011cc,Giordano:2015vla, Giordano:2016cjs,
  Vig:2021oyt,Giordano:2013taa,Nishigaki:2013uya,Giordano:2014qna,
  Ujfalusi:2015nha} have so far remained quite elusive. In this paper
I have discussed the effects of a finite density of localised
near-zero modes in the chiral limit on the pseudoscalar-pseudoscalar
correlator and on the massless Goldstone excitations expected from
Goldstone's theorem at finite temperature~\cite{Lange:1965zz,
  Kastler:1966wdu,Swieca:1966wna,Morchio:1987wd,Strocchi:2008gsa}. These
effects were discussed first in Refs.~\cite{giordano_GT_lett,
  Giordano:2021nat}, of which this work is the completion, including a
number of technical details omitted there. In summary, the main result
is that if a finite density of localised near-zero modes is present in
the chiral limit, and if the corresponding mobility edge vanishes more
slowly than the fermion mass, then no massless excitation are found in
this limit, in contrast with one's expectations from Goldstone's
theorem.  If the mobility edge vanishes as fast as the fermion mass,
massless excitations are present but the coefficient of the
corresponding delta-function term in the axial-vector-pseudoscalar
spectral function is reduced. To my knowledge, these are so far the
only direct (although in the chiral limit) physical consequences of
localisation of Dirac modes that have been found.

The argument presented here and, in shortened form, in
Refs.~\cite{giordano_GT_lett,Giordano:2021nat}, requires a number of
intermediate results that I believe are valuable for their own sake:
\begin{itemize}
\item a proof and an extension of Goldstone's theorem at finite
  temperature, in the case of the non-singlet axial flavour symmetry
  of gauge theories with massless fermions, based on the corresponding
  Ward-Takahashi identity in Euclidean space;
\item a detailed calculation of the contribution of localised modes to
  the pseudoscalar-pseudo\-scalar correlator,
  showing that they can lead to a $1/m$ divergence in the fermion mass
  if the mobility edge, $\lambda_c$, vanishes as fast as or more
  slowly than $m$;
\item a proof of the renormalisation-group-invariance of the mobility
  edge in units of the fermion mass,  $\lambda_c/m$.
\end{itemize}

The extension of Goldstone's theorem at finite temperature closes a
loophole in the usual proof (see
Refs.~\cite{Lange:1965zz,Kastler:1966wdu,Swieca:1966wna,Morchio:1987wd,
  Strocchi:2008gsa}), related to the possibility that the relevant
symmetry remains effectively explicitly broken in the chiral limit.
In this case the axial flavour symmetry remains not conserved even in
the chiral limit, and the usual Goldstone's theorem at finite
temperature is evaded. There is therefore no contradiction between
this theorem and the possible disappearance of massless excitations
mentioned above, in spite of the chiral condensate being nonzero due
to the presence of a finite density of near-zero modes.

A term explicitly breaking the symmetry, referred to as the anomalous
remnant in this paper, is present in the relevant Ward-Takahashi
identity in the chiral limit if near-zero modes are localised and have
a finite spectral density, and if their mobility edge does not vanish
faster than the fermion mass $m$. This results, in a manner similar to
the formation of anomalies, from the cancellation of the
symmetry-breaking parameter $m$ in the Ward-Takahashi identity against
a $1/m$ divergence in the pseudoscalar-pseudoscalar correlator. This
is the origin of the effects discussed above on the massless Goldstone
modes that are usually expected in the chiral limit.

The renormalisation-group-invariance of the ratio of the mobility edge
and the fermion mass was suggested in Ref.~\cite{Kovacs:2012zq}, and
is supported by numerical evidence in lattice QCD (also at finite
imaginary chemical potential) showing that this ratio has limited
sensitivity to the lattice spacing~\cite{Kovacs:2012zq,
  Cardinali:2021fpu}. The proof presented here puts this on firmer
ground, and further supports the fact that localisation of the low
Dirac modes is a genuine physical effect in Euclidean spacetime, and
not a lattice artefact.

The main result of this paper would be of more limited interest if one
could prove that localised modes must not be present in the chiral
limit, lest one violates some general property expected of a decent
quantum field theory at finite temperature.  Nonetheless, in this case
it could be reversed to show that no $1/m$ divergence should appear in
the pseudoscalar-pseudoscalar correlator, and that the massless
Goldstone modes cannot be removed if a nonzero condensate is
present. However, I am not aware of any general property of
finite-temperature quantum field theory in contrast with the
assumption of localised near-zero modes in the chiral limit.

The most pessimistic scenario, from the theoretical point of view, is
that while not violating any general principle, localised near-zero
modes are in practice not found in any useful model. It is then
important to follow up on the interesting clues about the possible
realisation of such a scenario, found in $N_f=2$ massless adjoint
QCD~\cite{Karsch:1998qj,Engels:2005te} and, most importantly, in QCD
towards the chiral limit~\cite{Dick:2015twa,Ding:2020xlj,
  Kaczmarek:2021ser}.

The connection between localisation and disappearance of Goldstone
modes has been known for a long time in condensed matter
physics~\cite{McKane:1980fs}, and for quite some time in
zero-temperature lattice field theory as well, although in an
unphysical setup~\cite{Golterman:2003qe} \kern -1.3pt (see also
Ref.~\cite{Giordano:2021nat}). The mechanism behind this connection is
the same in those cases and in finite-temperature quantum field
theory, and boils down to how the susceptibility appearing in the
integrated version of a suitable Ward-Takahashi identity blows up in
the limit in which the symmetry-breaking parameter is sent to zero. In
the case at hand this is the non-singlet pseudoscalar susceptibility,
which diverges like $1/m$ in the chiral limit. However, in the usual
case where low Dirac modes are delocalised this happens only
\textit{after} the spacetime integration that connects the
susceptibility to the pseudoscalar-pseudoscalar correlator. Instead,
if low Dirac modes are localised the $1/m$ divegence is already
present in the correlator \textit{before} spacetime integration. This
mechanism is identical to the one already discussed in
Refs.~\cite{McKane:1980fs, Golterman:2003qe}.  The major differences
between Refs.~\cite{McKane:1980fs,Golterman:2003qe} and this work and
Refs.~\cite{giordano_GT_lett,Giordano:2021nat} are on the one hand
technical, due to the reduced amount of symmetry in finite-temperature
quantum field theory in the imaginary-time formulation stemming from
the presence of a compactified direction; and on the other hand of
practical relevance (in principle\ldots), as this work is concerned
with physically more realistic theories.

In this respect, it would be interesting to work out the consequences
for the finite-mass theory of the realisation of a non-standard
scenario in the chiral limit, with Goldstone modes removed or at least
modified by the presence of localised modes. This would require to
preliminarily understand what kind of finite-temperature transition
one would find in the chiral limit, separating a low-temperature phase
where low modes are delocalised from a phase at higher temperature
where they are localised, but with a finite density on both sides of
the transition.\footnote{One should mention the claim of
  Refs.~\cite{Alexandru:2021pap,Alexandru:2021xoi} that modes in the
  immediate vicinity of the origin are {\it critical}, i.e., with
  nontrivial localisation properties, rather than localised, in
  high-temperature SU(3) pure gauge theory probed with overlap
  fermions. However, this claim has still to be fully confirmed by a
  finite size scaling analysis, and it is unclear how it would affect
  the opposite, chiral limit of massless fermions.}  What follows is
largely speculative.

Although it may seem natural to expect that the mobility edge is
vanishingly small at such a transition, it is not necessarily so, as
displayed by SU(3) pure gauge theory where a finite mobility edge
appears at the transition in the trivial Polyakov-loop sector (see
Ref.~\cite{Kovacs:2021fwq}). This may be a consequence of the
first-order nature of the transition.  Results at the first-order
reconfinement transition in trace-deformed SU(3) gauge theory at high
temperature are also compatible with a finite mobility edge at the
transition~\cite{Bonati:2020lal}.  Independently of this, the order of
the chiral transition is determined by the behaviour of the chiral
condensate rather than that of the mobility edge. In this respect, the
radical change in the nature of the low modes would more naturally
suggest a finite discontinuity in the condensate, and so a first-order
transition, but a continuous behaviour cannot be excluded.  In any
case, the fact that the anomalous remnant is identically zero in one
phase and nonzero in the other indicates the presence of a
non-analyticity in the partition function (as a function of an
extended set of variables including a suitable chemical potential),
and so that the phase transition is genuine. This is the case as long
as a nonzero anomalous remnant appears, irrespectively of whether it
is able to remove the Goldstone excitations from the spectrum or not.

It would be interesting to study how localisation of finitely-dense
low modes could be included in the theoretical analysis of phase
transitions in gauge systems, extending the standard analysis of
Ref.~\cite{Pisarski:1983ms}. This predicts a second-order chiral
transition for $N_f=2$ and a first-order one for $N_f\ge 3$ (see,
however, the recent analysis of Ref.~\cite{Fejos:2022mso} claiming the
possibility of a continuous phase transition also in this
case). According to the standard lore, isolated first-order points are
not expected, and a line of first-order transitions should reach out
from zero mass to a critical endpoint at some nonzero fermion mass
where the transition is second order. However, no second-order
endpoint has been observed so far for $N_f=3$, and there are recent
claims that no such critical endpoint is present up to $N_f=6$
flavours of light fermions~\cite{Cuteri:2021ikv}. The inclusion of
localisation effects may lead to revise one's expectations.

A transition from delocalised to localised finitely-dense near-zero
modes could be followed by a second transition at a higher
temperature, where the density of near-zero modes goes to zero and
chiral symmetry gets fully restored. This kind of scenario would fit
what is known about $N_f=2$ massless adjoint
QCD~\cite{Karsch:1998qj,Engels:2005te}: one would find Goldstone modes
being ``weakened'' at the deconfinement transition by the formation of
an anomalous remnant [like in scenario (ii) discussed in Section
\ref{sec:fate_goldstone}], and then gradually disappearing from the
spectrum until full restoration at higher temperature.  It is also
possible that extending the study of
Refs.~\cite{Karsch:1998qj,Engels:2005te} to larger volumes the effects
of localised near-zero modes become fully visible, and Goldstone modes
get entirely removed from the spectrum at deconfinement [scenario
(iii) in Section \ref{sec:fate_goldstone}].  In either case, if
something similar happened with massless fundamental fermions, then
the fact that in QCD (approximate) deconfinement and chiral
restoration take place together in the same relatively narrow
temperature range would be a consequence of the ``blurring'' of the
two separate sharp transitions into a single analytic crossover.

Whether any of the non-standard scenarios discussed in this paper
applies to QCD depends on the behaviour of the finite peak of
near-zero localised modes observed at near-physical quark
masses~\cite{Dick:2015twa}. Such a peak is known to be present in pure
gauge theory~\cite{Vig:2021oyt}, and so in the limit of large quark
masses; and to survive at the lower-than-physical quark masses
investigated so far (although localisation properties were not
studied)~\cite{Ding:2020xlj,Kaczmarek:2021ser,Ding:2021gdy,
  Ding:2021jtn}. If it survived the chiral limit, this peak would then
be a common feature in the whole quark mass range, most likely
appearing at the same temperature where near-zero modes first become
localised. This is the case in pure gauge theory, where this
temperature also coincides with the deconfinement
temperature~\cite{Kovacs:2017uiz,Vig:2020pgq}. Moreover, if the peak
consisted of localised modes all the way to the chiral limit, then the
temperature at which it appears there would mark a genuine phase
transition, distinct from the chiral transition, even if the peak had
a vanishing width in this limit (as long as this does not vanish too
fast with $m$). This would naturally, albeit unexpectedly, suggest to
identify the deconfinement transition in the presence of dynamical
fermions as the transition to localised near-zero modes. Presumably,
for any finite quark mass the near-zero peak disappears at some
sufficiently high temperature, corresponding to a second
pseudocritical temperature above the crossover one for physical
masses. In the chiral limit, this should extrapolate to the critical
temperature of a second transition, where the near-zero peak
disappears and chiral symmetry is restored.

The crucial question is whether or not the two transitions coincide in
the chiral limit. If they do, then localisation of near-zero modes and
chiral restoration take place at the same temperature, and no peak of
localised near-zero modes ever forms at $m=0$. This would fit with the
disordered medium scenario of Ref.~\cite{Diakonov:1995ea}, according
to which the accumulation of Dirac eigenvalues near the origin takes
place together with the delocalisation of the corresponding
eigenmodes. If the two transitions do not coincide, an intermediate
phase would appear where chiral symmetry is broken and Goldstone
excitations are weakened or absent altogether. Obviously, this is
possible only if the temperature where near-zero modes localise is
below the chiral transition temperature. A recent estimate of the
latter is $T_c^0\simeq 132\,{\rm MeV}$~\cite{HotQCD:2019xnw}, obtained
by extrapolating observables to zero light-quark mass according to
O(4) scaling. This suggests the presence of massless Goldstone modes
below $T_c^0$, which can be reconciled with the presence of an
intermediate phase in the chiral limit if scenario (ii) with weakened
Goldstone excitations is realised. On the other hand, the localisation
temperature is known only at the physical point, where it is obtained
by extrapolation from higher temperatures~\cite{Kovacs:2012zq}. While
the result clearly falls within the crossover range, somewhere above
the position of the peak of the chiral susceptibility, the accuracy of
the extrapolation is not fully under control. To this end, it would be
interesting to improve on the results of Ref.~\cite{Kovacs:2012zq} and
determine the temperature at which localisation appears by a direct
study near the crossover at physical and lower-than-physical quark
masses.

It should be noted that the existence of an intermediate phase in QCD,
between the crossover temperature and a (much) higher one, has been
suggested several times in the literature, although with various
different motivations (see, e.g.,
Refs.~\cite{Glozman:2019fku,Alexandru:2019gdm} and the discussion in
Ref.~\cite{Cardinali:2021mfh}). It would be interesting to investigate
what happens to the near-zero peak at the higher temperatures where
the conjectured intermediate phase should end.

\acknowledgments{I thank A.~Portelli for getting me interested in the
  issue of localisation in the chiral limit, C.~Bonati, M.~D'Elia,
  S.~D.~Katz, D.~N\'ogr\'adi, A.~P\'asztor, and Zs.~Sz\'ep for useful
  discussions, G.~Morchio and F.~Strocchi for correspondence, and
  T.~G.~Ko\-v\'acs for discussions and for a careful reading of the
  manuscript. I also thank an anonymous referee for helpful
  suggestions. This work is partially supported by the NKFIH grant
  KKP-126769.}

\appendix

\section{Analyticity and reality properties of two-point functions}
\label{sec:anrealcorr}

\begin{figure}[t]
  \centering
  \includegraphics[width=0.475\textwidth]{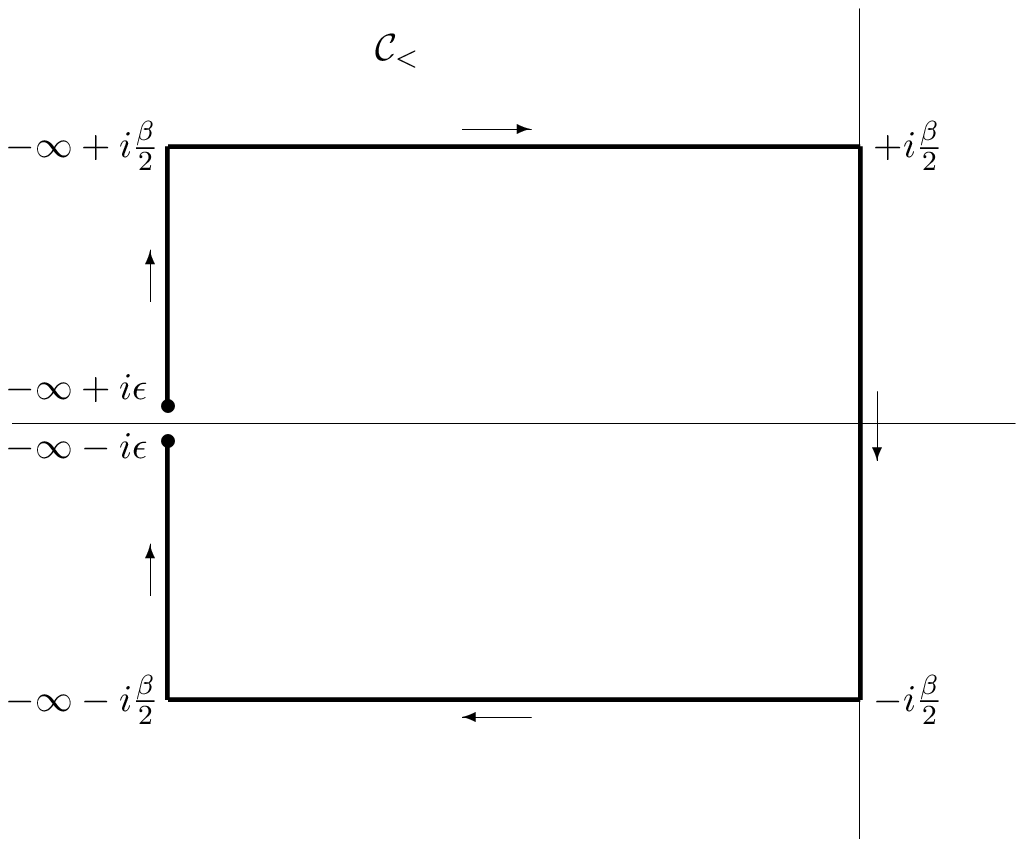}
  \hfil\hfil
  \includegraphics[width=0.475\textwidth]{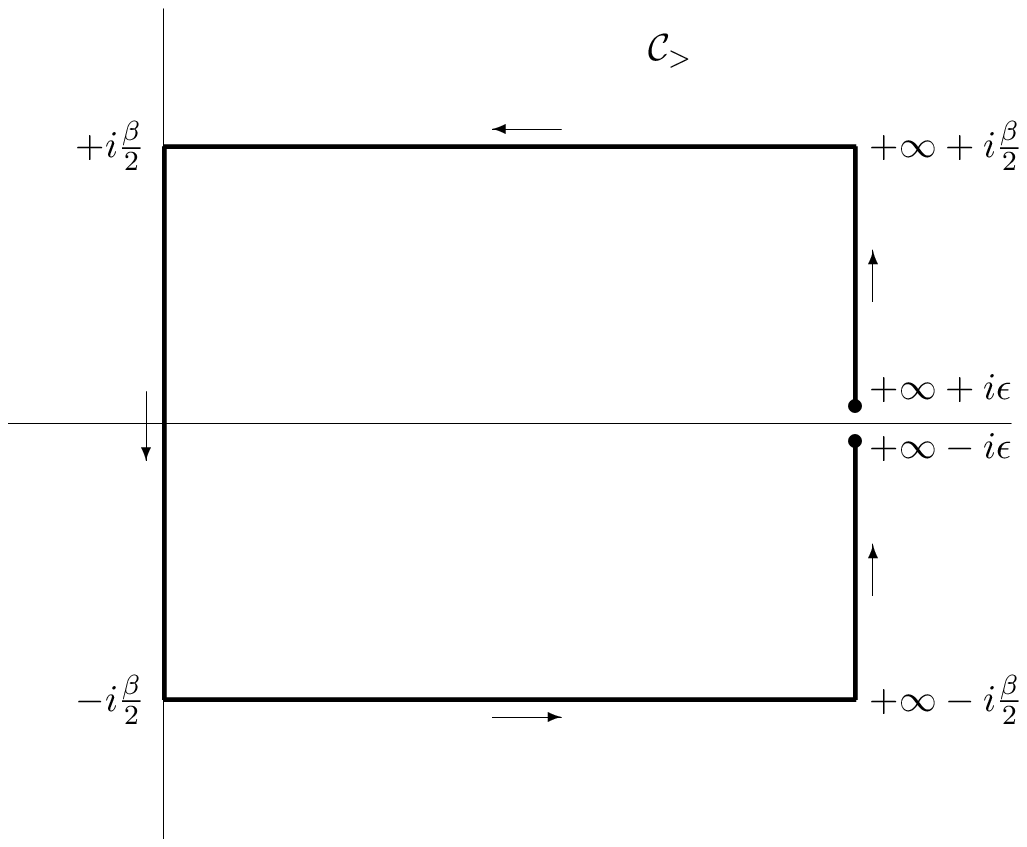}
  \caption{Integration paths in Eq.~\protect{\eqref{eq:GT23}}.}
  \label{fig:int_path}
\end{figure}

In this Appendix I provide further details on the analytic
continuation of two-point functions and on their reality properties.

\paragraph{Analytic continuation relations}

The analytic continuation relations, Eqs.~\eqref{eq:GT20} and
\eqref{eq:spec_func_def3}, between the Fourier coefficients of the
Euclidean correlator, Eq.~\eqref{eq:corrfunc9ter}, and the retarded
and advanced propagators, Eq.~\eqref{eq:GT19}, can be obtained
following the approach of Ref.~\cite{Bros:1996mw} based on the
deformation of the integration path of a suitable complex
integral. One defines the integrals
\begin{equation}
  \label{eq:GT23}
  I_\gtrless(\omega,\vec{k}\,) \equiv  i\int_{{\cal C}_\gtrless} dz \int d^3x\,e^{i(\omega
    z - \vec{k}\cdot\vec{x})}\Gc_{\phi_1\phi_2}(z,\vec{x})  \,,
\end{equation}
for $\Im \omega \gtrless 0$, respectively, where the paths
${\cal C}_\gtrless$ are shown in Fig.~\ref{fig:int_path}.  The
contribution of the paths along the imaginary direction at large
$|\Re z|$ are suppressed exponentially as the paths are pushed to
infinity and so can be discarded. Independently of $\omega$, the paths
can be shrunk to run close to the real axis, leading to
\begin{equation}
  \label{eq:GT24}
  \begin{aligned}
    I_>(\omega,\vec{k}\,) & = \tilde{r}_{\phi_1\phi_2}(\omega,\vec{k})
    \,, \quad\Im \omega> 0\,; &&& I_<(\omega,\vec{k}\,) & =
    \tilde{a}_{\phi_1\phi_2}(\omega,\vec{k}) \,, \quad\Im \omega< 0\,.
  \end{aligned}
\end{equation}
On the other hand, for the special values $\omega=i\omega_n$,
$n\neq 0$, one finds that thanks to periodicity the contributions from
the paths reaching to infinity cancel each other out, so that
\begin{equation}
  \label{eq:GT25biss}
  I_>(i\omega_n,\vec{k}\,)=
  \tilde{\Gsc}_{\phi_1\phi_2}(\omega_n,-\vec{k})\,,\quad n>0\,;
  \qquad
  I_<(i\omega_n,\vec{k}\,)=
  \tilde{\Gsc}_{\phi_1\phi_2}(\omega_n,-\vec{k})\,, \quad n<0\,.
\end{equation}
Comparing Eqs.~\eqref{eq:GT24} and \eqref{eq:GT25biss}, one obtains
Eq.~\eqref{eq:GT20}, as in Ref.~\cite{Bros:1996mw}.  The case $n=0$,
instead, was not discussed in detail there.  In this case one starts
from $I_>(i\epsilon,\vec{k}\,)$ and $I_<(-i\epsilon,\vec{k}\,)$,
eventually taking $\epsilon\to 0$.  One finds
\begin{equation}
  \label{eq:GT25_n0_long}
  \begin{aligned}
    I_{\gtrless}(\pm i\epsilon,\vec{k}\,) &=
    \int_{-\f{\beta}{2}}^{\f{\beta}{2}} d\tau \int d^3x\,e^{-i
      \vec{k}\cdot\vec{x}} \Gc_{\phi_1\phi_2}(-i\tau,\vec{x}) \\
    &\phantom{=} -\epsilon \beta \int d^4x\,\theta(\pm t) e^{\mp
      \epsilon t} e^{-i \vec{k}\cdot\vec{x}}
    \Gc_{\phi_1\phi_2}(t-i\tf{\beta}{2},\vec{x})\,,
\end{aligned}
\end{equation}
up to terms that vanish in the limit $\epsilon\to 0$, and so
\begin{equation}
  \label{eq:GT25_n0}
  \begin{aligned}
\lim_{\epsilon\to 0}    I_{\gtrless}(\pm i\epsilon,\vec{k}\,) &=
\tilde{\Gsc}_{\phi_1\phi_2}(0,-\vec{k}) -
    A_{\phi_1\phi_2}^{(\pm)}(\vec{k})\,,
\end{aligned}
\end{equation}
where
\begin{equation}
  \label{eq:tp_plus}
  A_{\phi_1\phi_2}^{(\pm)}(\vec{k}) \equiv \lim_{t\to \pm\infty} \beta\int
  d^3x\, e^{-i  \vec{k}\cdot\vec{x}}   \Gc_{\phi_1\phi_2}(t-i\tf{\beta}{2},\vec{x})\,,
\end{equation}
assuming that this quantity is finite.

\paragraph{Transport peak}

As discussed in Refs.~\cite{Meyer:2010ii,Meyer:2011gj},
$A_{\phi_1\phi_2}^{(\pm)}$ in Eq.~\eqref{eq:GT25_n0} can be identified
with the coefficient of the transport peak.  If
$\tilde{\rho}_{\phi_1\phi_2}/\omega$ has a Dirac delta singularity at
the origin [see Eq.~\eqref{eq:spec_func_def2}],
\begin{equation}
  \label{eq:tp_def}
  \tilde{\rho}_{\phi_1\phi_2}(\omega,\vec{k})= 2\pi
  A_{\phi_1\phi_2}(\vec{k})\omega\delta(\omega) +B_{\phi_1\phi_2}(\omega,\vec{k})\,,
\end{equation}
with $B_{\phi_1\phi_2}$ regular, one finds\footnote{While only the
  result for the retarded propagator is discussed in
  Refs.~\cite{Meyer:2010ii,Meyer:2011gj}, the one for the advanced
  propagator can be obtained by a simple extension of the
  calculation.}
\begin{equation}
  \label{eq:tp_def2}
  \tilde{\Gsc}_{\phi_1\phi_2}(0,-\vec{k})-\tilde{r}_{\phi_1\phi_2}(i\epsilon,\vec{k})
  = 
  \tilde{\Gsc}_{\phi_1\phi_2}(0,-\vec{k})-\tilde{a}_{\phi_1\phi_2}(-i\epsilon,\vec{k})
  = A_{\phi_1\phi_2}(\vec{k})\,,
\end{equation}
where the limit $\epsilon\to 0$ is understood, so that
$A_{\phi_1\phi_2}=A_{\phi_1\phi_2}^{(+)}=A_{\phi_1\phi_2}^{(-)}$.
Notice that if the Euclidean fields $\phi_{E1,2}$ corresponding to
$\phi_{1,2}$ have simple transformation properties under
time-reflection (see Section \ref{sec:syman}),
$\phi_{E1,2}(t,\vec{x})\to
\varsigma_{1,2}\phi_{E1,2}(\beta-t,\vec{x})$, $\varsigma_{1,2}=\pm 1$,
then
$ A_{\phi_1\phi_2}^{(\pm)}(\vec{k})=\varsigma_1\varsigma_2
A_{\phi_1\phi_2}^{(\mp)}(\vec{k})$, and so a transport peak can only
be present if $\varsigma_1\varsigma_2=1$.

\paragraph{Reality of Euclidean correlators}

The correlators $\Gsc_{AP\,\mu}(t,\vec{x})$ and
$\Gsc_{PP}(t,\vec{x})$, and more generally the Euclidean correlation
function of an arbitrary number of non-singlet axial-vector and
pseudoscalar densities, are real functions. To see this, consider the
generating functional
\begin{equation}
  \label{eq:anp1}
  \begin{aligned}
    Z_{A,P}[j] &\equiv \int_\beta [\Ds\psi][\Ds\bar{\psi}]
    e^{-\int_\beta d^4x\,\bar{\psi}K[j]\psi}\,, &&& K[j] &\equiv \slaD
    + m\mathbf{1} + j^a_{A\mu}\gamma_\mu\gamma_5 t^a + j^a_P\gamma_5
    t^a \,,
  \end{aligned}
\end{equation}
where $j$ denotes collectively the set of real sources $j^a_{A\mu}(x)$
and $j^a_P(x)$.  Performing the Grassmann integral one obtains 
\begin{equation}
  \label{eq:anp2}
  Z_{A,P}[j] = {\rm Det}\, K[j]\,.
\end{equation}
Since $ K[j]^\dag = \gamma_5 K[j] \gamma_5$, one finds that
$Z_{A,P}[j]$ is real,
\begin{equation}
  \label{eq:anp3}
  \begin{aligned}
 Z_{A,P}[j]^*    =  ({\rm Det}\, K[j])^* &=  {\rm Det}\, K[j]^\dag 
  = {\rm Det}\, K[j] =  Z_{A,P}[j]\,.
  \end{aligned}
\end{equation}
Taking functional derivatives with respect to the currents, setting
$j=0$, and averaging over gauge fields as in
Eq.~\eqref{eq:gauge_corr_func1},\footnote{In Lorenz gauge, the gauge
  action and integration measure are manifestly real.} the claimed
result follows. This formal derivation holds also in a regularised,
lattice version of the theory as long as the discretised Dirac
operator satisfies the $\gamma_5$-Hermiticity property
$\gamma_5\slaD\gamma_5= \slaD^\dag$.

\paragraph{Reality properties of thermal expectation values}

A consequence of the reality property discussed above is that the
analytic extensions\footnote{While contact terms may be present in the
  Euclidean correlation functions, it is understood that they are not
  involved in the process of analytic continuation.} of
$\Gsc_{AP\,\mu}$ and $\Gsc_{PP}$ in the complex-$t$ plane satisfy the
Schwarz reflection principle, i.e.,
\begin{equation}
  \label{eq:schwarz_princ}
  \Gsc_{AP\,\mu}(z^*,\vec{x})= \Gsc_{AP\,\mu}(z,\vec{x})^*\,,\qquad
  \Gsc_{PP}(z^*,\vec{x})= \Gsc_{PP}(z,\vec{x})^*\,. 
\end{equation}
This can be combined with the symmetry properties and the periodicity
of the correlators as follows. Setting $z=\epsilon -it$, one finds,
using the analytic extension of the relation
Eq.~\eqref{eq:sym_2point}, that
\begin{equation}
  \label{eq:anp5}
  \begin{aligned}
    \Gsc_{AP\,\mu}(\epsilon+it,\vec{x})^*&=
    \Gsc_{AP\,\mu}(\epsilon-it,\vec{x}) =
    \zeta_\mu\Gsc_{AP\,\mu}(\beta-\epsilon+it,\vec{x}) 
    = \zeta_\mu\Gsc_{AP\,\mu}(-\epsilon+it,\vec{x}) \,.
  \end{aligned}
\end{equation}
Using the analytic continuation relations
Eq.~\eqref{eq:GT19ext_1_main}, one sees that these are relations
between the real-time correlation functions of the axial-vector and
pseudoscalar Minkowskian operators, that can be summarised as
\begin{equation}
  \label{eq:anp6}
  \begin{aligned}
    \lla \hat{\Ac}^a_{\mu}(t,\vec{x})i\hat{\Poc}^b(0)\rra_\beta^* &=
    \lla i\hat{\Poc}^b(0)\hat{\Ac}^a_{\mu}(t,\vec{x})\rra_\beta\,.
  \end{aligned}
\end{equation}
Similarly, using Eq.~\eqref{eq:sym_2point3}, one finds
\begin{equation}
  \label{eq:anp9}
  \begin{aligned}
    \Gsc_{PP}(\epsilon+it,\vec{x})^*&= \Gsc_{PP}(\epsilon-it,\vec{x})
    = \Gsc_{PP}(\beta-\epsilon+it,\vec{x}) =
    \Gsc_{PP}(-\epsilon+it,\vec{x}) \,,
  \end{aligned}
\end{equation}
and, using again Eq.~\eqref{eq:GT19ext_1_main},
\begin{equation}
  \label{eq:anp10}
  \begin{aligned}
    \lla \hat{\Poc}^{a}(t,\vec{x})\hat{\Poc}^b(0)\rra_\beta^* &= \lla
    \hat{\Poc}^b(0) \hat{\Poc}^{a}(t,\vec{x})\rra_\beta\,.
\end{aligned}
\end{equation}
Equations \eqref{eq:anp6} and \eqref{eq:anp10} imply
\begin{equation}
  \label{eq:anp11}
  \begin{aligned}
    \lla [\hat{\Ac}_\mu^{a}(t,\vec{x}),\hat{\Poc}^b(0)]\rra_\beta^* &=
    \lla [\hat{\Ac}_\mu^{a}(t,\vec{x}),\hat{\Poc}^b(0)]\rra_\beta\,,
    \\
    \lla [\hat{\Poc}^{a}(t,\vec{x}),\hat{\Poc}^b(0)]\rra_\beta^* &=
    -\lla [\hat{\Poc}^{a}(t,\vec{x}),\hat{\Poc}^b(0)]\rra_\beta\,,
  \end{aligned}
\end{equation}
for the thermal expectation values of the commutators, and in turn
\begin{equation}
  \label{eq:anp11_bis}
  \begin{aligned}
    \tilde{c}(\omega,\vec{k})^* &= \tilde\rho_{A^{0a}
      P^a}(\omega,\vec{k})^* = \tilde\rho_{A^{0a}
      P^a}(-\omega,-\vec{k})
    = \tilde{c}(-\omega,-\vec{k})\,,\\
    \tilde{c}^P(\omega,\vec{k})^* &= \tilde\rho_{P^a
      P^a}(\omega,\vec{k})^* = - \tilde\rho_{P^a
      P^a}(-\omega,-\vec{k})= -\tilde{c}^P(-\omega,-\vec{k})\,,
  \end{aligned}
\end{equation}
for the relevant spectral functions [see
Eq.~\eqref{eq:GT39ext2_def_0}].

\section{Non-singlet axial Ward-Takahashi identities}
\label{sec:app_WTid}

In this Appendix I review the derivation of the non-singlet axial
Ward-Takahashi identities. In general, Ward-Takahashi identities are
obtained observing that a change of integration variables in the path
integral trivially leaves the result unchanged, and applying this
observation to a change of variables through the infinitesimal form of
some continuous symmetry but with $x$-dependent parameters.

In the case at hand, one upgrades the infinitesimal form of the
non-singlet axial symmetry transformations, Eq.~\eqref{eq:sua}, to the
$x$-dependent transformations $\psi\to \psi + \delta_{A}\psi$ and
$\bar{\psi}\to \bar{\psi} + \delta_{A}\bar{\psi}$, with
\begin{equation}
  \label{eq:sua_inf_app}
  \delta_A\psi(x) = i\epsilon_a(x) t^a\gamma_5\psi(x)\,, \qquad
  \delta_A\bar{\psi}(x) = i\epsilon_a(x)\bar{\psi}(x) t^a\gamma_5\,,
\end{equation}
with infinitesimal $\epsilon_a(x)$. Since the functional integration
measure is invariant thanks to $\tr t^a=0$, one obtains for any
observable $\Oc$
\begin{equation}
  \label{eq:WI3}
  \la \delta_{A}\Oc\ra_\beta = \la\Oc\delta_{A}S_{\rm F}\ra_\beta\,,
  \qquad   \delta_A S_{\rm F}
  = i\int_\beta d^4x\,\epsilon_a(x) \left(-\de_\mu\Ac^a_\mu(x)      +
    2m \Poc^a(x) 
  \right) \,, 
\end{equation}
with $\Ac^a_\mu$ the flavour non-singlet axial-vector currents, and
$\Poc^a$ the flavour non-singlet pseudoscalar densities, defined in
Eq.~\eqref{eq:bil}.  Since $\epsilon_a(x)$ is infinitesimal but
otherwise arbitrary, one finds Eq.~\eqref{eq:WI4},
\begin{equation}
  \label{eq:WI4_app}
  \left\la \left( -\de_\mu \Ac^a_\mu(x) + 2m\Poc^a(x)   \right) \Oc\right\ra_\beta
  = \left\la -i\f{\delta_A
      \Oc}{\delta\epsilon_a(x)}\right\ra_\beta\,. 
\end{equation}
In the case $\Oc = \Poc^b(y)$, a straightforward calculation shows
that
\begin{equation}
  \label{eq:WI5}
  -i\f{\delta_A \Poc^b(y)}{\delta\epsilon_a(x)} =
  \delta^{(4)}(x-y)\left(\f{1}{N_f} \delta^{ab} \Sc(y) + d^{abc}
    \Sc^c(y)\right)\,, 
\end{equation}
where the flavour singlet and flavour non-singlet scalar densities
$\Sc$ and $\Sc^a$ read
\begin{equation}
  \label{eq:bil3}
  \Sc(x) \equiv\bar{\psi}(x)  \psi(x)        \,, \qquad 
  \Sc^a(x) \equiv\bar{\psi}(x) t^a    \psi(x)        \,,
\end{equation}
with the totally symmetric symbol $d^{abc}$ defined through
\begin{equation}
  \label{eq:dabc}
  \{t^a,t^b\} = \f{1}{N_f}\delta^{ab} + d^{abc}t^c\,.
\end{equation}
The four-dimensional Dirac delta in Eq.~\eqref{eq:WI5} is periodic in
time, see Eq.~\eqref{eq:Ddelta}. Invariance under vector flavour
transformations implies that
\begin{equation}
  \label{eq:WI7}
  \la  \bar{\psi}_{f_1}(0)\psi_{f_2}(0) \ra_\beta
  \equiv \delta_{f_1 f_2} \Sigma\,,
\end{equation}
with $\Sigma$ the chiral condensate, and so
\begin{equation}
  \label{eq:app_condensate}
  \la  \Sc(0) \ra_\beta = N_f\Sigma
  \,, \qquad    \la  \Sc^a(0) \ra_\beta = 0 \,. 
\end{equation}
One then finds the Ward-Takahashi identity Eq.~\eqref{eq:WI6},
\begin{equation}
  \label{eq:WI6_app}
  -\de_\mu \la   \Ac^a_\mu(x)  \Poc^b(0)\ra_\beta
  +  2m \la \Poc^a(x)  \Poc^b(0)\ra_\beta
  = \delta^{(4)}(x)\delta^{ab}\Sigma\,.
\end{equation}
Further exploiting vector flavour invariance, one has
\begin{equation}
  \label{eq:sym_2point_def_app}
  \begin{aligned}
    \la \Ac^a_\mu(x)\Poc^b(0) \ra_\beta & \equiv
    \delta^{ab}\Gsc_{AP\,\mu}(x) \,, &&& \la \Poc^a(x) \Poc^b(0)
    \ra_\beta & \equiv \delta^{ab}\Gsc_{PP}(x) \,,
  \end{aligned}
\end{equation}
and so Eqs.~\eqref{eq:WI8} and \eqref{eq:WI8_mom} follow,
\begin{align}
  \label{eq:WI8_app}
  &
    -\de_\mu \Gsc_{AP\,\mu}(x)
    +  2m \Gsc_{PP}(x)
    = \delta^{(4)}(x)  \Sigma\,,\\
    \label{eq:WI8_2app}
 &     i\omega_n \widetilde{\Gsc}_{AP\,4}(\omega_n,\vec{k})
  +     i\vec{k}\cdot\vec{\mkern 0mu\widetilde{\Gsc}}_{AP}(\omega_n,\vec{k})
    +  2m \widetilde{\Gsc}_{PP}(\omega_n,\vec{k})
    =   \Sigma\,.
\end{align}
For future utility one defines also the flavour non-singlet vector
currents, $\Vc^a_\mu$, as
\begin{equation} 
  \label{eq:bil_bis}
  \Vc^a_\mu(x) \equiv\bar{\psi}(x) \gamma_\mu t^a    \psi(x)
  \,.
\end{equation}
Identities involving $\Oc=\Ac_\nu^b(y),\Vc_\nu^b(y)$, see
Eqs.~\eqref{eq:aa_WI} and \eqref{eq:ren2}, are obtained using
\begin{equation}
  \label{eq:WI5_A}
  -i\f{\delta_A \Ac_\nu^b(y)}{\delta\epsilon_a(x)} =
  -\delta^{(4)}(x-y)if^{abc}
  \Vc_\nu^c(y)\,,
  \qquad
  -i\f{\delta_A \Vc_\nu^b(y)}{\delta\epsilon_a(x)} =
  -\delta^{(4)}(x-y)if^{abc}
  \Ac_\nu^c(y)\,.
\end{equation}
In particular, Eq.~\eqref{eq:aa_WI} follows since
$\la\Vc_\nu^c\ra_\beta=0$ due to vector flavour symmetry (or to
rotation and reflection symmetries).

A second set of Ward-Takahashi identities is obtained
starting from the vector symmetry transformations Eq.~\eqref{eq:suv}.
Changing variables according to $\psi\to \psi + \delta_{V}\psi$ and
$\bar{\psi}\to \bar{\psi} + \delta_{V}\bar{\psi}$, with
\begin{equation}
  \label{eq:suv_inf}
  \delta_V\psi(x) = i\epsilon_a(x) t^a\psi(x)\,, \qquad
    \delta_V\bar{\psi}(x) = -i\epsilon_a(x) \bar{\psi}(x)t^a\,, 
\end{equation}
since the functional integration measure is invariant one obtains
for any observable $\Oc$
\begin{equation}
  \label{eq:WI3_bis}
  \la \delta_{V}\Oc\ra_\beta =\la  \Oc\delta_{V}S_{\rm F}\ra_\beta\,,
  \qquad  \delta_V S_{\rm F}
  = i\int_\beta d^4x\,\epsilon_a(x)\left(-\de_\mu \Vc^a_\mu(x)
  \right) 
  \,. 
\end{equation}
Since $\epsilon_a(x)$ is infinitesimal but otherwise arbitrary, one
finds
\begin{equation}
  \label{eq:WI4_bis}
  \la  -\de_\mu \Vc^a_\mu(x)     \Oc\ra_\beta
  = \left\la -i\f{\delta_V
      \Oc}{\delta\epsilon_a(x)}\right\ra_\beta\,.
\end{equation}

\section{Renormalisation of the Ward-Takahashi identity}
\label{sec:app_renorm}

In this Appendix I exploit the Ward-Takahashi identities
Eq.~\eqref{eq:WI4_app} [Eq.~\eqref{eq:WI4}] to discuss the
renormalisation of the relevant composite operators.  The point of
view is that explained in Section \ref{sec:reno}: correlation
functions are regularised by cutting off their Dirac mode
decomposition at some UV cutoff $\Lambda$, which leads to violations
of the Ward-Takahashi identities that nonetheless will vanish as
$\Lambda\to\infty$. This is guaranteed by the existence of
chiral-symmetry-respecting regularisations. One can then use the
Ward-Takahashi identities to constrain and relate the various UV
divergences. Since UV divergences are the same at zero and nonzero
temperature (see, e.g.,~\cite{Norton:1974bm,Kislinger:1975ab,
  Pisarski:1987wc}), renormalisation of the zero-temperature theory is
sufficient to make the finite-temperature theory finite as well. In
what follows $T=0$, i.e., $\beta=\infty$, unless specified otherwise.
The line of reasoning is standard (see, e.g.,
Refs.~\cite{Hasenfratz:1998jp,Vladikas:2011bp}). It is assumed that
the usual mass and coupling constant multiplicative renormalisations
(additive mass renormalisation being forbidden by chiral symmetry)
have already been carried out. In particular, the bare mass $m_B$ and
the renormalised mass $m$ are related by $m_B=Z_m m$.

\paragraph{Multiplicative renormalisation}

In the vector identities Eq.~\eqref{eq:WI4_bis}, with $\Oc$ an
arbitrary string of renormalised fundamental fields, one finds on the
right-hand side only a sum of finite contact terms, and so the
divergence $\de_\mu\Vc^a_{B\mu}$ of the bare vector currents is
finite. Since $\Vc^a_{B\mu}$ cannot mix with other operators of equal
or lower dimension for symmetry reasons, the renormalised currents are
simply $\Vc^a_{B\mu}=Z_V\Vc^a_{\mu}$, and finiteness of
$\de_\mu\Vc^a_{B\mu}$ implies finiteness of $Z_V$, that can be set to
$Z_V=1$.  In general, vector flavour invariance implies independence
of the flavour index $a$, and SO(4) invariance implies independence of
$\mu$.

The bare non-singlet axial currents $\Ac_{B\mu}^a$ and non-singlet
pseudoscalar densities $\Poc_B^a$, appearing in the axial
Ward-Takahashi identities Eq.~\eqref{eq:WI4_app}, cannot mix with
other operators, again due to symmetry reasons, so only multiplicative
renormalisation may be required. Additive divergences, however, can
still appear on the left-hand side of the identities in the form of
contact terms at $x=0$. These are discussed below in the case of
interest.  Let $\Ac^a_{B\mu} = Z_A \Ac^a_{\mu}$ and
$\Poc_B^a = Z_P \Poc^a$ relate the bare and the renormalised
axial-vector currents and pseudoscalar densities. Using
Eq.~\eqref{eq:WI4_app} with $\Oc$ a string of fundamental fields,
finiteness of the contact terms on the right-hand side implies that
$Z_A^{-1}Z_mZ_P$ is finite and can be set to 1. Alternatively, one
could use Eq.~\eqref{eq:WI6_app} at $x\neq 0$ to obtain the same
result. Taking instead $\Oc=\Vc^b_{B\nu}(y)\Ac^c_{B\rho}(z)$, one
finds the following identity for the renormalised fields,
\begin{equation}
  \label{eq:ren2}
  \begin{aligned}
    & Z_A^2\la \left(-\de_\mu \Ac^a_{\mu}(x) + 2m \Poc^a(x)\right)
    \Vc^b_{\nu}(y) \Ac^c_{\rho}(z)\ra_\beta \\ & \phantom{diogordo}
    = -Z_A^2 \delta^{(4)}(x-y) if^{abd} \la \Ac^d_{\nu}(y)
    \Ac^c_{\rho}(z) \ra_\beta - \delta^{(4)}(x-z) if^{acd} \la
    \Vc^b_{\nu}(y) \Vc^d_{\rho}(z) \ra_\beta\,,
  \end{aligned}
\end{equation}
valid for generic $\beta$ and in particular for $\beta=\infty$.
Taking $y\neq z$ and integrating in $x$ over a domain containing $z$
but not $y$, one finds a finite right-hand side, which implies that
$Z_A$ must be finite, and can be set to 1. One then concludes
$Z_V=Z_A=Z_m Z_P=1$.

\paragraph{Contact terms}

Renormalisation of the Ward-Takahashi identity Eq.~\eqref{eq:WI6_app}
requires also the subtraction of divergent contact terms at $x=0$.  In
general (at least in perturbation theory), UV divergences must be
polynomial in $m$ as a consequence of locality. Dimensional analysis
and the symmetries of the theory then constrain the divergent contact
terms $\delta^{ab}{\rm CT}_{AP}$ and $\delta^{ab}{\rm CT}_{PP}$,
appearing respectively in $ \la \Ac^a_{B\mu}(x)\Poc_B^b(0) \ra$ and
$ \la \Poc_B^a(x)\Poc_B^b(0) \ra$, to be of the following form,
\begin{equation}
  \label{eq:ren3}
  \begin{aligned}
    {\rm CT}_{AP}(x) &= \de_\mu
    \delta^{(4)}(x) m K_{AP}\,,\\
    {\rm CT}_{PP}(x) &= \left( \delta^{(4)}(x) \left(\Lambda^2
        K_{PP}^{(1)} + m^2 K_{PP}^{(2)}\right) +\Box \delta^{(4)}(x)
      K_{PP}^{(3)}\right)\,,
  \end{aligned}
\end{equation}
with $K_{AP}$, $K_{PP}^{(1,2,3)}$ dimensionless quantities, depending
logarithmically on $\Lambda$. Since the pseudoscalar-pseudoscalar
correlator appears multiplied by $m$, all these contact terms drop
from Eq.~\eqref{eq:WI6_app} in the chiral limit.  The spacetime and
flavour structure are determined by SO(4) and reflection invariance,
and by the unbroken vector flavour symmetry, respectively. The
dependence on $m$ is dictated by the fact that a
``${\cal R}_5$-parity'' transformation~\cite{Frezzotti:2003ni},
\begin{equation}
  \label{eq:R5_sym}
  \psi \to \gamma_5\psi\,, \qquad \bar{\psi}\to -\bar{\psi}\gamma_5\,,
\end{equation}
which is an element of the non-anomalous
${\rm SU}(N_f)_V\times {\rm SU}(N_f)_A$ symmetry group in the chiral
limit, is equivalent to changing the sign of the fermion mass. This
requires the expectation value of operators that are even
(respectively odd) under ${\cal R}_5$-parity to be even (respectively
odd) under $m\to -m$. Since the axial currents are odd while the
pseudoscalar densities are even, the mass dependence in
Eq.~\eqref{eq:ren3} follows.

The contact term on the right-hand side of Eq.~\eqref{eq:WI6_app} is
proportional to the chiral condensate, i.e., the expectation value of
the scalar density. This operator can mix with the identity operator,
and so requires both additive and multiplicative renormalisation,
$\Sc_B=Z_S\Sc + Z_{1}\mathbf{1}$.  The divergent part of the mixing
coefficient, $Z_{1}$, is determined by the same type of argument used
above to be of the form
\begin{equation}
  \label{eq:ren4}
  Z_{1} =  N_f\left( m \Lambda^2 K_S^{(1)} + m^3 K_S^{(2)} \right)\,,
\end{equation}
with dimensionless coefficients $K_S^{(1,2)}$ (again depending
logarithmically on $\Lambda$), since the scalar density is odd under
${\cal R}_5$-parity.  Also these terms drop from
Eq.~\eqref{eq:WI6_app} in the chiral limit, while matching the two
sides at finite $m$ one finds the relations $K_{AP}=2K_{PP}^{(3)}$,
$2K_{PP}^{(1)}=K_S^{(1)}$, and $2K_{PP}^{(2)}=K_S^{(2)}$.

\paragraph{Renormalised correlation functions}

Defining now the fully renormalised correlation functions and chiral
condensate via
\begin{equation}
  \label{eq:ren8}
  \begin{aligned}
    Z_AZ_P\Gsc_{AP\,\mu}(x) \delta^{ab}&= Z_AZ_P \la
    \Ac^a_{\mu}(x)\Poc^b(0) \ra_{\beta} = \la
    \Ac^a_{B\mu}(x)\Poc_B^b(0) \ra_{\beta}- \delta^{ab}{\rm CT}_{AP}(x)\,, \\
    Z_P^2\Gsc_{PP}(x)\delta^{ab}&= Z_P^2 \la \Poc^a(x)\Poc^b(0)
    \ra_{\beta} = \la \Poc_B^a(x)\Poc_B^b(0) \ra_{\beta} -
    \delta^{ab}{\rm CT}_{PP}(x)
    \,,\\
    Z_S\Sigma &= \f{1}{N_f} Z_S\la\Sc(0)\ra_\beta =
    \f{1}{N_f}(\la\Sc_B(0)\ra_\beta - Z_{1})\,,
  \end{aligned}
\end{equation}
one can use again the Ward-Takahashi identity to fix the
multiplicative renormalisation constant $Z_S$.  After renormalisation
and integration over spacetime, one finds from Eq.~\eqref{eq:WI6_app}
\begin{equation}
  \label{eq:ren9}
  Z_m Z_P^2 \, 2 m  \int_\beta d^4x \,  \la \Poc^a(x)\Poc^b(0) \ra_{\beta} =
  \delta^{ab} Z_S\Sigma\,,
\end{equation}
since for finite $m$ the integral of the divergence of the axial
currents gives zero contribution. One then concludes
$Z_S=Z_P=Z_m^{-1}$.

It should be noted that subleading terms in the regularised bare
chiral condensate $\f{1}{N_f}\la \Sc_B(0) \ra_{\beta}$, that vanish as
$\Lambda\to \infty$, could conspire with the UV divergences in $Z_S$
to give a finite contribution, generating further finite but more
singular contact terms on the right-hand side of
Eq.~\eqref{eq:WI6_app}.  However, the only other possible term allowed
by locality, ${\rm SO}(4)$ invariance, behaviour under
${\cal R}_5$-parity and dimensional analysis is
$m\Box\delta^{(4)}(x)$, which vanishes in the chiral limit and is
therefore irrelevant for the purposes of this paper.\footnote{At
  finite $m$, in momentum space this term becomes simply
  $m (\omega^2 + \vec{k}^{\,2})$ and so vanishes in the limit of zero
  frequency and zero spatial momentum.}

\section{Euclidean Goldstone theorem in coordinate space}
\label{sec:gt_coord}

In this Appendix I give another proof of Goldstone's theorem at finite
temperature, based on the Ward-Takahashi identity Eq.~\eqref{eq:WI6}
[Eq.~\eqref{eq:WI6_app}], obtained by working in coordinate space. To
this end, one defines the integrated correlation functions
\begin{equation}
  \label{eq:gt_coord_intcorr}
  \begin{aligned}
    \Gco(t) &\equiv \lim_{V\to\infty}\int_V d^3x \,
    \Gsc_{AP\,4}(t,\vec{x})\,,
    \\
    \Bco(t)&\equiv \lim_{V\to\infty} \int_V d^3x \,
    \vec{\nabla}\cdot\vec{\Gsc}_{AP}(t,\vec{x}) = \lim_{V\to\infty}
    \int_{\de V} d^2\vec{\Sigma}\cdot\vec{\Gsc}_{AP}(t,\vec{x}) \,,
    \\
    \Rco(t) &\equiv \lim_{V\to\infty} \int_V d^3x\,
    2m\Gsc_{PP}(t,\vec{x})= \lim_{V\to\infty} \int_V d^3x\,
    \Po(t,\vec{x})\,.
  \end{aligned}
\end{equation}
An infrared cutoff is imposed on the spatial integral in the form of a
finite volume $V$ with boundary $\de V$ (with outward-oriented
infinitesimal surface element $d^2\vec{\Sigma}$), which is removed
only at the end of the calculation. In particular, in the chiral limit
this is done \textit{after} the limit $m\to 0$ has already been
taken. Notice that this is only a cutoff on the integral and not on
the full theory, which is defined in infinite volume. Integrating
Eq.~\eqref{eq:WI6_app} over space one finds
\begin{equation}
  \label{eq:GT_coord_integrated}
  -\de_t \Gco(t) - \Bco(t) + \Rco(t) = \delta_P(t)
  \Sigma\,,
\end{equation}
where the periodic Dirac delta is defined in Eq.~\eqref{eq:Ddelta}.

\paragraph{Continuity properties of the integrated correlators}
  
To proceed further one needs to discuss first the continuity
properties at $t=0$ of the integrated correlators,
Eq.~\eqref{eq:gt_coord_intcorr}. Here the symmetry properties of the
Euclidean correlators under time reflection,
Eqs.~\eqref{eq:sym_2point} and \eqref{eq:sym_2point3}, are used. For
$\vec{x}\neq \vzero$, $\Gsc_{AP\,\mu}(z,\vec{x})$ is analytic for
complex $z=t-i\tau $ also for $t=0$ if $|\tau|< |\vec{x}|$, so in
particular $\Gsc_{AP\,\mu}(t,\vec{x})$ is continuous at $t=0$.  For
$\Gsc_{AP\,4}$ at $t=0$ one finds
\begin{equation}
  \label{eq:sym_2point4}
  \Gsc_{AP\,4}(0,\vec{x}) = -\Gsc_{AP\,4}(\beta,\vec{x})  \,.
\end{equation}
Combining Eq.~\eqref{eq:sym_2point4} with periodicity one concludes
$\Gsc_{AP\,4}(0,\vec{x})=0$ for $\vec{x}\neq \vzero$, while for
$\vec{x}= \vzero$ this does not follow since continuity at $t=0$ is
not guaranteed. In general then $\Gco(t)$ satisfies
$\Gco(\beta-t)=-\Gco(t)$ but need not be continuous at $t=0$.  On the
other hand, $\Gsc_{AP\,\mu}$ is continuous (in fact, analytic) at
$t=\f{\beta}{2}$ for any $\vec{x}$, and so
\begin{equation}
  \label{eq:sym_2point5}
  \Gsc_{AP\,4}\big(\tf{\beta}{2},\vec{x}\big) =
  -\Gsc_{AP\,4}\big(\tf{\beta}{2},\vec{x}\big) = 0  \,, \qquad
  \Gco(\tf{\beta}{2})=-\Gco(\tf{\beta}{2})=0\,.
\end{equation}
For the spatial components $\vec{\Gsc}_{AP}$,
Eq.~\eqref{eq:sym_2point} and periodicity imply
$\vec{\Gsc}_{AP}(t,\vec{x})=\vec{\Gsc}_{AP}(\beta
-t,\vec{x})=\vec{\Gsc}_{AP}(-t,\vec{x})$ for all $\vec{x}$, which
already follows from continuity for $\vec{x}\neq 0$. If the limits for
$t\to 0^\pm$ exist also for $\vec{x}=\vzero$, then
$\vec{\Gsc}_{AP}(t,\vec{0})$ must be continuous at $t=0$.

More interestingly, and independently of time-reflection symmetry, if
$\vec{\Gsc}_{AP}$ vanishes sufficiently fast at spatial infinity then
$\Bco(t)$ is continuous at $t=0$, since the point $\vec{x}=\vzero$
is not involved in the integral.  Moreover, as a consequence of the
regularity condition Eq.~\eqref{eq:GT21} discussed in Section
\ref{sec:reg_cond}, one finds that $\Bco(t)$ is constant in
(Euclidean) time. In fact, $\Bco(t)$ is obtained by summing
$\vec{k}\cdot\vec{\mkern 0mu\widetilde{\Gsc}}_{AP}$ over Matsubara
frequencies and taking the limit $\vec{k} \to 0$ [see
Eq.~\eqref{eq:corrfunc9bis}],
\begin{equation}
  \label{eq:B_const}
  \Bco(t) =  \lim_{\vec{k}\to 0}\f{1}{\beta}\sum_n e^{-i\omega_n t}
  (-i\vec{k})\cdot\vec{\mkern 0mu\widetilde{\Gsc}}_{AP}(\omega_n,\vec{k})
  =-\f{i}{\beta} \lim_{\vec{k}\to 0}\vec{k}\cdot\vec{\mkern
    0mu\widetilde{\Gsc}}_{AP}(0,\vec{k})\,, 
\end{equation}
which is time-independent.  Notice that for $\omega_0=0$ there is no
requirement from Eq.~\eqref{eq:GT21}, so that $\Bco(t)=\Bco$ need not
be zero in general.

Finally, Eq.~\eqref{eq:sym_2point3} implies that
$\Gsc_{PP}(t,\vec{x})$ is continuous at $t=0$ for
$\vec{x}\neq \vzero$, and also for $\vec{x}= 0$ if the limits
$t\to 0^\pm$ exist, like for $\vec{\Gsc}_{AP}$.  The quantity
$\Rco(t)$ may not be continuous at $t=0$ if $\Gsc_{PP}$ is divergent
at $x=0$, but for our purposes it suffices to assume that at $t=0$ it
develops at most an integrable singularity in $t$.

\paragraph{Chiral limit}

The properties discussed above are expected to hold also in the chiral
limit. In particular, as in Section \ref{sec:reg_cond}, it is assumed
that the regularity condition holds as $V\to\infty$ in the spatial
integral (i.e., $\vec{k}\to 0$ in momentum space) also after the
chiral limit has been taken.  Taking now the chiral limit in the
integrated Ward-Takahashi identity,
Eq.~\eqref{eq:GT_coord_integrated}, one finds
\begin{equation}
  \label{eq:GT_coord_remnant2}
  \begin{aligned}
    -\de_t \Gco_*(t) - \Bco_* + \Rco_*(t) &= \delta_P(t) \Sigma_*\,,
  \end{aligned}
\end{equation}
where I used Eq.~\eqref{eq:B_const}, and all quantities are computed
in the chiral limit (taken before $V\to\infty$ in the integral), as
denoted by the subscript $*$.  Further integrating over time between
$\f{\beta}{2}$ and $t\in(0,\beta)$, and using $\Gco(\f{\beta}{2})=0$
following from antisymmetry [see Eq.~\eqref{eq:sym_2point5}], one
finds
\begin{equation}
  \label{eq:GT_coord_remnant3}
  \Gco_*(t) = - \Bco_*\cdot\left(t-\f{\beta}{2}\right) +
  \int_{\f{\beta}{2}}^t dt'\, \Rco_*(t')\,, \qquad t\in(0,\beta)\,.
\end{equation}
This is then repeated periodically as
$\Gco_*(t+n\beta)=\Gco_*(t)$, $\forall n\in\mathbb{Z}$. 

The value of $\Bco_*$ is determined by the contact term at $t=0$ in
Eq.~\eqref{eq:GT_coord_remnant2}. Integrating this equation in an
infinitesimal neighbourhood $[-\epsilon,\epsilon]$ of $t=0$, using the
property $\Gco_*(-t)=\Gco_*(\beta-t)=-\Gco_*(t)$ which follows from
periodicity and antisymmetry under temporal reflection, and using
integrability of $\Rco_*$ at $t=0$, one finds
\begin{equation}
  \label{eq:GT_coord_remnant4}
\lim_{\epsilon\to 0}  - [ \Gco_*(\epsilon) - \Gco_*(-\epsilon)]
= \lim_{\epsilon\to 0} -2\Gco_*(\epsilon) =\Sigma_*\,.
\end{equation}
One the other hand, using the solution
Eq.~\eqref{eq:GT_coord_remnant3} and the property
$\Rco_*(t)=\Rco_*(\beta-t)$ one obtains
\begin{equation}
  \label{eq:GT_coord_remnant5}
- \Sigma_* = \lim_{\epsilon\to 0} 2\Gco_*(\epsilon) =   \Bco_*\beta +
2\int_{\f{\beta}{2}}^0 dt'\, \Rco_*(t')=   \Bco_*\beta -
2\int_0^{\beta} dt'\, \Rco_*(t')  \equiv  \Bco_*\beta -\Xi_*\,.
\end{equation}
Plugging this into Eq.~\eqref{eq:GT_coord_remnant3} one finally
obtains
\begin{equation}
  \label{eq:GT_coord_remnant6}
  \Gco_*(t) = - (\Sigma_* -\Xi_*)\left(\f{t}{\beta}-\f{1}{2}\right) +
  \int_{\f{\beta}{2}}^t dt'\, \Rco_*(t')\,, \qquad t\in(0,\beta)\,.
\end{equation}
In the standard case $\Rco_*=0$, the resulting function manifestly
satisfies the analyticity, continuity and bounded-growth hypotheses of
the reconstruction theorem of real-time Green functions discussed in
Ref.~\cite{Cuniberti:2001hm}.  This will hold also if $\Rco_*\neq 0$
under reasonable analyticity and bounded-growth assumptions for this
quantity.

\paragraph{Analytic continuation}

The analytic continuation required to obtain the spectral function is
most clearly done exploiting again the antisymmetry of $\Gco$,
extended by analytic continuation. One finds
\begin{equation}
  \label{eq:GT_coord_remnant7}
  \begin{aligned}
    \Gco_*(\epsilon+it)-\Gco_*(-\epsilon+it) &=
    \Gco_*(\epsilon+it)+ \Gco_*(\epsilon-it) \\ &= -(\Sigma_* -\Xi_*)
    + \int_{\f{\beta}{2}}^{\epsilon+it} dt'\, \Rco_*(t') +
    \int_{\f{\beta}{2}}^{\epsilon-it} dt'\, \Rco_*(t')\,,
  \end{aligned}
\end{equation}
where the complex paths appearing in the last two terms are chosen to
run along the real $t'$ axis from $\f{\beta}{2}$ to $\epsilon$, and
then on the axis $z = \epsilon + it'$ in the positive or negative $t'$
direction. If $\Rco_*(z)$ is free of singularities for
$\Re z\in (0,\beta)$, then the analytic continuation is unambiguous in
this strip. Exploiting the symmetry properties of $\Rco_*$, one finds
for the last two terms in Eq.~\eqref{eq:GT_coord_remnant7}
\begin{equation}
  \label{eq:GT_coord_remnant8}
  \begin{aligned}
    \int_{\f{\beta}{2}}^{\epsilon+it} dt'\, \Rco_*(t') +
    \int_{\f{\beta}{2}}^{\epsilon-it} dt'\, \Rco_*(t') 
    &= -\Xi_* + i\int_{0}^{t} dt'\, \left[\Rco_*(\epsilon +
      it')-\Rco_*(\epsilon - it')\right] \\ &= -\Xi_* + i\int_{0}^{t}
    dt'\, \left[\Rco_*(\epsilon + it')-\Rco_*(-\epsilon + it')\right]
    \\ &= -\Xi_* + i\int_{0}^{t} dt'\,\Lambda_*(t')\,,
  \end{aligned}
\end{equation}
where [recall Eq.~\eqref{eq:GT19ext_1_main}]
\begin{equation}
  \label{eq:def_lambda}
  \Lambda_*(t) \equiv \Rco_*(\epsilon + it')-\Rco_*(-\epsilon + it')=
  \lim_{V\to\infty}\int_V d^3x\, \lim_{m\to 0} 2m \lla
  [\hat{\Poc}^a(t,\vec{x}),\hat{\Poc}^a(0)]
  \rra_\beta\,.
\end{equation}
One then obtains [recall again Eq.~\eqref{eq:GT19ext_1_main}]
\begin{equation}
  \label{eq:GT_coord_remnant9}
  \begin{aligned}
    \Gco_*(\epsilon+it)-\Gco_*(-\epsilon+it) &=
    \lim_{V\to\infty}\int_V d^3x\, \lim_{m\to 0} \lla
    [\hat{\Ac}^{0a}(t',\vec{x}),\hat{\Poc}^a(0)] \rra_\beta \\ &=
    -\Sigma_* + i\int_{0}^{t} dt'\,\Lambda_*(t')\,.
  \end{aligned}
\end{equation}
In the standard case $\Rco_*=0$ (i.e., $\Po_*=0$, or no $1/m$
divergence in the pseudoscalar-pseudoscalar correlator), it is easy to
see that the axial-vector-pseudoscalar spectral function contains a
Dirac-delta term proportional to $\Sigma_*$:
\begin{equation}
  \label{eq:spec_dens_coord_app}
  \lim_{\vec{k}\to 0}  \tilde{c}_*(\omega,\vec{k}) =
  \lim_{V\to\infty}  \int dt \int_V d^3x\, e^{i\omega t} \lim_{m\to 0}\lla
  [\hat{\Ac}^{0a}(t,\vec{x}),\hat{\Poc}^a(0)] \rra_\beta = - 2\pi
  \Sigma_* \delta(\omega)\,.
\end{equation}
In the case $\Rco_*\neq 0$, one still generally finds a Dirac-delta
contribution to the spectral function, whose coefficient depends on
the large-time behaviour of the integral in
Eq.~\eqref{eq:GT_coord_remnant9}. This can be related back to the
Euclidean correlator by means of a path deformation argument analogous
to the one used in Ref.~\cite{Bros:1996mw} and above in Appendix
\ref{sec:anrealcorr}.  Consider
\begin{equation}
  \label{eq:GT_coord_remnant10}
  I=   \int_{{\cal C}_>} dz\,  \Rco_*(iz)\,, 
\end{equation}
where the path ${\cal C}_>$ is shown in Fig.~\ref{fig:int_path}.
Shrinking the path towards the imaginary axis one finds
\begin{equation}
  \label{eq:GT_coord_remnant11}
  I = i\int_0^\infty dt\, \left[\Rco_*(\epsilon +it)-\Rco_*(-\epsilon +it)\right]
  = i \int_0^\infty dt \,\Lambda_*(t)\,.
\end{equation}
On the other hand, periodicity implies that the integrals on the parts
of the path reaching to infinity cancel each other out, and so
\begin{equation}
  \label{eq:GT_coord_remnant12}
  I = \int_{-\f{\beta}{2}}^{\f{\beta}{2}}dt \,\Rco_*(t) =
  \int_{0}^{\beta}dt \,\Rco_*(t) = \Xi_*\,,
\end{equation}
assuming that $\Rco_*(z)$ vanishes for $\Im z\to \infty$, so that the
part of ${\cal C}_>$ at large $\Im z$ does not contribute. Then
\begin{equation}
  \label{eq:GT_coord_remnant13}
  \lim_{t\to +\infty}\Gco_*(\epsilon+it)-\Gco_*(-\epsilon+it) = -(\Sigma_*-\Xi_*)\,,
\end{equation}
and by antisymmetry
\begin{equation}
  \label{eq:GT_coord_remnant13_min}
  \lim_{t\to -\infty} \Gco_*(\epsilon+it)-\Gco_*(-\epsilon+it) =
  -      \lim_{t\to +\infty} \Gco_*(-\epsilon+it)-\Gco_*(\epsilon+it) =
  -(\Sigma_*-\Xi_*)\,.
\end{equation}
Isolating the contribution of the constant large-$|t|$ behaviour one
concludes that
\begin{equation}
  \label{eq:spec_dens_coord_app2}
  \begin{aligned}
    \lim_{\vec{k}\to 0} \tilde{c}_*(\omega,\vec{k}) &=
    \lim_{V\to\infty} \int dt \int_V d^3x\, e^{i\omega t} \lim_{m\to
      0}\lla [\hat{\Ac}^{0a}(t,\vec{x}),\hat{\Poc}^a(0)] \rra_\beta \\
    &= - 2\pi (\Sigma_*-\Xi_*) \delta(\omega) + \text{less
      singular}\,.
  \end{aligned}
\end{equation}
Since $\Xi_* = {\rm R}_*(0)$ in the notation of Section
\ref{sec:egt_ms} [see Eqs.~\eqref{eq:GT32} and
\eqref{eq:GT_neweqs_2}], the result Eq.~\eqref{eq:GTrev5_quater} is
reproduced.

If $\Rco_*(z)$ has a finite nonzero limit as $\Im z\to \infty$, then
under suitable boundedness conditions in the strips
$\Re z \in (0,\f{\beta}{2}]$ and $\Re z \in [-\f{\beta}{2},0)$, the
limits $\lim_{\tau\to\infty}\Rco_*(\pm t + i\tau)$,
$t\in (0,\f{\beta}{2}]$, are independent of $t$ due to the
Phragm\'en-Lindel\"of theorem (see, e.g.,
Ref.~\cite{titchmarsh1939theory}), and they also do not depend on the
sign due to periodicity fixing
$ \lim_{\tau\to\infty}\Rco_*(-\f{\beta}{2} + i\tau)=
\lim_{\tau\to\infty}\Rco_*(\f{\beta}{2} + i\tau)$.  One then finds an
extra contribution to $I$, corresponding to the effect of a transport
peak in the spectral density [see Eq.~\eqref{eq:tp_plus} and following
discussion, and footnote \ref{foot:transpeak}],
\begin{equation}
  \label{eq:gt_tp}
  \begin{aligned}
    I-\Xi_* &= -\beta \Rco_*(\tf{\beta}{2} + i\infty) = -\beta \int
    d^3x \,\lim_{m\to 0} 2m \Gsc_{PP} (\tf{\beta}{2} + i\infty)
    \\
    & = - \lim_{\vec{k}\to 0} \lim_{m\to 0} 2m A_{PP}(\vec{k}) =
    -A_*\,,
   \end{aligned}
\end{equation}
and so
$\lim_{\vec{k}\to 0} \tilde{c}_*(\omega,\vec{k})|_{\text{singular}}=
-2\pi (\Sigma_*-\Xi_*+A_*) \delta(\omega)$ (see footnote
\ref{foot:transpeak}).

\section{Exponentially localised modes}
\label{sec:expoenv}

In this Appendix I repeat the large-volume estimate of the
contributions of localised modes to $\spcorr_{s\,V,\,\Lambda}^\Gamma$
done in Section \ref{sec:pplargev}, using the more accurate
characterisation of localised modes as exponentially decreasing in
space, rather than strictly confined in a finite region. This improves
over the $x$-independent estimate
Eq.~\eqref{eq:large_v_estimate_same}, showing that an exponential
decay at large separation is to be expected.

Assume that localised modes are bounded by an envelope that
exponentially decays in space starting from some localisation centre
$\vec{x}_{0n}$,
\begin{equation}
  \label{eq:largev2}
  \Vert\psi_n(x) \Vert^2  \le 
  \f{K_n}{\ell_n^3} e^{-\f{|\vec{x}-\vec{x}_{0n}|}{\ell_n}}\,,
\end{equation}
with $\ell_n$ the localisation length of the mode, and $K_n$ a
positive constant. Using the bound [see the second line of
Eq.~\eqref{eq:largev_corrbound}]
\begin{equation}
  \label{eq:exploc_app1}
  \begin{aligned}
    \left| \la \Oc^\Gamma_{nn}(x) \Oc^\Gamma_{nn}(0) \ra_\beta\right|
    &\le \la \left|\Oc^\Gamma_{nn}(x) \right| \left|\Oc^\Gamma_{nn}(0)
    \right|\ra_\beta \le \la \Vert \psi_n(x)\Vert^2\Vert
    \psi_n(0)\Vert^2 \ra_\beta \\ &= \f{1}{\beta V}\int_\beta d^4y \,
    \la \Vert\psi_{n}(x+y)\Vert^2 \Vert\psi_{n}(y)\Vert^2 \ra_\beta
    \,,
  \end{aligned}
\end{equation}
where in the last passage I used translation invariance, together with
Eq.~\eqref{eq:largev2}, one finds
\begin{equation}
  \label{eq:exploc_app2}
  \begin{aligned}
    \left| \la \Oc^\Gamma_{nn}(x) \Oc^\Gamma_{nn}(0) \ra_\beta\right|
    &\le \f{1}{\beta V}\int_\beta d^4y \, \left\la \f{K_n^2}{\ell_n^6}
      e^{-\f{|\vec{x}+\vec{y}-\vec{x}_{0n}|+|\vec{y}-\vec{x}_{0n}|}{\ell_n}}\right\ra_\beta
    = \f{1}{V} \la K_n^2 R(\vec{x},\ell_n)\ra_\beta\,,
  \end{aligned}
\end{equation}
where
\begin{equation}
  \label{eq:exploc_app3}
  R(\vec{x},\ell)\equiv     \f{1}{\ell^6}\int d^3y \,
  e^{-\f{|\vec{x}+\vec{y}|+|\vec{y}|}{\ell}}
  \,.
\end{equation}
An explicit calculation shows that
\begin{equation}
  \label{eq:exploc_app4}
  R(\vec{x},\ell) =
  \f{\pi}{\ell^3}   e^{-\f{|\vec{x}|}{\ell}}\left(1+\f{|\vec{x}|}{\ell}
    +\f{1}{3}\f{|\vec{x}|^2}{\ell^2}\right)\,.
\end{equation}
One then qualitatively expects an exponential suppression of the
spectral correlators at large distance.

Under the further assumption that the support of the local
probability distribution of the localisation length at a given point
in the spectrum is bounded from above by some $\xi(\lambda)$, and if
$K_n$ is also locally bounded by some $K(\lambda)$, then since
\begin{equation}
  \label{eq:exploc_app5}
\f{\de}{\de\ell} \f{1}{\ell^3}e^{-\f{|\vec{x}|}{\ell}}\left(1+\f{|\vec{x}|}{\ell}
  +\f{1}{3}\f{|\vec{x}|^2}{\ell^2}\right) =
\f{1}{3\ell^4}e^{-\f{|\vec{x}|}{\ell}}\left(
\f{|\vec{x}|^3}{\ell^3} -
2\f{|\vec{x}|^2}{\ell^2}
-9\f{|\vec{x}|}{\ell}-9\right)\,,
\end{equation}
which is positive for $|\vec{x}|\ge c_0 {\ell}$, $c_0\simeq 4.466$,
one obtains the exponential bound
\begin{equation}
  \label{eq:exploc_app6}
  \begin{aligned}
    \left| \left\la \sum_n\delta(\lambda-\lambda_n) \Oc^\Gamma_{nn}(x)
        \Oc^\Gamma_{nn}(0) \right\ra_\beta\right| &\le
    \rho_{B,V}(\lambda) \f{\pi K(\lambda)^2}{\xi(\lambda)^3}
    e^{-\f{|\vec{x}|}{\xi(\lambda)}}\left(1+\f{|\vec{x}|}{\xi(\lambda)}
      +\f{1}{3}\f{|\vec{x}|^2}{\xi(\lambda)^2}\right)\,,
  \end{aligned}
\end{equation}
valid at large $|\vec{x}|\ge c_0\xi(\lambda)$.

\section{Pseudoscalar-pseudoscalar correlator in
  the chiral limit}
\label{sec:int_genarg}

In this Appendix I discuss the evaluation of the various contributions
to the pseudoscalar-pseudoscalar correlator in the chiral limit. For
the contributions involving large eigenvalues, I show that
independently of the subtraction procedure employed to deal with
additive divergences, the remaining finite contributions as
$\Lambda\to \infty$ stay finite also as $m\to 0$. I then identify what
contributions from the small eigenvalues can lead to a finite $\Po_*$
in the chiral limit.

It is assumed that finite limits exist as the argument $\lambda$ of
the spectral correlators $\spcorr_{\rm loc}^\Gamma$, and either or
both of the arguments $\lambda,\lambda'$ of the spectral correlators
$\bar{\spcorr}^\Gamma$, tend to zero. It is also assumed that such
limits are approached at least as fast as some power law. Finally, it
is assumed that the limit $m\to 0$ of the spectral correlators is well
defined.

\paragraph{Bounds on spectral integrals}

The following inequalities are used to derive bounds on the
asymptotic $m$-dependence of the relevant spectral integrals as $m\to 0$,
\begin{align}
  \label{eq:schwarz1}
  \left|  \int_0^\mu d\lambda\,\f{f(\lambda)}{\lambda^2+m^2} \right|&\le
                                                                      \left(\f{\pi}{2m} \right)^{1-\f{1}{2^N}}\left(\int_0^\mu
                                                                      d\lambda\,\f{|f(\lambda)|^{2^N}}{\lambda^2+m^2}\right)^{\f{1}{2^N}}
                                                                      \,, \\
    \label{eq:schwarz1_prime}
  \left|  \int_0^\mu d\lambda\,\f{f(\lambda)}{(\lambda^2+m^2)^2} \right|&\le
                                                                          \left(\f{1}{2\mu m^2}+\f{\pi}{4m^3} \right)^{1-\f{1}{2^N}}\left(\int_0^\mu
                                                                          d\lambda\,\f{|f(\lambda)|^{2^N}}{(\lambda^2+m^2)^2}\right)^{\f{1}{2^N}}
                                                                          \,,  
\end{align}
as well as
\begin{equation}
  \label{eq:schwarz1_again}
  \begin{aligned}
    \bigg| \int_0^\mu d\lambda\int_0^\mu d\lambda'\, &
    \f{F(\lambda,\lambda')}{(\lambda^2+m^2)(\lambda^{\prime \,2}+m^2)}
    \bigg| \\ &\le \left(\f{\pi}{2m}
    \right)^{2\left(1-\f{1}{2^N}\right)}\left(\int_0^\mu
      d\lambda\int_0^\mu
      d\lambda'\,\f{|F(\lambda,\lambda')|^{2^N}}{(\lambda^2+m^2)(\lambda^{\prime
          \,2}+m^2)}\right)^{\f{1}{2^N}} \,,
  \end{aligned}
\end{equation}
where $N$ is an arbitrary non-negative integer number.  These
inequalities follow from repeated application of the Cauchy-Schwarz
inequality, and from the following elementary results,
\begin{equation}
  \label{eq:schwarz23_bis0}
  \begin{aligned}
    \int_0^\mu d\lambda\,\f{1}{\lambda^2+m^2} &= \f{1}{m}\arctan\f{\mu}{m}\le \f{\pi}{2m}\,,    \\
    \int_0^\mu d\lambda\,\f{1}{(\lambda^2+m^2)^2} &=
    \f{1}{2m^2}\left(\f{\mu}{\mu^2+m^2} + \f{1}{m}\arctan\f{\mu}{m}
    \right) \le \f{1}{2m^3}\left( \f{\pi}{2} + \f{m}{\mu}\right) \,.
\end{aligned}
\end{equation}
As an example, Eq.~\eqref{eq:schwarz1} is obtained by noticing that
the case $N=0$ is obvious, and using the Cauchy-Schwarz inequality to
show that
\begin{equation}
  \label{eq:schwarz3_main}
  \begin{aligned}
    \left(\int_0^\mu
      d\lambda\,\f{|f(\lambda)|^{2^{N}}}{\lambda^2+m^2}\right)^{\f{1}{2^{N}}}
    & \le \left[ \left(\int_0^\mu
        d\lambda\,\f{|f(\lambda)|^{2^{N+1}}}{\lambda^2+m^2}\right)^{\f{1}{2}}
      \left(\int_0^\mu
        d\lambda\,\f{1}{\lambda^2+m^2}\right)^{\f{1}{2}}\right]^{\f{1}{2^N}}
    \\ & \le \left(\f{\pi}{2m}\right)^{\f{1}{2^{N+1}}}
    \left(\int_0^\mu
      d\lambda\,\f{|f(\lambda)|^{2^{N+1}}}{\lambda^2+m^2}\right)^{\f{1}{2^{N+1}}}
    \,,
  \end{aligned}
\end{equation}
from which the result follows since
$1-2^{-N}+2^{-(N+1)}= 1-2^{-(N+1)}$.

Let now $f(\lambda)$ vanish at least as fast as $\lambda^\gamma$ for
$\lambda\to 0$, i.e.,
$\lim_{\lambda\to 0}\lambda^{-\gamma}f(\lambda)<\infty$, with $\gamma >0$.  Using
Eq.~\eqref{eq:schwarz1}, one finds
\begin{equation}
  \label{eq:bound1}
  \begin{aligned}
    \left| \int_0^\mu d\lambda\,\f{m f(\lambda)}{\lambda^2+m^2}\right|
    & \le m \left(\f{\pi}{2m} \right)^{1-\f{1}{2^N}}\left(\int_0^\mu
      d\lambda\,\f{|f(\lambda)|^{2^N}}{\lambda^2+m^2}\right)^{\f{1}{2^N}}\\
    & \le m^{\f{1}{2^N}} \left(\f{\pi}{2}
    \right)^{1-\f{1}{2^N}}\left(\int_0^\mu
      d\lambda\,\f{|f(\lambda)|^{2^N}}{\lambda^2}\right)^{\f{1}{2^N}}
    = o(1)\,,
  \end{aligned}
\end{equation}
if $N$ is chosen such that $2^N \gamma > 1$, so that the last integral
is convergent. Similarly, using Eq.~\eqref{eq:schwarz1_prime}, one
finds
\begin{equation}
  \label{eq:bound2}
  \begin{aligned}
    \left| \int_0^\mu d\lambda\,\f{m^3
        f(\lambda)}{(\lambda^2+m^2)^2}\right| & \le m^3
    \left(\f{1}{2\mu m^2}+\f{\pi}{4m^3}
    \right)^{1-\f{1}{2^N}}\left(\int_0^\mu
      d\lambda\,\f{|f(\lambda)|^{2^N}}{(\lambda^2+m^2)^2}\right)^{\f{1}{2^N}}\\
    & \le m^{\f{3}{2^N}} \left(\f{m}{2\mu}+\f{\pi}{4}
    \right)^{1-\f{1}{2^N}}\left(\int_0^\mu
      d\lambda\,\f{|f(\lambda)|^{2^N}}{\lambda^4}\right)^{\f{1}{2^N}}
    = o(1)\,,
  \end{aligned}
\end{equation}
provided $2^N \gamma > 3$.  Finally, for $F(\lambda,\lambda')$
vanishing at least as fast as $\lambda^\gamma$ (respectively
$\lambda^{\gamma'}$) for $\lambda\to 0$ (respectively
$\lambda'\to 0$), using Eq.~\eqref{eq:schwarz1_again} one finds
\begin{equation}
  \label{eq:bound3}
  \begin{aligned}
    \bigg| \int_0^\mu d\lambda\,\int_0^\mu d\lambda'\, & \f{m^2
      F(\lambda,\lambda')}{(\lambda^2+m^2)(\lambda^{\prime \,2}+m^2)}
    \bigg| \\ &\le m^{\f{1}{2^{N-1}}} \left(\f{\pi}{2}
    \right)^{2\left(1-\f{1}{2^N}\right)}\left(\int_0^\mu
      d\lambda\int_0^\mu
      d\lambda'\,\f{|F(\lambda,\lambda')|^{2^N}}{(\lambda^2+m^2)(\lambda^{\prime
          \,2}+m^2)}\right)^{\f{1}{2^N}} \\ &\le m^{\f{1}{2^{N-1}}}
    \left(\f{\pi}{2}
    \right)^{2\left(1-\f{1}{2^N}\right)}\left(\int_0^\mu
      d\lambda\int_0^\mu
      d\lambda'\,\f{|F(\lambda,\lambda')|^{2^N}}{\lambda^2\lambda^{\prime
          \,2}}\right)^{\f{1}{2^N}} = o(1)\,,
  \end{aligned}
\end{equation}
provided $2^N \min(\gamma,\gamma') > 1$.

\paragraph{Large-$\lambda$ contributions}

Additive divergences originating from the large-eigenvalue region
appear in $\Gsc_{PP\,B}^{(12)}$, $\Gsc_{PP\,B}^{(21)}$, and
$\Gsc_{PP\,B}^{(22)}$. These are removed by subtracting the leading
contributions from the factors $(\lambda^2 + m^2)^{-1}$ and
$(\lambda^{\prime\, 2} + m^2)^{-1}$ appearing in
Eq.~\eqref{eq:mod_dec7_resplit_mult_reno} when the integration range
is $I_2=[\mu,\Lambda/Z_m]$. One has
\begin{equation}
  \label{eq:deno_expansion}
  \begin{aligned}
    \f{1}{\lambda^2 + m^2} = \left(
      \sum_{n=0}^{N-1}(-1)^n\f{m^{2n}}{\lambda^{2(n+1)}}\right) +
    \f{\left(-m^2\right)^{N}}{\lambda^{2N}(\lambda^2+m^2)}\,,
  \end{aligned}
\end{equation}
with the case of no subtraction corresponding to setting $N=0$.  This
shows that terms more suppressed in $\lambda^2$ contain higher powers
of $m^2$, and that the remainder is suppressed by powers of $m^2$ as
well. For sufficiently large $N$ the integral containing the remainder
is convergent. Independently of the specific procedure employed to
subtract the divergent part, the remaining finite contributions coming
from the $N$ leading terms in the expansion
Eq.~\eqref{eq:deno_expansion} are then suppressed by powers of $m$,
and can produce at most a constant term in the chiral limit. One
similarly shows that the same holds true for the contribution of the
remainders. It is assumed that the residual $m$ dependence of the
remaining integrals through the spectral correlators is regular, with
finite limits as $m\to 0$.\footnote{More precisely, the requirement is
  that the coefficients of the expansion of $Z_m^2\Gsc_{PP\,B}^{(ij)}$
  in powers of $\Lambda$ and of nested logarithms of $\Lambda/\mu$
  (including negative powers) remain finite as $m\to 0$.}

To show this in detail, for $\Gsc_{PP\,B}^{(12)}=\Gsc_{PP\,B}^{(21)}$
one writes
\begin{equation}
  \label{eq:largelambda_g12}
  \begin{aligned}
    -Z_m^2\Gsc_{PP\,B}^{(12)} &= \sum_{n=0}^{N-1}(-1)^nm^{2n}
    \int_0^\mu d\lambda\int_\mu^{\f{\Lambda}{Z_m}} d\lambda'\, \f{ m^2
      \bar{\spcorr}^{+}(\lambda, \lambda';x;m) +
      \lambda\lambda'\bar{\spcorr}^{-}( \lambda, \lambda';x; m)
    }{(\lambda^2 +m^2)\lambda^{\prime \, 2(n+1)}} \\ &\phantom{=} +
    m^{2N}\int_0^\mu d\lambda\int_\mu^\infty d\lambda'\, \f{ m^2
      \bar{\spcorr}^{+}(\lambda, \lambda';x;m) + \lambda\lambda'
      \bar{\spcorr}^{-}( \lambda, \lambda';x; m) }{(\lambda^2
      +m^2)(\lambda^{\prime\, 2} +m^2)\lambda^{\prime\,2N}}\,,
  \end{aligned}
\end{equation}
where I set
$\bar{\spcorr}^{\pm}\equiv \bar{\spcorr}^{\mathbf{1}}\pm
\bar{\spcorr}^{\gamma_5}$, and terms that vanish as
$\Lambda\to \infty$ have been omitted.  The integral over $\lambda'$
in the first $N$ terms is divergent as $\Lambda\to\infty$, but it does
not introduce any further mass dependence besides that of the spectral
correlators, which is assumed to be sufficiently regular. The further
mass dependence introduced by the integral over $\lambda$ is harmless,
since
\begin{equation}
  \label{eq:g12_bound0_alt}
  \begin{aligned}
    &\left| \int_0^\mu d\lambda\int_\mu^{\f{\Lambda}{Z_m}} d\lambda'\,
      \f{ m^2 \bar{\spcorr}^{+}(\lambda,\lambda';x;m) +
        \lambda\lambda' \bar{\spcorr}^{-}(\lambda,\lambda';x;m)
      }{(\lambda^2
        +m^2)\lambda^{\prime\,2(n+1)}} \right| \\
    &\le \int_0^\mu d\lambda\int_\mu^{\f{\Lambda}{Z_m}} d\lambda' \,
    \f{ |\bar{\spcorr}^{+}(\lambda,\lambda';x;m)|
    }{\lambda^{\prime\,2(n+1)}} + \int_0^\mu
    d\lambda\int_\mu^{\f{\Lambda}{Z_m}} d\lambda' \, \f{
      |\bar{\spcorr}^{-}(\lambda,\lambda';x;m)|
    }{\lambda\lambda^{\prime\,2n-1}} \,,
\end{aligned}
\end{equation}
where the integral over $\lambda$ in the second term on the second
line is convergent, since one has
$\bar{\spcorr}^{-}(0, \lambda';x;m) = 0$ [see
Eq.~\eqref{eq:sym_rel_sp_corr}], and since it is assumed that the
limit $\lambda\to 0$ is reached at least as fast as some power law.
These contributions have then a regular, at most $O(1)$ chiral
limit.\footnote{One can actually show that the contribution from
  $\bar{\spcorr}^{+}$ is $O(m)$.}  This must hold separately for the
divergent and finite parts, whose contributions to
$Z_m^2\Gsc_{PP\,B}^{(12)}$ are then constant or suppressed by powers
of $m$ in the chiral limit. The exact same argument with $n\to N$
works for the remainder term, since one can bound
$(\lambda^{\prime\,2} + m^2)^{-1} \le \lambda^{\prime\,-2}$.

For $\Gsc_{PP\,B}^{(22)}$, one splits
$\Gsc_{PP\,B}^{(22)}=\Gsc_{PP\,B}^{(22),1}+\Gsc_{PP\,B}^{(22),2}$,
with
\begin{equation}
  \label{eq:largelambda_g22_loc}
  \begin{aligned}
    -Z_m^2\Gsc_{PP\,B}^{(22),1} &= \sum_{n=0}^{N-1}(-1)^n m^{2n}
    \int_\mu^{\f{\Lambda}{Z_m}} d\lambda\, \left( \f{\spcorr_{{\rm
            loc}}^{\mathbf{1}}(\lambda;x;m)}{\lambda^{2(n+1)}} +
      \f{(m^2-\lambda^2) \spcorr_{{\rm
            loc}}^{\gamma_5}(\lambda;x;m)}{\lambda^{2(n+1)}(\lambda^2
        +m^2)} \right) \\ &\phantom{=}+ m^{2N} \int_\mu^{\infty}
    d\lambda\, \left( \f{\spcorr_{{\rm
            loc}}^{\mathbf{1}}(\lambda;x;m)}{\lambda^{2N}(\lambda^2
        +m^2)} + \f{(m^2-\lambda^2) \spcorr_{{\rm
            loc}}^{\gamma_5}(\lambda;x;m)}{\lambda^{2N}(\lambda^2
        +m^2)^2} \right)\,,
  \end{aligned}
\end{equation}
where I expanded only one power of $(\lambda^2+m^2)^{-1}$ in the
second term under the integrals, and
\begin{equation}
  \label{eq:largelambda_g22}
  \begin{aligned}
    -Z_m^2\Gsc_{PP\,B}^{(22),2} &= \sum_{n=0}^{N-1}
    \sum_{n'=0}^{N-1}(-1)^{n+n'}m^{2(n+n')}
    \int_\mu^{\f{\Lambda}{Z_m}} d\lambda\int_\mu^{\f{\Lambda}{Z_m}}
    d\lambda'\, \f{ W( \lambda, \lambda';x; m)
    }{\lambda^{2(n+1)} \lambda^{\prime \, 2(n'+1)}} \\
    &\phantom{=} + \sum_{n=0}^{N-1}(-1)^nm^{2n}
    \int_\mu^{\f{\Lambda}{Z_m}} d\lambda\int_\mu^{\f{\Lambda}{Z_m}}
    d\lambda'\, \f{ W( \lambda, \lambda';x; m)+ W( \lambda',
      \lambda;x; m)
    }{\lambda^{2(n+1)}(\lambda^{\prime\, 2} +m^2)\lambda^{\prime\,2N}} \\
    &\phantom{=} + m^{4N} \int_\mu^\infty d\lambda\int_\mu^\infty
    d\lambda'\, \f{ W( \lambda, \lambda';x; m) }{(\lambda^2
      +m^2)(\lambda^{\prime\, 2}
      +m^2)\lambda^{2N}\lambda^{\prime\,2N}}\,,
  \end{aligned}
\end{equation}
where I set
$W=m^2 \bar{\spcorr}^{+}+ \lambda\lambda'\bar{\spcorr}^{-}$ for
brevity and exploited the symmetry under
$\lambda\leftrightarrow\lambda'$ of the spectral correlators.  Terms
that vanish as $\Lambda\to\infty$ are again ignored.  In principle
different choices of $N$ may be needed for the two contributions from
localised modes; the generalisation is straightforward.  Clearly, all
integrals are convergent (at finite $\Lambda$) if one lets $m\to 0$ in
the denominators, and so $O(1)$ in the chiral limit. This holds for
divergent and finite parts separately, whose contributions to
$Z_m^2\Gsc_{PP\,B}^{(22)}$ are then constant or suppressed by powers
of $m$ in the chiral limit. One then concludes that, after subtraction
of the divergent parts and multiplicative renormalisation, the
contributions of $\Gsc_{PP\,B}^{(12)}$, $\Gsc_{PP\,B}^{(21)}$, and
$\Gsc_{PP\,B}^{(22)}$ to the renormalised pseudoscalar-pseudoscalar
correlator tend to a constant in the chiral limit [see
Eq.~\eqref{eq:CT_chiral}].

Notice that for finite $m$, the finite parts in the subtraction terms
used for $\Gsc_{PP}$ should be matched with those employed in the
renormalisation of the chiral condensate and of $\Gsc_{AP\,\mu}$, in
order to ensure that the Ward-Takahashi identity
Eq.~\eqref{eq:WI6_app} [Eq.~\eqref{eq:WI6}] holds after
renormalisation. This also determines how far in the expansion
Eq.~\eqref{eq:deno_expansion} one has to go. In any case, these finite
terms remain finite also as $m\to 0$ and so are irrelevant for the
chiral limit of $\Po = 2m\Gsc_{PP}$.

\paragraph{Small-$\lambda$ contributions}

I now discuss the contributions of small eigenvalues
($|\lambda|\le\mu$) to $\Po_*$ under the assumptions of Section
\ref{sec:ppchlim}, see Eq.~\eqref{eq:CT_chiral2_spec_again_0}.  In
this case one finds three types of contributions. The first
contribution, Eq.~\eqref{eq:CT_chiral2_spec_again_1}, comes from
localised near-zero modes. One sets
\begin{equation}
  \label{eq:chi_contrib0}
  \begin{aligned}
    \spcorr_{{\rm loc}}^{\Gamma}(\lambda;x;m) &= \spcorr_{{\rm
        loc}}^{\Gamma}(0;x;m) + \left(\spcorr_{{\rm
          loc}}^{\Gamma}(\lambda;x;m)- \spcorr_{{\rm
          loc}}^{\Gamma}(0;x;m)\right) \\
    &= \spcorr_{{\rm loc}}^{\mathbf{1}}(0;x;m) +
    f^\Gamma(\lambda;x;m)\,,
  \end{aligned}
\end{equation}
and assuming that $f^\Gamma$ vanishes at least as fast as some power
law as $\lambda\to 0$, one finds
\begin{equation}
  \label{eq:chi_contrib1}
  \begin{aligned}
    \int_0^{\lambda_c} d\lambda\, \f{\spcorr_{{\rm
          loc}}^{\mathbf{1}}(\lambda;x;m)}{\lambda^2 +m^2} &=
    \spcorr_{{\rm loc}}^{\mathbf{1}}(0;x;m) \int_0^{\lambda_c}
    d\lambda\, \f{1}{\lambda^2 +m^2} + \f{1}{m}\int_0^{\lambda_c}
    d\lambda\,
    \f{mf^{\mathbf{1}}(\lambda;x;m)}{\lambda^2 +m^2} \\
    &=\spcorr_{{\rm loc}}^{\mathbf{1}}(0;x;m) \f{1}{m}
    \arctan\f{\lambda_c}{m} + o(1/m)\,,
  \end{aligned}
  \end{equation}
having used Eq.~\eqref{eq:bound1}, and
\begin{equation}
  \label{eq:chi_contrib2}
  \begin{aligned}
    \int_0^{\lambda_c} d\lambda\, \f{(m^2-\lambda^2)\spcorr_{{\rm
          loc}}^{\gamma_5}(\lambda;x;m)}{(\lambda^2 +m^2)^2} &=
    \spcorr_{{\rm loc}}^{\gamma_5}(0;x;m) \int_0^{\lambda_c}
    d\lambda\, \f{m^2-\lambda^2}{(\lambda^2 +m^2)^2}\\ &\phantom{=} -
    \f{1}{m}\int_0^{\lambda_c} d\lambda\,
    \f{mf^{\gamma_5}(\lambda;x;m)}{\lambda^2 +m^2} + \f{2}{m}
    \int_0^{\lambda_c} d\lambda\,
    \f{m^3f^{\gamma_5}(\lambda;x;m)}{(\lambda^2 +m^2)^2}\\ & =
    \spcorr_{{\rm loc}}^{\gamma_5}(0;x;m) \f{1}{m}
    \f{\f{\lambda_c}{m}}{1+\f{\lambda_c^2}{m^2}} + o(1/m)\,,
  \end{aligned}
\end{equation}
having used Eq.~\eqref{eq:bound2}.  The $o(1/m)$ estimate for the
behaviour of the omitted terms is correct even if $\lambda_c$ is not
finite but vanishes in the chiral limit. In fact, since the omitted
terms are already $o(1/m)$ if $\lambda_c$ is treated as an independent
variable, then taking into account its dependence on $m$ can only make
them less divergent, as they involve integrals that vanish as
$\lambda_c\to 0$. This means that they cannot produce a divergence as
strong as or stronger than $1/m$ in any case, and a contribution to
$\Po_*$ can only come from the explicitly computed terms.

On the other hand, if $\lambda_c\to 0$ as $m\to 0$, it is not
guaranteed that these terms are actually the leading terms for the
pseudoscalar-pseudoscalar correlator.  In general, for
$\lambda\in[0,\lambda_c]$ one can bound
$|f^\Gamma|\le a_\Gamma\lambda^{\tau_\Gamma}$, for some $a_\Gamma$
independent of $m$ for sufficiently small $m$ and some
$\tau_\Gamma>0$.  One has
\begin{equation}
  \label{eq:chi_contrib2_extra}
  \begin{aligned}
    \left| \int_0^{\lambda_c} d\lambda\,
      \f{mf^{\mathbf{1}}(\lambda;x;m)}{\lambda^2 +m^2} \right| & \le
    \f{a_{\mathbf{1}}}{\tau_{\mathbf{1}}+1}
    \lambda_c^{\tau_{\mathbf{1}}}\f{\lambda_c}{m}\,,
    \\
    \left| \int_0^{\lambda_c} d\lambda\, \f{(m^2-\lambda^2)m
        f^{\gamma_5}(\lambda;x;m)}{(\lambda^2 +m^2)^2}\right| & \le
    \f{a_{\gamma_5}}{\tau_{\gamma_5}+1} \lambda_c^{\tau_{\gamma_5}}
    \f{\lambda_c}{m}\left(1+ \f{\lambda_c^2}{m^2}\right)\,,
\end{aligned}
\end{equation}
and so if $\lambda_c$ vanishes and $\lambda_c/m$ vanishes or remains
constant, one has that the omitted terms are actually
$o(\lambda_c/m^2)$, and so negligible compared to the explicitly
computed ones.  If instead $\lambda_c/m$ diverges in the chiral limit,
the first term in Eq.~\eqref{eq:chi_contrib1} is precisely $O(1/m)$,
and so the omitted terms are surely subleading. The first term in
Eq.~\eqref{eq:chi_contrib2}, instead, diverges only like $1/\lambda_c$
in this case, and so more slowly than $1/m$, thus not contributing to
$\Po_*$. While an explicit estimate shows that the omitted terms may
actually be leading in Eq.~\eqref{eq:chi_contrib2}, they are
nonetheless inconsequential for $\Po_*$, and for the qualitative fact
that the pseudoscalar-pseudoscalar correlator diverges as $m\to 0$, as
this follows already from the explicitly computed term.

The second contribution, Eq.~\eqref{eq:CT_chiral2_spec_again_1}, is
from localised modes separated from the origin of the spectrum.  One
easily finds
\begin{equation}
  \label{eq:chi_contrib3}
  \begin{aligned}
    &\left| \int_{\lambda_c'}^\mu d\lambda\,\left( \f{\spcorr_{{\rm
              loc}}^{\mathbf{1}}(\lambda;x;m)}{\lambda^2 +m^2} +
        \f{(m^2-\lambda^2)\spcorr_{{\rm
              loc}}^{\gamma_5}(\lambda;x;m)}{(\lambda^2
          +m^2)^2}\right)\right| \le 2\int_{\lambda_c'}^\mu
    d\lambda\,\f{\spcorr_{{\rm
          loc}}^{\mathbf{1}}(\lambda;x;m)}{\lambda^2}\,,
  \end{aligned}
\end{equation}
which is at most $O(1)$ under the assumption that $\lambda_c'$ remains
separated from the origin in the chiral limit.  This follows from
$|\spcorr_{{\rm loc}}^{\gamma_5}|\le \spcorr_{{\rm
    loc}}^{\mathbf{1}}$, which is a consequence of the bound
Eq.~\eqref{eq:largev0}. Therefore, these modes cannot contribute to
$\Po_*$.

Finally, the third contribution,
Eq.~\eqref{eq:CT_chiral2_spec_again_1}, comes from both localised and
delocalised modes. One sets
\begin{equation}
  \label{eq:chi_contrib4}
  \begin{aligned}
    \bar{\spcorr}^{\,\Gamma}(\lambda,\lambda';x;m) &= c(x;m) +
    f(\lambda;x;m) +
    f(\lambda';x;m)+F^\Gamma(\lambda,\lambda';x;m)\,,\\
    c(x;m)&=    \bar{\spcorr}^{\,\Gamma}(0,0;x;m)\,,\\
    f(\lambda;x;m)&=\bar{\spcorr}^{\,\Gamma}(\lambda,0;x;m)
    -\bar{\spcorr}^{\,\Gamma}(0,0;x;m)=\bar{\spcorr}^{\,\Gamma}(0,\lambda;x;m)
    -\bar{\spcorr}^{\,\Gamma}(0,0;x;m)\,,\\
    F^\Gamma(\lambda,\lambda';x;m) &=
    \bar{\spcorr}^{\,\Gamma}(\lambda,\lambda';x;m) - f(\lambda) -
    f(\lambda')\,,
  \end{aligned}
\end{equation}
where $c$ and $f$ do not depend on the choice of $\Gamma$ due to
Eq.~\eqref{eq:sym_rel_sp_corr}, and again it is assumed that the
limits of vanishing $\lambda$ or $\lambda'$ are approached at least as
fast as some power law. One then finds
\begin{equation}
  \label{eq:chi_contrib5_0}
  \begin{aligned}
    & \int_0^{\mu} d\lambda\int_0^{\mu}
    d\lambda'\,\f{(m^2+\lambda\lambda')
      \bar{\spcorr}^{\mathbf{1}}(\lambda,\lambda';x;m)+(m^2-\lambda\lambda')
      \bar{\spcorr}^{\gamma_5}(\lambda,\lambda';x;m)}{(\lambda^2
      +m^2)(\lambda^{\prime\,2} +m^2)} = I_0 + I_1 + I_+ + I_-\,,
\end{aligned}
\end{equation}
where
\begin{equation}
  \label{eq:chi_contrib5_1}
  \begin{aligned}
    I_0&= 2\,c(x;m)\left( \int_0^{\mu} d\lambda\,\f{ m}{\lambda^2
        +m^2}\right)^2
    =2\,\bar{\spcorr}^{\mathbf{1}}(0,0;x;m)\left(\arctan\f{\mu}{m}\right)^2
    \,,
\end{aligned}
\end{equation}
is obtained explicitly and is $O(1)$,
\begin{equation}
  \label{eq:chi_contrib5_2}
  \begin{aligned}
    I_1&= 4 \int_0^{\mu} d\lambda\,\f{ m}{\lambda^2 +m^2}\int_0^{\mu}
    d\lambda'\,\f{ m f(\lambda';x;m) }{\lambda^{\prime\,2} +m^2} =
    4\arctan\f{\mu}{m}\cdot o(1) = o(1)\,,
  \end{aligned}
\end{equation}
having used Eq.~\eqref{eq:bound1}
\begin{equation}
  \label{eq:chi_contrib5_3}
  \begin{aligned}
    I_+ &= \int_0^{\mu} d\lambda\int_0^{\mu} d\lambda'\,\f{ m^2\left(
        F^{\mathbf{1}}(\lambda,\lambda';x;m)+
        F^{\gamma_5}(\lambda,\lambda';x;m) \right)}{(\lambda^2
      +m^2)(\lambda^{\prime\,2} +m^2)} = o(1)\,,
  \end{aligned}
\end{equation}
having used Eq.~\eqref{eq:bound3}, and
\begin{equation}
  \label{eq:chi_contrib5_4}
  \begin{aligned}
    I_- &= \int_0^{\mu} d\lambda\int_0^{\mu} d\lambda'\,\f{
      \lambda\lambda'\left( F^{\mathbf{1}}(\lambda,\lambda';x;m)-
        F^{\gamma_5}(\lambda,\lambda';x;m) \right)}{(\lambda^2
      +m^2)(\lambda^{\prime\,2} +m^2)} = O(1)\,, 
  \end{aligned}
\end{equation}
since 
\begin{equation}
  \label{eq:chi_contrib6}
  F^{\mathbf{1}}(\lambda,\lambda';x;m)-
  F^{\gamma_5}(\lambda,\lambda';x;m)=
  \bar{\spcorr}^{\mathbf{1}}(\lambda,\lambda';x;m)-
  \bar{\spcorr}^{\gamma_5}(\lambda,\lambda';x;m)\,,
\end{equation}
and so the integral remains convergent if one sets $m=0$ in the
denominator. Also in this case there is no contribution to $\Po_*$.

\section{Renormalisation of the spectral correlators}
\label{sec:lc_ren}

In this appendix I show that the renormalised spectral correlators
[see Eqs.~\eqref{eq:spcorr_1} and \eqref{eq:spec_corr_reno2}],
\begin{equation}
  \label{eq:ssv1}
  C^{\Gamma}(\lambda,\lambda';x;m)\equiv \lim_{\Lambda\to\infty}\lim_{V\to\infty} Z_m^2
  C_{\Lambda,\,V}^{\Gamma}(Z_m\lambda,Z_m\lambda';x;Z_m m)    \,,
  \qquad \Gamma=\mathbf{1},\gamma_5\,, 
\end{equation}
are finite functions of $\lambda$, $\lambda'$, and of the renormalised
mass $m$ (and of $x$).  The proof follows closely the strategy of
Refs.~\cite{DelDebbio:2005qa,Giusti:2008vb}, and relies on the
renormalisation properties of the so-called density chain correlation
functions. As mentioned in Appendix \ref{sec:app_renorm},
renormalisation properties at finite temperature are identical to
those of the zero-temperature theory.

\paragraph{Density chain correlation functions}

In a partially quenched gauge theory with $2\bar{N}_f$ extra pairs of
``valence'' fermion fields $\psi_i,\bar{\psi}_i$,
$i=1,\ldots,2\bar{N}_f$, and corresponding $2\bar{N}_f$ pseudofermion
fields exactly cancelling out their contribution to the fermionic
determinant, all with the same mass as the original $N_f$ fermions,
one defines the (bare) density operators
\begin{equation}
  \label{eq:ssv2}
  X_{ij\,B}^{\Gamma} \equiv   \bar{\psi}_i(x)\Gamma\psi_j(x)\,,  \qquad
  \Gamma=\mathbf{1},\gamma_5\,,\quad i,j=1,\ldots, 2\bar{N}_f\,.
\end{equation}
These composite fields renormalise in the standard way, i.e., for
$i\neq j$,
\begin{equation}
  \label{eq:ssv3}
  X_{ij\,B}^\Gamma(x)  = Z_\Gamma X_{ij}(x) \,,
\end{equation}
with flavour-independent renormalisation constants $Z_\Gamma$ that can
be taken equal to those obtained in the $\bar{N}_f=0$
case~\cite{Bernard:1993sv}, i.e., $Z_{\mathbf{1}}=Z_S$ and
$Z_{\gamma_5}=Z_P$. In a regularisation that preserves (some form of)
chiral symmetry, one further has $Z_S = Z_P = Z_m^{-1}$, so
$Z_\Gamma = Z_m^{-1}$. As shown in Refs.~\cite{Luscher:2004fu,
  DelDebbio:2005qa,Giusti:2008vb,Cichy:2014yca}, the density-chain
correlation functions,
\begin{equation}
  \label{eq:ssv4}
  \Xsc^{\Gamma_1\ldots\Gamma_n}(x_1,\ldots,x_{n-1})\equiv  \la
  X^{\Gamma_1}_{n 1\,B}(x_1)X^{\Gamma_2}_{12\,B}(x_2)\ldots
  X^{\Gamma_{n-1}}_{n-2\, n-1\,B}(x_{n-1})
  X^{\Gamma_n}_{n-1\,n\,B}(0)\ra_\beta\,, 
\end{equation}
renormalise multiplicatively, i.e.,
\begin{equation}
  \label{eq:ssv5}
  \begin{aligned}
    Z_{\Gamma_1}^{-1}\ldots Z_{\Gamma_n}^{-1}
    \Xsc^{\Gamma_1\ldots\Gamma_n}(x_1,\ldots,x_{n-1})
     =
    (Z_m)^n \Xsc^{\Gamma_1\ldots\Gamma_n}(x_1,\ldots,x_{n-1}) 
   \end{aligned}
\end{equation}
is a renormalised quantity after the usual mass and coupling
renormalisation.  Moreover, their short-distance singularities when
the $x_i$ get close to each other or to 0 are integrable if $n\ge 5$.

\paragraph{Spectral correlators from density chain correlation
  functions}

The relevant density-chain correlation functions for the problem at
hand are the bare quantities
\begin{equation}
  \label{eq:ssv6}
  \begin{aligned}
    \Mc_{\ell_1\ell_2\,B}^\Gamma(x;m_B) &\equiv \int d^4x_1\ldots \int
    d^4x_{2(\ell_1+\ell_2)+1} \,\delta^{(4)}(x-x_{2\ell_2+1})\\ &
    \phantom{\equiv \int d^4x_1\ldots \int d^4x_{2(\ell_1+\ell_2)+1}}
    \times\Xsc^{\overbrace{\scriptstyle \gamma_5\ldots
        \gamma_5}^{2\ell_2} \Gamma
      \overbrace{\scriptstyle\gamma_5\ldots\gamma_5}^{2\ell_1}
      \Gamma}(x_1,\ldots, x_{2(\ell_1+\ell_2)+1})\,,
  \end{aligned}
\end{equation}
defined in a finite volume and in the UV-regularised theory, and their
renormalised counterparts $\Mc_{\ell_1\ell_2}^\Gamma$,
\begin{equation}
  \label{eq:ssv7}
  \Mc_{\ell_1\ell_2}^\Gamma(x;m)=
  \lim_{\Lambda\to\infty}\lim_{V\to\infty}  Z_m^{2(\ell_1+\ell_2+1)}
  \Mc_{\ell_1\ell_2\,B}^\Gamma(x;Z_m m) \,. 
\end{equation}
In a finite volume the Dirac spectrum is discrete, and one can use the
decomposition of the quark propagator in Dirac eigenmodes to get with
a straightforward calculation\footnote{The exchange of the order of
  the various integrals has been justified in Section
  \ref{sec:anomalous_rem}.}
\begin{equation}
  \label{eq:ssv8}
  \Mc_{\ell_1\ell_2\,B}^\Gamma(x;m_B)     = - \int_{-\Lambda}^\Lambda
  d\lambda
  \int_{-\Lambda}^\Lambda d\lambda'
  \f{(m_B^2-\lambda\lambda')C_{V,\,\Lambda}^{\Gamma}(\lambda,\lambda';x;m_B)}{(\lambda^2+m_B^2)^{1+\ell_1}  
    (\lambda^{\prime\,2}+m_B^2)^{1+\ell_2}}    
  \,.
\end{equation}
Here $C_{V,\,\Lambda}^{\Gamma}$ is computed in the partially quenched
theory with $\bar{N}_f=\ell_1+\ell_2+1$, but since it is obtained by
averaging the Dirac spectrum over gauge configurations only, and since
the weight of a configuration is independent of $\bar{N}_f$ due to
partial quenching, one obtains the same result as in the standard
($\bar{N}_f=0$) theory. One can then vary the number of extra valence
quarks as demanded by the left-hand side of Eq.~\eqref{eq:ssv8} in
order to freely vary $\ell_{1,2}$, without changing
$C_{V,\,\Lambda}^{\Gamma}$ appearing on the right-hand side.

Next, one defines the resolvents,
\begin{equation}
  \label{eq:ssv11}
  R^\Gamma_{r\,B}(z,z';x;m_B) \equiv \int d\lambda \int  d\lambda'\,
  \f{(m_B^2-\lambda\lambda')C_{V,\,\Lambda}^{\Gamma}(\lambda,\lambda';x;m_B)}{(\lambda^2+m_B^2)^r
    (\lambda^{\prime\,2}+m_B^2)^r}\f{1}{\lambda^2+m_B^2-z}\f{1}{\lambda^{\prime\,2}+m_B^2-z'}
  \,.
\end{equation}
These functions are analytic in $z$ and $z'$ in the entire complex
plane, except for cuts at $|\Re z|,|\Re z'|\ge m_B$.  The integer $r$
may have to be set to a nonzero value to guarantee convergence at
large $\lambda,\lambda'$, and to avoid non-integrable short distance
singularities.  It is simple to show that
$C_{V,\Lambda}^{\Gamma}(\lambda,\lambda';x;m_B)$ can be recovered from
the following discontinuity of the resolvents,
\begin{equation}
  \label{eq:ssv12}
  \Ds_{r\,B}^\Gamma(\lambda,\lambda';x;m_B) 
  \equiv \lim_{\epsilon\to
    0}\phantom{+}\sum_{\sigma,\sigma'=\pm 1}
  \sigma\sigma'  R^\Gamma_{r\,B}(m_B^2+\lambda^2+i\sigma\epsilon,m_B^2
  +\lambda^{\prime\,2}+i\sigma'\epsilon;x;m_B)\,.
\end{equation}
One has explicitly for $\lambda\lambda'\ge 0$
\begin{equation}
  \label{eq:ssv13}
  \begin{aligned}
    C_{V,\,\Lambda}^{\Gamma}(\lambda,\lambda';x;m_B) & =
    F^{-}_r(\lambda,\lambda';x;m_B)
    \Ds^{\Gamma}(\lambda,\lambda';x;m_B) \\ & \phantom{=}
    +F^{+}_r(\lambda,\lambda';x;m_B)
    \Ds^{\gamma_5\Gamma}(\lambda,\lambda';x;m_B) \,,
  \end{aligned}
\end{equation}
with
\begin{equation}
  \label{eq:ssv14}
  F_r^\pm(\lambda,\lambda';x;m_B)=
  -\f{1}{8\pi^2m_B^2}\left(\lambda^2+m_B^2\right)^r
  \left(\lambda^{\prime\,2}+m_B^2\right)^r\left(\lambda\lambda'\pm
    m_B^2\right)\,.
\end{equation}
The case $\lambda\lambda'< 0$ is obtained using the symmetry property
$C_{V,\,\Lambda}^{\Gamma}(-\lambda,\lambda') =
C_{V,\,\Lambda}^{\gamma_5\Gamma}(\lambda,\lambda')$,
Eq.~\eqref{eq:sym_rel_sp_corr}. Expanding the resolvent in powers of
$z,z'$ one finds
\begin{equation}
  \label{eq:ssv16}
  R^\Gamma_{r\,B}(z,z';x;m_B) 
  = \sum_{\ell,\ell'=0}^\infty z^\ell z^{\prime\,\ell'}\Mc_{r+\ell\,\,r+\ell'\,B}^\Gamma(x;m_B)\,,
\end{equation}
and using Eq.~\eqref{eq:ssv7} one finds that the quantity
\begin{equation}
  \label{eq:ssv18}
  \begin{aligned}
    R^\Gamma_{r}(\zeta,\zeta';x;m) &\equiv \lim_{\Lambda\to\infty}
    \lim_{V\to\infty} Z_m^{2(2r+1)}
    R^\Gamma_{r\,B}(Z_m^2\zeta,Z_m^2\zeta';x;Z_m m) \\
    &=\sum_{\ell,\ell'=0} \zeta^\ell \zeta^{\prime\,\ell'}
    \Mc_{r+\ell\,\,r+\ell'}^\Gamma(x;m)
  \end{aligned}
\end{equation}
is a finite function of $\zeta$, $\zeta'$, and $m$.  Since the number
of density operators $2(2r + \ell + \ell' + 1)\ge 4r +1$, $r=1$
suffices to ensure the absence of non-integrable short-distance
singularities. Equation \eqref{eq:ssv1} then follows from
Eqs.~\eqref{eq:ssv12}--\eqref{eq:ssv14} and \eqref{eq:ssv18}.

\paragraph{Renormalisation of the mobility edge}

In the mode-sum representation of $C^{\Gamma}$, it is convenient to
separate the contribution from $\lambda_n=\pm\lambda_{n'}$ from the
rest and write\footnote{Accidental degeneracies of eigenvalues can be
  ignored. See footnotes \ref{foot:degmodes0} and
  \ref{foot:degmodes}.} [see Eqs.~\eqref{eq:spcorr_more_definitions}
and \eqref{eq:spcorr_1_reprise}]
\begin{equation}
  \label{eq:ssv20}
  \begin{aligned}
    C_{V,\,\Lambda}^{\Gamma}(\lambda,\lambda';x;m_B) &=
    \delta(\lambda-\lambda')
    C_{s\,V,\,\Lambda}^{\Gamma}(\lambda;x;m_B) +
    \delta(\lambda+\lambda')
    C_{s\,V,\,\Lambda}^{\gamma_5\Gamma}(\lambda;x;m_B) \\ &\phantom{=}+
    \bar{C}_{V,\,\Lambda}^{\Gamma}(\lambda,\lambda';x;m_B)\,,
  \end{aligned}
\end{equation}
with
\begin{equation}
  \label{eq:ssv20_bis}
  \begin{aligned}
    C_{s\,V,\,\Lambda}^{\Gamma}(\lambda;x;m_B) &\equiv \bigg\la
    \sum_{n} \delta(\lambda-\lambda_n) \Oc_{nn}^{\Gamma}(x)
    \Oc_{nn}^{\Gamma}(0)    \bigg\ra_{\beta}\,,\\
    \bar{C}_{V,\,\Lambda}^{\Gamma}(\lambda,\lambda';x;m_B) &\equiv
    \bigg\la \sum_{\substack{n,n'\\ n\neq \pm n'}}
    \delta(\lambda-\lambda_n)\delta(\lambda'-\lambda_{n'})
    \Oc_{n'n}^{\Gamma}(x) \Oc_{nn'}^{\Gamma}(0) \bigg\ra_{\beta}\,,
  \end{aligned}
\end{equation}
with $\bar{C}_{V,\,\Lambda}^{\Gamma}$ regular as
$\lambda\to\pm\lambda'$. Using Eq.~\eqref{eq:ssv1} one finds
\begin{equation}
  \label{eq:ssv21}
  \begin{aligned}
    C^{\Gamma}(\lambda,\lambda';x;m) &= \lim_{\Lambda\to\infty}
    \lim_{V\to\infty}\Big\{ \delta(\lambda-\lambda') Z_m
    C^{\Gamma}_{s\,V,\,\Lambda}(Z_m\lambda;x;Z_mm) \\
    &\phantom{=\lim_{\Lambda\to\infty} \lim_{V\to\infty}\Big\{ }+
    \delta(\lambda+\lambda') Z_m
    C^{\gamma_5\Gamma}_{s\,V,\,\Lambda}(Z_m\lambda;x;Z_mm) \\
    &\phantom{=\lim_{\Lambda\to\infty} \lim_{V\to\infty}\Big\{ } +
    Z_m^2\bar{C}_{V,\,\Lambda}^{\Gamma}(Z_m\lambda,Z_m\lambda';x;Z_mm)\Big\}\\
    &\equiv \delta(\lambda-\lambda')
    C^{\Gamma}_{\mathrm{loc}}(\lambda;x;m) + \delta(\lambda+\lambda')
    C^{\gamma_5\Gamma}_{\mathrm{loc}}(\lambda;x;m) +
    \bar{C}^{\Gamma}(\lambda,\lambda';x;m)\,,
   \end{aligned}
\end{equation}
with each term on the right-hand side separately finite due their
different degree of singularity for $\lambda\to \pm\lambda'$.

In Eq.~\eqref{eq:ssv21} the results of Section \ref{sec:pplargev} have
been used, that show that $C^{\Gamma}_{{\rm loc}}$ has support only in
spectral regions where modes are localised.  These are separated by
mobility edges $\lambda^{(i)}_{c\,B}$ ($i=1,\ldots$) from regions
where modes are delocalised.  The renormalisation properties of
$C^{\Gamma}_{\mathrm{loc}}$ then imply that
$ \lambda^{(i)}_{c} \equiv Z_m^{-1}\lambda^{(i)}_{c\,B}$ are finite,
renormalised quantities.  Formally, the unrenormalised spectral
correlator in infinite volume,
$\spcorr_{{\rm loc}\,\Lambda}^\Gamma(\lambda;x;m_B)$, reads
\begin{equation}
  \label{eq:lc_reno_1_app}
  \spcorr_{{\rm loc}\,\Lambda}^\Gamma(\lambda;x;m_B) = \sum_{i}
  \chi_{I_B^{(i)}}(\lambda) f^\Gamma_{(i)\,\Lambda}(\lambda;x;m_B)\,,
\end{equation}
with $\chi_{I_B^{(i)}}$ the characteristic functions of the disjoint
spectral regions
$I_B^{(i)}=[\lambda_{c\,B}^{(2i+1)},\lambda_{c\,B}^{(2i+2)}]$
($i=0,\ldots$) where modes are localised, delimited by the lower and
upper mobility edges $\lambda_{c\,B}^{(2i+1)}$ and
$\lambda_{c\,B}^{(2i+2)}$. After renormalisation
\begin{equation}
  \label{eq:lc_reno_2_app}
  \begin{aligned}
    \spcorr_{{\rm loc}}^\Gamma(\lambda;x;m) &= \lim_{\Lambda\to\infty}
    \sum_{i} \chi_{I_B^{(i)}}( Z_m\lambda) Z_m
    f^\Gamma_{(i)\,\Lambda}( Z_m\lambda;x;
    Z_mm)\\
    &= \lim_{\Lambda\to\infty} \sum_{i} \chi_{Z_m^{-1}
      I_B^{(i)}}(\lambda) Z_m f^\Gamma_{(i)\,\Lambda}( Z_m\lambda;x;
    Z_mm)\,,
  \end{aligned}
\end{equation}
and finiteness of the left-hand side requires that the quantities
\begin{equation}
  \label{eq:lc_reno_3_app}
   f^\Gamma_{(i)}(\lambda;x;m)=\lim_{\Lambda\to\infty}Z_m
f^\Gamma_{(i)\,\Lambda}(Z_m\lambda;x;Z_mm)
\end{equation}
are finite for each $i$, since they have disjoint support.  More
importantly, it also implies that the renormalised spectral regions
\begin{equation}
  \label{eq:lc_reno_4_app}
  I^{(i)}\equiv [\lambda_{c}^{(2i+1)},\lambda_{c}^{(2i+2)}] =
  \lim_{\Lambda\to\infty} Z_m^{-1} I_B^{(i)} =
  \lim_{\Lambda\to\infty}[Z_m^{-1}\lambda_{c\,B}^{(2i+1)},Z_m^{-1}\lambda_{c\,B}^{(2i+2)}]
\end{equation}
are delimited by finite renormalised mobility edges
$\lambda_{c}^{(i)}\equiv \lim_{\Lambda\to\infty}
Z_m^{-1}\lambda_{c\,B}^{(i)}$.  In other words, the mobility edges
renormalise like the fermion mass, so that $\lambda_{c\,B}^{(i)}/m_B$
are re\-nor\-mal\-i\-sa\-tion-group-invariant quantities, up to terms
that vanish as $\Lambda\to\infty$.

\section{Bound for localised modes}
\label{sec:app_bound}

In this Appendix I argue that the $1/m$-divergent part of the
pseudoscalar-pseudoscalar correlator, that originates from localised
near-zero modes and leads to a finite $\Poc_*$ in chiral limit, is
expected to have fast decay properties at large spatial distance, that
allow to interchange spacetime integration and chiral limit.  Since
\begin{equation}
  \label{eq:cgamma5_bound}
    \begin{aligned}
      & \left| \int_0^{\lambda_c} d\lambda\,\left(
          \f{(m^2-\lambda^2)\spcorr_{{\rm
                loc}}^{\gamma_5}(\lambda;x;m)}{(\lambda^2
            +m^2)^2}\right) \right| \le \int_0^{\lambda_c} d\lambda\,
      \f{\spcorr_{{\rm loc}}^{\mathbf{1}}(\lambda;x;m)}{\lambda^2
        +m^2} \,,
  \end{aligned}
\end{equation}
it suffices to check the contribution of the spectral correlator
$\spcorr_{{\rm loc}}^{\mathbf{1}}$.  It is reasonable to assume that
it decays exponentially with the spatial distance, up to power-law
corrections that do not affect the following argument (see Appendix
\ref{sec:expoenv}).  Since the temporal direction is compact, one
could maximise the right-hand side of Eq.~\eqref{eq:cgamma5_bound}
over time. However, since I am only interested in justifying the
exchange of chiral limit and spacetime integration, and integration
over time causes no problem there, it suffices and is practically more
convenient to integrate over time. One then considers
\begin{equation}
  \label{eq:fast_decay_def}
  {\rm C}_{\rm loc}(\vec{x};m)  \equiv
  \int_0^{\lambda_c}d\lambda\, \f{2m}{\lambda^2 + m^2}\,{\rm L}_{\rm loc}(\lambda;\vec{x};m) \,,\qquad
  {\rm L}_{\rm loc}(\lambda;\vec{x};m)\equiv
  \int_0^\beta dt \,\spcorr_{{\rm
      loc}}^{\mathbf{1}}(\lambda;x;m)\,,
\end{equation}
from which ${\rm R}_*(0)$ is obtained as follows,
\begin{equation}
  \label{eq:fast_decay_def_again}
{\rm R}_*(0) = \int d^3x\,\lim_{m\to 0}  {\rm C}_{\rm loc}(\vec{x};m)  \,.
\end{equation}
I now assume that ${\rm L}_{\rm loc}$ takes the following form,
\begin{equation}
  \label{eq:fast_decay2}
  {\rm L}_{\rm loc}(\lambda;\vec{x};m)  
= A(\lambda)e^{-\f{|\vec{x}|}{\xi(\lambda)}}\,,
\end{equation}
where $\xi(\lambda)$ is a $\lambda$-dependent (and possibly
$m$-dependent) correlation length. As discussed in Section
\ref{sec:anomalous_rem}, interchanging spacetime integration and the
various limits is expected to be justified at finite fermion mass, and
in this case using Eq.~\eqref{eq:ano_rem2_noch} one finds that
\begin{equation}
  \label{eq:fast_decay4}
  \rho(\lambda,m) = \int_\beta d^4x\, \spcorr_{{\rm loc}}^{\mathbf{1}}(\lambda;x;m)
  = \int d^3x \,   {\rm L}_{\rm loc}(\lambda;\vec{x};m)  
  =A(\lambda) 8\pi \xi(\lambda)^3\,,
\end{equation}
and so
\begin{equation}
  \label{eq:fast_decay5}
  {\rm C}_{\rm loc}(\vec{x};m) = \int_0^{\lambda_c}d\lambda\,\rho(\lambda,m)
  \f{2m}{\lambda^2 + m^2}\,\f{1}{8\pi\xi(\lambda)^3}e^{-\f{|\vec{x}|}{\xi(\lambda)}}\,.
\end{equation}
Equation \eqref{eq:fast_decay5} is expected to provide at least a
qualitative understanding of how the localisation properties of the
modes are transferred to the correlator.

Further assumptions are needed to proceed.  I consider first the case
of a maximal correlation length. Since
\begin{equation}
  \label{eq:fast_decay_ximax}
  \f{\de}{\de\xi}
  \f{1}{\xi^3}e^{-\f{|\vec{x}|}{\xi}} =
  \f{1}{\xi^4}e^{-\f{|\vec{x}|}{\xi}}\left(\f{|\vec{x}|}{\xi}-3\right)\,, 
\end{equation}
if $\xi(\lambda)\le \xi_{\rm max}$ for $\lambda\in[0,\lambda_c]$ (at
least for sufficiently small $m$), then for
$|\vec{x}|\ge 3\xi_{\rm max}$ one has
\begin{equation}
  \label{eq:fast_decay5_ximax}
  {\rm C}_{\rm loc}(\vec{x};m) \le
\f{1}{8\pi\xi_{\rm max}^3}e^{-\f{|\vec{x}|}{\xi_{\rm max}}}
  \int_0^{\lambda_c}d\lambda\,\rho(\lambda,m)
  \f{2m}{\lambda^2 + m^2}\mathop{\to}_{m\to 0}
\f{\rho(0;0)}{4\pi\xi_{\rm max}^3}e^{-\f{|\vec{x}|}{\xi_{\rm max}}}
\arctan\tf{\lambda_c}{m}
  \,.
\end{equation}
In this case, the fast decay of Eq.~\eqref{eq:fast_decay5_ximax}
allows one to use the dominated convergence theorem to justify
interchanging spacetime integration and chiral limit.

On the other hand, since the localisation length diverges at the
mobility edge~\cite{thouless1974electrons,lee1985disordered,
  kramer1993localization,Evers:2008zz,anderson50,manybody}, one could
find that also $\xi(\lambda)$ diverges. As an alternative
possibility, I now assume that
\begin{equation}
  \label{eq:fast_decay3}
  \xi(\lambda)=\xi_0 \left|1-\tf{\lambda}{\lambda_c}\right|^{-\nu}  \,,
\end{equation}
with $\nu$ a suitable exponent, which could be, e.g., the localisation
length critical exponent appropriate for the symmetry class of the
Dirac operator for the given gauge group~\cite{Verbaarschot:2000dy}. I
also assume that $\xi_{0}$ remains finite in the chiral limit: in this
way, even if $\lambda_c\to 0$, it still makes sense to speak of
localised modes at the origin. Notice that while the localisation
length of the localised modes $\psi_n$ diverges at $\lambda_c$, the
correlation length of
$\la \Vert\psi_n(x)\Vert^2 \Vert\psi_n(0)\Vert^2\ra_\beta$ could
remain finite, due to other long-distance effects related to the
averaging over gauge fields; and if it diverges, an algebraic decay is
expected at criticality~\cite{Evers:2008zz}. Using
Eq.~\eqref{eq:fast_decay3} should then lead at least to an upper bound
on the large-distance behaviour of ${\rm C}_{\rm loc}$.\footnote{It
  should also be noted that the localisation length, characterising
  the exponential fall-off of localised modes and diverging at the
  Anderson transition, is in general different from the typical size
  of a localised mode (see, e.g., Ref.~\cite{kramer1993localization}),
  and so even assuming that Eq.~\eqref{eq:fast_decay3} holds, it is
  not clear what exponent one should use. However, this does not
  affect the qualitative features of the following argument.}

Working under the assumption Eq.~\eqref{eq:fast_decay3}, if
$\lambda_c$ remains finite or vanishes more slowly than $m$, so that
$\lambda_c/m\to\infty$ as $m\to 0$, one can change variables to
$\lambda=m z$ and write
\begin{equation}
  \label{eq:fast_decay6}
  \begin{aligned}
    {\rm C}_{\rm loc}(\vec{x};m) & =
    \f{1}{4\pi\xi_0^3}\int_0^{\f{\lambda_c}{m}}dz\,\rho(mz ,m)
    \f{\left|1-\f{m}{\lambda_c}z\right|^{\nu}}{z^2 + 1}\,
    e^{-\f{|\vec{x}|}{\xi_0}\left|1-\f{m}{\lambda_c}z\right|^{\nu}}
    \mathop\to_{m\to 0} \f{\rho(0
      ,0)}{8\xi_0^3}e^{-\f{|\vec{x}|}{\xi_0}}\,,
  \end{aligned}
\end{equation}
and so ${\rm C}_{\rm loc}$ can again be bounded exponentially in
$|\vec{x}|$, uniformly in $m$ for sufficiently small $m$, and
dominated convergence allows one to exchange chiral limit and
spacetime integration.  On the other hand, setting
$\chi=(|\vec{x}|/\xi_0)^{1/\nu}$, one can generally change variables
to $\lambda = \lambda_c(1-z/\chi)$ to write
\begin{equation}
  \label{eq:fast_decay7}
  \begin{aligned}
    {\rm C}_{\rm loc}(\vec{x};m) &=
    \f{\lambda_c}{m}\f{1}{4\pi\xi_0^3}{\left(\f{\xi_0}{|\vec{x}|}\right)^{3+\f{1}{\nu}}}
    \int_0^{\chi}dz\, \f{\rho\left(\lambda_c
        \left(1-\tf{z}{\chi}\right),m\right)}{1+
      \f{\lambda_c^2}{m^2}\left(1-\tf{z}{\chi}\right)^2}z^{3\nu}e^{-z^\nu}\\
    &\le \f{\lambda_c}{m}\f{\rho_{\rm
        max}}{4\pi|\vec{x}|^3}\left(\f{\xi_0}{|\vec{x}|}\right)^{\f{1}{\nu}}
    \int_0^{\chi}dz\,z^{3\nu}e^{-z^\nu} \le
    \f{\lambda_c}{m}\f{\rho_{\rm
        max}}{4\pi|\vec{x}|^3}\left(\f{\xi_0}{|\vec{x}|}\right)^{\f{1}{\nu}}
    \f{1}{\nu}\Gamma\left(3+\f{1}{\nu}\right)\,,
  \end{aligned}
\end{equation}
where $\rho_{\rm max}$ is a bound on the mode density in the spectral
region of localised modes for small mass.  The bound is integrable (at
large distances) in three spatial dimensions, and so, if $\lambda_c/m$
does not diverge, it can be used to invoke dominated convergence.

While only qualitative, the estimates above are quite robust under
refinements. For example, adding power corrections $(|\vec{x}|/\xi)^k$
to the exponential behaviour of the spectral correlator only modifies
the large-$|\vec{x}|$ behaviour of ${\rm C}_{\rm loc}$ by similar
power corrections $(|\vec{x}|/\xi_0)^k$ in
Eq.~\eqref{eq:fast_decay5_ximax} (at sufficiently large distance) and
in Eq.~\eqref{eq:fast_decay6}.  In Eq.~\eqref{eq:fast_decay7},
instead, it only modifies the integrand in the last passage, and so
the numerical prefactor, but not the large-$|\vec{x}|$ behaviour of
${\rm C}_{\rm loc}$. The exchange of chiral limit and spacetime
integration seems then justified in the calculation of the anomalous
remnant ${\rm R}_*(0)$ originating from localised near-zero modes.

It is worth commenting on how Eqs.~\eqref{eq:fast_decay6} and
\eqref{eq:fast_decay7} compare with the scenarios discussed in Section
\ref{sec:ppchlim}.  Equation \eqref{eq:fast_decay7} leads one to
expect that if $\lambda_c/m$ vanishes in the chiral limit, so does
${\rm C}_{\rm loc}$. One would then expect no anomalous remnant, in
agreement with what was stated above in Section \ref{sec:ppchlim} for
scenario (i).  Moreover, Eq.~\eqref{eq:fast_decay7} shows that if
$\lambda_c/m$ vanishes faster than $m$, then one expects the
pseudoscalar correlator to be regular in the chiral limit, as in
scenario (i-a), while if it vanishes more slowly than $m$ one expects
a divergent correlator, as in scenario (i-b).  If $\lambda_c/m$
remains constant in the chiral limit, corresponding to scenario (ii),
then Eq.~\eqref{eq:fast_decay7} still allows to bound
${\rm C}_{\rm loc}$ with an algebraic but integrable decay
$|\vec{x}|^{-3-\f{1}{\nu}}$ instead of an exponential one in the
chiral limit.  This may actually be the qualitative behaviour of the
correlator in this case, and so a distinguishing feature for scenario
(ii) as compared to the standard scenario, if the correlation length
diverges at the mobility edge. This is clearly speculative at this
stage, and a better understanding of the spectral correlators is
required to make a definite statement. Finally, scenario (iii) is
covered by Eq.~\eqref{eq:fast_decay6}, showing that an exponential
bound should be expected.

\bibliographystyle{JHEP}
\bibliography{references_gt}

\end{document}